\documentclass[11pt,onecolumn,nofootinbib,aps,superscriptaddress,prd]{revtex4}
\usepackage{graphicx,dcolumn,bm,amsfonts,amsmath,color,xcolor,epsfig}
\usepackage{slashed}
\usepackage{graphics}
\usepackage{tikz}
\usepackage{multirow}
\usepackage{amsmath}
\usepackage{gensymb}
\usepackage{epstopdf}
\usepackage{amssymb}
\usepackage[colorlinks,
            citecolor=blue,
            anchorcolor=blue,
            menucolor=blue,
            linkcolor=blue,
            filecolor=blue,
            runcolor=blue,
            urlcolor=blue,
            frenchlinks=blue]{hyperref}
\allowdisplaybreaks

\begin{document}

\title{Study of the mass spectra of doubly heavy $\Xi_{QQ^{\prime}}$ and $\Omega_{QQ^{\prime}}$ baryons}

\author{Ji-Hai Pan}
\email{Tunmnwnu@outlook.com}
\address{Department of Mathematics and Physics, Liuzhou Institute of Technology, Liuzhou 545616, China}
\author{Ji-Si Pan\footnote{Corresponding author}}
\email{panjisi@gxstnu.edu.cn}
\address{School of Physics and Information Engineering, Guangxi Science $\&$ Technology Normal University,
Laibin 546199, China}

\begin{abstract}
LHCb Collaboration first observed a doubly charmed baryon $\Xi^{++}_{cc}$ in the $\Lambda^{+}_{c}K^{-}\pi^{+}\pi^{+}$ decay with a mass of $3621.40\pm0.72\pm0.27$ MeV. In this paper, we enumerated the mass spectra of the radial and orbital excited states for the doubly heavy $\Xi_{QQ^{\prime}}$ and $\Omega_{QQ^{\prime}}$ baryons using the Regge trajectory model and the scaling rules. Our studies suggest that $\Xi^{++}_{cc}$ can be grouped into the $1S$-wave state with the spin-parity quantum number $J^{P} = 1/2^{+}$. On the other hand, the mass of $\Xi_{cc}$ state with $J^{P} = 3/2^{+}$ is predicted to be $3699.69 \pm 4.59$ MeV. We also predict the mass spectra of the unknown ground and excited states for the doubly heavy $\Xi_{QQ^{\prime}}$ and $\Omega_{QQ^{\prime}}$ baryons, which provide useful references for the experimental test in the future.

\end{abstract}

\maketitle

\section{introduction}

During the past two decades, the study of the spectroscopy of heavy baryons is an interesting and important issue in hadronic physics. It enables us to gain a deeper understanding of the internal structure and strong interactions of hadrons. Doubly heavy baryon is composed of one light quark ($q$ = $u$, $d$, or $s$), and two heavy quarks ($QQ^{\prime}$ = $cc$, $bc$, and $bb$), that the two heavy quarks can be defined as the heavy diquark. The doubly heavy baryons have $\Xi_{QQ^{\prime}}$ and $\Omega_{QQ^{\prime}}$ family. The $\Xi_{QQ^{\prime}}$ family has up quark ($u$) or down quark ($d$), namely, $\Xi_{cc}$, $\Xi_{bc}$, and $\Xi_{bb}$ baryons, while the $\Omega_{QQ^{\prime}}$ family has one strange quark ($s$), namely, $\Omega_{cc}$, $\Omega_{bc}$, and $\Omega_{bb}$ baryons. As a component of hadron, it has been explored by various methods and aroused wide interest.

In 2017, LHCb Collaboration has observed a doubly charmed baryon $\Xi^{++}_{cc}$ in the $\Lambda^{+}_{c}K^{-}\pi^{+}\pi^{+}$ mass spectra \cite{Aaije:A11}, the measured mass is $3621.40\pm0.72\pm0.27$ MeV. Later, the LHCb Collaboration confirmed the existence of this state in the $\Xi^{+}_{c}\pi^{+}$ decay \cite{Aaije:AA11}. In 2020, precision measurement of the $\Xi^{++}_{cc}$ mass is given by $3621.55 \pm 0.23 \pm 0.30$ MeV \cite{Aaije:A111}. However, the spin-parity quantum number $J^{P}$ of the $\Xi^{++}_{cc}$ state have not been determined. Our calculation indicated that can be regarded as a $1S$ state with $J^{P} = 1/2^{+}$.

In 2002, the SELEX collaboration reported the first observation of a candidate of a double charmed baryon state $\Xi^{+}_{cc}$ in the charged decay mode $\Xi^{+}_{cc}$ $\rightarrow$ $\Lambda^{+}_{c}K^{-}\pi^{+}$ \cite{Mattson:A11}. The observed mass of $\Xi^{+}_{cc}$ state is $3519 \pm 1$ MeV. In 2005, the SELEX collaboration reported again the doubly charmed baryon $\Xi^{+}_{cc}$ in the charged decay mode $\Xi^{+}_{cc}$ $\rightarrow$ $pD^{+}K^{-}$ \cite{Ocherashvili:A11}, but could not be confirmed by Belle, $BABAR$, CLEO, and LHCb collaborations \cite{Science:A11}. However, our knowledge of the ground and higher excited states for the doubly heavy $\Xi_{QQ^{\prime}}$ and $\Omega_{QQ^{\prime}}$ baryons remain unclear.

The properties of doubly heavy baryons have been extensively studied using a variety of approaches, such as the relativistic quark model \cite{Ebert:A11, Zhong:A11}, nonrelativistic quark model \cite{Karlinere:A11, Albertu:A11, GhalenoviS:A11}, effective field theory \cite{PSoto:A11}, hypercentral constituent quark model (hCQM) \cite{Shah:A11, Mohajery:A11, Salehi:A11}, QCD sum rule \cite{ZhangH:A11, ZGWang:A11, AlievB:A11, ShekariA:A11}, relativistic quark-diquark picture \cite{EFG:C10}, compact diquark model \cite{Ali:PP888}, lattice QCD \cite{ZBrown:A11, MPadmanath:A11, Lewis:A11, MathurM:A11}, bag model \cite{HeQian:A11}, Salpeter model \cite{Giannuzzi:A11}, and Faddeev method \cite{ValcarceG:A11}. In Ref. \cite{Pacheco:A11}, the $\Xi^{++}_{cc}$ was assigned as the ground state in the framework of a non-relativistic harmonic oscillator quark model. In the framework of Regge phenomenology \cite{Oudichhyahr:A11}, the author used the relations between slope ratios, intercepts, and baryon masses to calculate the ground and excited state masses of doubly heavy $\Xi_{cc}$, $\Xi_{bc}$, $\Omega_{cc}$, and $\Omega_{bc}$ baryons. They confirmed the spin parity of $\Xi^{++}_{cc}$ with $J^{P} = 1/2^{+}$. For more extra references, see recent review Refs. \cite{WengDeng:A11, Shahr:A11, LiYuWang:A11, ChenLuo:H11}.

In this paper, we use the Regge trajectory to complete the spin-average masses of the $\Xi_{QQ^{\prime}}$ and $\Omega_{QQ^{\prime}}$ baryons. We also study the effective masses of the heavy diquark and light quark with the kinetic energy. In addition, to obtain the mass shifts, we exploit the scaling relations to calculate the spin-coupling parameters. For the unknown doubly heavy baryons in experiments, the mass spectra of the radial excited and orbital excited states are predicted.

The structure of this paper is as follows. In Sec. II, we study the effective masses of heavy diquark and light quark. We analyze the spin-average masses of the doubly heavy baryons in Sec. III. We use a all-$JLS$ coupling to compute the mass spectra expressions of doubly heavy baryons in Sec. IV. We discuss the scaling relations in Sec. V. In Sec. VI, we calculate the mass spectra of the $\Xi_{QQ^{\prime}}$ baryons. In Sec. VII, a similar mass analysis is given by the scaling relations for the $\Omega_{QQ^{\prime}}$ baryons. We end this paper with a conclusion in Sec. VIII.

\section{The effective mass of quark}\label{Sec.II}

Generally, the effective mass of the quark is a concept in quantum chromodynamics (QCD) that arises from the confinement of quarks within hadrons \cite{Griffiths:PP888, Close:A11}. Accordingly, the quark effective mass varies across different hadronic states. This variation arises because the quark effective mass is inherently dependent on the choice of radial and orbital quantum numbers. Exploring this phenomenon in baryon system presents an intriguing avenue for further research.

By including relativistic effects, one can obtain the effective masses $M_{1}$ and $M_{2}$ of the quarks for the heavy baryons given by
\begin{eqnarray}
M_{1} &=& \frac{m_{1}}{\sqrt{1-v^{2}_{1}}},  \label{OPP1} \\
M_{2} &=& \frac{m_{2}}{\sqrt{1-v^{2}_{2}}}, \label{OPP2}
\end{eqnarray}
where $m_{1}$, $m_{2}$ are the current masses of the quarks, $v_{1}$ and $v_{2}$ are the velocities of the quarks. For simplicity, we have chosen the velocity of light c = 1. To obtain the values of the quark velocities $v_{i}$ $(i=1, 2)$ in the heavy-light system we exploit the kinetic energy $T_{i}$ = $\frac{1}{2} m_{i} v_{i}^{2}$ with the current mass $m_{i}$. Then, by taking the average of the square velocity $v_{i}^{2}$, we get
\begin{eqnarray}
\langle v_{i}^{2}\rangle = \langle \frac{2}{m_{i}}T_{i} \rangle. \label{nhf11}
\end{eqnarray}
Here, $\langle T_{i} \rangle$ can be interpreted as the Virial theorem in the spherical coordinates $r$,
\begin{eqnarray}
\langle T_{i} \rangle = \frac{1}{2}\langle rV^{\prime}\rangle.  \label{OPP11}
\end{eqnarray}
The prime denotes differentiation with respect to $r$. Considering the short-range interactions in the three-body quark system, we choose the derivative of the Coulomb potential $V$ = $-4\alpha_{s}/3r$, where $\alpha_{s}$ is the running coupling constant of the heavy baryons.

Combining the momentum conservation $m_{1}v_{1}$ = $m_{2}v_{2}$, the square velocities $\langle v_{i}^{2}\rangle$ with $\langle 1/r\rangle$ = $1/a_{B}N^{2}$ are
\begin{eqnarray}
\langle v_{1}^{2}\rangle &=& \frac{1}{m_{1}}\frac{4\alpha_{s}}{3}\frac{1}{a_{B}N^{2}},  \label{OPP5} \\
\langle v_{2}^{2}\rangle &=& \frac{m_{1}}{m^{2}_{2}}\frac{4\alpha_{s}}{3}\frac{1}{a_{B}N^{2}}, \label{OPP6}
\end{eqnarray}
where $a_{B}$ = $1/\mu$ is the Bohr radius related to the reduced mass $\mu$ = $m_{1}m_{2}/(m_{1}+m_{2})$ in the baryon system. $N = n+L+1$ is the principle quantum number of the heavy baryons with the radial quantum number $n$ $(n = 0, 1, 2 ,3, \cdot\cdot\cdot)$ and the orbital quantum number $L$ $(L = 0, 1, 2 ,3, \cdot\cdot\cdot)$. Using the above Eqs. (\ref{OPP5}) and (\ref{OPP6}), Eqs. (\ref{OPP1}) and (\ref{OPP2}) become
\begin{eqnarray}
M_{1} &=& \frac{m_{1}}{\sqrt{1-\frac{1}{m_{1}}\frac{4\alpha_{s}}{3}\frac{\mu}{N^{2}}}},  \label{OPP7} \\
M_{2} &=& \frac{m_{2}}{\sqrt{1-\frac{m_{1}}{m^{2}_{2}}\frac{4\alpha_{s}}{3}\frac{\mu}{N^{2}}}}. \label{OPP8}
\end{eqnarray}

\section{The spin-average masses of doubly heavy baryons}\label{Sec.III}

In the heavy-light quark picture, we consider the baryon system as a bound state of three constituent quarks under the strong interaction. The doubly heavy baryon can be done by viewing a QCD rotating-string tied to the heavy diquark $(QQ^{\prime})$ with spin-1 as one end and to the light quark $(q)$ with spin-1/2 as the other end, see Fig. 1. Based on this model, we make an attempt to investigate the Regge trajectory behavior of the hadronic system.
\begin{figure*}[htbp]
\centering
\begin{minipage}[ht]{0.49\textwidth}
\includegraphics[width=6.5cm]{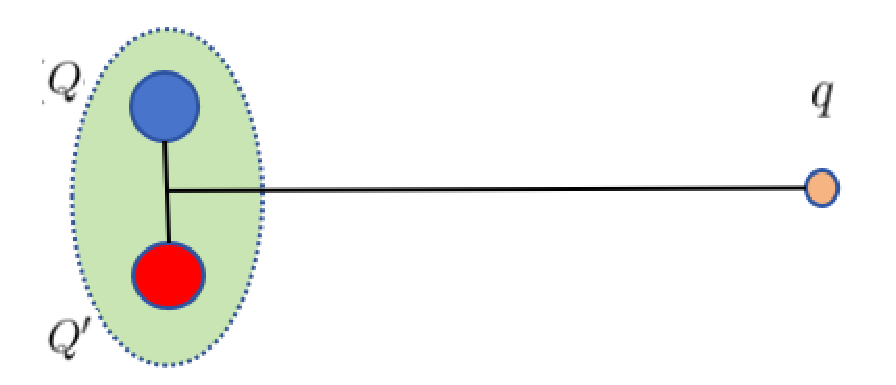}
\caption{Schematic diagram of doubly heavy baryon in the heavy-light quark picture.}
\end{minipage}
\end{figure*}

For the orbital excitations of the doubly heavy baryons, the spin-average mass $\bar M$ and the angular momentum $L$ following the equations are given by \cite{LaCourse:A13, ChenWei:A13}
\begin{equation}
\bar M=\frac{m_{QQ^{\prime}}}{\sqrt{1-{v_{QQ^{\prime}}^{2}}}}+\frac{\alpha}{\omega}\int_{0}^{v_{QQ^{\prime}}}\frac{d u}{\sqrt{1-{u}^{2}}}+\frac{m_{ q}}{\sqrt{1-{v_{q}^{2}}}}+\frac{\alpha}{\omega}\int_{0}^{v_{q}}\frac{d u}{\sqrt{1-{u}^{2}}} , \label{ppu1}
\end{equation}
\begin{equation}
L=\frac{m_{QQ^{\prime}}v_{QQ^{\prime}}^{2}}{\sqrt{1-{{v_{QQ^{\prime}}^{2}}}}}+\frac{\alpha}{\omega^{2}}\int_{0}^{v_{QQ^{\prime}}}\frac{u^{2}d u}{\sqrt{1-{u}^{2}}}+\frac{m_{q}v_{q}^{2}}{\sqrt{1-{{v_{q}^{2}}}}}+\frac{\alpha}{\omega^{2}}\int_{0}^{v_{q}}\frac{u^{2}d u}{\sqrt{1-{u}^{2}}} , \label{ppu2}
\end{equation}
where $m_{QQ^{\prime}}$ and $m_{q}$ are the current masses of the heavy diquark and light quark, respectively. $\alpha$ is the QCD string tension coefficient. In Ref. \cite{EFG:A11}, the slope of the Regge trajectory $\alpha^{\prime}$ is mainly determined by the effective mass $M_{QQ^{\prime}}$ of the heavy diquark,
\begin{equation}
\alpha^{\prime}=\frac{1}{2\pi \alpha} \propto \frac{1}{\sqrt{M_{QQ^{\prime}}}}.
\end{equation}
By introducing the coefficient $k$, the string tension coefficient $\alpha$ is defined as
\begin{eqnarray}
\alpha = \frac{k}{2\pi} (M_{QQ^{\prime}})^{\frac{1}{2}}.  \label{pptt1}
\end{eqnarray}
In addition, the velocity of the quark in the doubly heavy baryon system is defined as $v_{i^{\prime}}=\omega r_{i^{\prime}}$ $(i^{\prime} = QQ^{\prime}, q)$, where $\omega$ and $r_{i^{\prime}}$ are the angular velocity and the position from the centre-of-mass, respectively. Applying Eqs. (\ref{OPP1}) and (\ref{OPP2}), we expand Eqs. (\ref{ppu1}) and (\ref{ppu2}) up to the second order in the parameter $v_{QQ^{\prime}}$,
\begin{equation}
\bar M=M_{QQ^{\prime}}+M_{q}+M_{QQ^{\prime}}v_{QQ^{\prime}}^{2}+\frac{\pi \alpha}{2\omega} , \label{pu0}
\end{equation}
\begin{equation}
L=\frac{1}{\omega}(M_{q}+M_{QQ^{\prime}}v_{QQ^{\prime}}^{2}+\frac{\pi \alpha}{4\omega}). \label{pppuuu2}
\end{equation}
For the string ending at the heavy diquark we utilize the boundary condition
\begin{eqnarray}
\frac{\alpha}{\omega} = \frac{m_{QQ^{\prime}}v_{QQ^{\prime}}}{1-v^{2}_{QQ^{\prime}}}\approx M_{QQ^{\prime}}v_{QQ^{\prime}}. \label{ff11}
\end{eqnarray}
By substituting Eq. (\ref{ff11}) into Eqs. (\ref{pu0}) and (\ref{pppuuu2}), and eliminating the angular velocity $\omega$, one has the spin-averaged mass in the orbital excited states,
\begin{equation}
(\bar M-M_{QQ^{\prime}})^{2}=\pi \alpha L+(M_{q}+M_{QQ^{\prime}}v_{QQ^{\prime}}^{2})^{2}. \label{ppp7uu1}
\end{equation}
Using Eq. (\ref{OPP6}), Eq. (\ref{ppp7uu1}) becomes
\begin{equation}
(\bar M-M_{QQ^{\prime}})^{2} = \pi \alpha L+\left( M_{q}+\frac{M_{QQ^{\prime}}m_{q}}{m^{2}_{QQ^{\prime}}}\frac{4\alpha_{s}}{3}\frac{1}{a_{B}N^{2}}\right)^{2}. \label{pp321}
\end{equation}

In order to obtain the spin-average masses of the radial excited states, we extended the Regge-like mass relation Eq. (\ref{pp321}) by replacing $\pi\alpha L$ with $\pi\alpha (L+1.37n+1)$ \cite{PP:A11}, gives
\begin{equation}
\bar M = M_{QQ^{\prime}}+\left(\frac{k}{2} (M_{QQ^{\prime}})^{\frac{1}{2}} (L+1.37n+1)+\left( M_{q}+\frac{M_{QQ^{\prime}}m_{q}}{m^{2}_{QQ^{\prime}}}\frac{4\alpha_{s}}{3}\frac{\mu}{(n+L+1)^{2}}\right) ^{2}\right)^{\frac{1}{2}}. \label{PP421}
\end{equation}

\section{The mass spectra expressions of doubly heavy baryons}\label{Sec.IV}

For the doubly heavy baryon states, we constructed the mass spectra expressions with the spin-coupling parameters as follows. The mass spectrum are  calculated from the spin-average mass $\bar M$ (\ref{PP421}) and the mass shift $\Delta M(J, j)$,
\begin{equation}
M(J, j)= \bar M + \Delta M(J, j), \label{qqb123}
\end{equation}
where $\mathbf{J}$ = $\mathbf{S}$ + $\mathbf{L}$ is the total angular momentum with the total spin $\mathbf{S}$ = $\mathbf{S}_{QQ^{\prime}}$ + $\mathbf{S}_{q}$, and $j$ = $L$ + $S_{QQ^{\prime}}$ is the total angular momentum of the heavy diquark.

To estimate the mass splitting $\Delta M(J, j)$ in Eq. (\ref{qqb123}), we use the spin-dependent Hamiltonian \cite{EFG:C10, KarlinerRP:PP888},
\begin{equation}
H^{SD}=a_{1}\mathbf{L}\cdot \mathbf{S}_{q}+a_{2}\mathbf{L}\cdot \mathbf{S}%
_{QQ^{\prime}}+b_{1}S_{12}+c_{1}\mathbf{S}_{QQ^{\prime}}\cdot \mathbf{S}_{q},  \label{PP5}
\end{equation}%
where $a_{1}$, $a_{2}$, $b_{1}$, and $c_{1}$ are the spin-coupling parameters. The first two terms are the spin-orbit interactions, the third is the tensor energy, and the last is the contact interaction between the heavy diquark spin $\mathbf{S}_{QQ^{\prime}}$ and light quark spin $\mathbf{S}_{q}$. Here, the tensor interaction $S_{12}$ can be given by  \cite{Landau:A11, KarlinerR:11, Ali:PP888}
\begin{eqnarray}
S_{12} &=& 3(\mathbf{S}_{q}\cdot \mathbf{\hat{r}})(\mathbf{S}_{QQ^{\prime}}\cdot \mathbf{\hat{r}})-\mathbf{S}_{QQ^{\prime}}\cdot \mathbf{S}_{q} \notag\\
&=&-\frac{3}{(2L-1)(2L+3)}[(\mathbf{L}\cdot \mathbf{S}_{q})(\mathbf{L}\cdot \mathbf{S}_{QQ^{\prime}})+(\mathbf{L}\cdot \mathbf{S}_{QQ^{\prime}})(\mathbf{L}\cdot \mathbf{S}_{q})-\frac{2}{3}L(L+1)(\mathbf{S}_{QQ^{\prime}}\cdot \mathbf{S}_{q})],  \label{PP010}
\end{eqnarray}%
where $\mathbf{\hat{r}}$ = $\vec{\mathbf{r}}/r$ is the unit vector of position pointing from the center of mass of the light quark to the heavy diquark. Eq. (\ref{PP010}) suggests $S_{12}$ varies with $L$.

Now, we discuss $\Delta M(J, j)$, $i.e.$, $H^{SD}$ in $S$-, $P$-, and $D$-wave states, respectively.

We analyze the $S$-wave states with $L$ = 0 of the doubly heavy baryons by using the Hamiltonian Eq. (\ref{PP5}). In this case, only the last term survives,
\begin{eqnarray}
H^{SD}(L = 0) = c_{1}\mathbf{S}_{QQ^{\prime}}\cdot \mathbf{S}_{q}.  \label{PP001}
\end{eqnarray}%
The spin of the heavy diquark $S_{QQ^{\prime}}$ = 1 can be coupled with the light quark $S_{q}$ = 1/2. Thus, there are two possibilities for the total spin $S$, one is 1/2 and the other is 3/2. More detailed calculations of the $S$-wave states are found in Appendix A. The expectation value of $\mathbf{S}_{QQ^{\prime}}\cdot\mathbf{S}_{q}$ is obtained by Eq. (\ref{pp8}),
\begin{eqnarray}
\langle\mathbf{S}_{QQ^{\prime}}\cdot \mathbf{S}_{q}\rangle = \left[
\begin{array}{cc}
-1 & 0 \\
0 & \frac{1}{2} \label{pp80}
\end{array}
\right].
\end{eqnarray}
The set of masses are
\begin{eqnarray}
M(1/2, 1)&=& \bar M-c_{1}, \notag \\
M(3/2, 1)&=& \bar M+\frac{1}{2}c_{1}. \label{PP121}
\end{eqnarray}%

For the $P$-wave baryon system with $L$ = 1. The spin of the diquark $S_{QQ^{\prime}}=1$ in $L$-$S$ coupling scheme can be coupled with the heavy quark spin $S_{q}=1/2$ and $L=1$ to the total angular momentum $J=1/2, 3/2$ or $1/2, 3/2, 5/2$ with negative parity $P=-1$. Here, in contrast to the scheme used in Ref. \cite{PP:A11}, we proposed a new scheme of states classification, named all-$JLS$ coupling. The Hamiltonian $H^{SD}(L = 1)$ is treated as a representation operator involving all terms in Eq. (\ref{PP5}). By using this coupling, one can obtain the five mass shifts $\Delta M$$(J, j)$ in Eq. (\ref{MM121}) for the $P$-wave states, and refer to Appendix B. From Eqs. (\ref{qqb123}) and (\ref{MM121}), we find the following mass spectra of the doubly heavy baryons,
\begin{eqnarray}
M(1/2,0)&=& \bar M+\frac{1}{4}\left(-a_{1}-6a_{2}-2b_{1}-c_{1}\right) \notag \\
               &-&\frac{1}{12}\sqrt{(9a_{1}-2a_{2}+10b_{1}-7c_{1})^{2}+8(2a_{2}-b_{1}-2c_{1})^{2}},  \notag \\
M(1/2,1)&=&\bar M+\frac{1}{4}\left(-a_{1}-6a_{2}-2b_{1}-c_{1}\right) \notag \\
               &+&\frac{1}{12}\sqrt{(9a_{1}-2a_{2}+10b_{1}-7c_{1})^{2}+8(2a_{2}-b_{1}-2c_{1})^{2}},  \notag \\
M(3/2,1)&=&\bar M+\frac{1}{20}\left(-5a_{1}+8b_{1}-5c_{1}\right) \notag \\
               &-&\frac{1}{60}\sqrt{(45a_{1}-40a_{2}-16b_{1}-5c_{1})^{2}+5(20a_{2}-10b_{1}-20c_{1})^{2}},  \notag \\
M(3/2,2)&=&\bar M+\frac{1}{20}\left(-5a_{1}+8b_{1}-5c_{1}\right) \notag \\
               &+&\frac{1}{60}\sqrt{(45a_{1}-40a_{2}-16b_{1}-5c_{1})^{2}+5(20a_{2}-10b_{1}-20c_{1})^{2}},  \notag \\
M(5/2,2)&=&\bar M+\frac{1}{2}a_{1}+a_{2}-\frac{1}{5}b_{1}+\frac{1}{2}c_{1}.  \label{MM111}
\end{eqnarray}

For $D$-wave baryon system with $L$ = 2, we further discuss the mass expressions for calculating the doubly heavy baryons. The spin $S_{QQ^{\prime}}$ = 1 in $L$-$S$ coupling scheme can be coupled with $S_{q}$ = 1/2 to determine the total spin $S$ = $1/2$, $3/2$. Coupling of $L=2$ give six states with $J$ = $3/2$, $5/2$ or $1/2$, $3/2$, $5/2$, $7/2$ with positive parity $P=+1$. Similarly, we also analyze the mass shifts $\Delta M$$(J, j)$ in Eq. (\ref{MM212}) for the $D$-wave states, and the Hamiltonian  $H^{SD}(L = 2)$ in Eq. (\ref{PP5}) is treated as a representation operator. Details of calculating $\Delta M$$(J, j)$ are presented in Appendix C. The mass spectra expressions can be written as
\begin{eqnarray}
M(1/2,1)&=&\bar M-\frac{3}{2}a_{1}-3a_{2}-b_{1}+\frac{1}{2}c_{1},    \notag \\
M(3/2,1)&=&\bar M+\frac{1}{4}\left(-a_{1}-8a_{2}-c_{1}\right) \notag \\
               &-&\frac{1}{20}\sqrt{(25a_{1}-16a_{2}+8b_{1}-9c_{1})^{2}+36(2a_{2}-b_{1}-2c_{1})^{2}}, \notag \\
M(3/2,2)&=&\bar M+\frac{1}{4}\left(-a_{1}-8a_{2}-c_{1}\right) \notag \\
               &+&\frac{1}{20}\sqrt{(25a_{1}-16a_{2}+8b_{1}-9c_{1})^{2}+36(2a_{2}-b_{1}-2c_{1})^{2}}, \notag \\
M(5/2,2)&=&\bar M+\frac{1}{28}\left(-7a_{1}+14a_{2}+10b_{1}-7c_{1}\right) \notag \\
               &-&\frac{1}{140}\sqrt{(175a_{1}-182a_{2}-34b_{1}+7c_{1})^{2}+2744(2a_{2}-b_{1}-2c_{1})^{2}},  \notag \\
M(5/2,3)&=&\bar M+\frac{1}{28}\left(-7a_{1}+14a_{2}+10b_{1}-7c_{1}\right) \notag \\
               &+&\frac{1}{140}\sqrt{(175a_{1}-182a_{2}-34b_{1}+7c_{1})^{2}+2744(2a_{2}-b_{1}-2c_{1})^{2}},  \notag \\
M(7/2,3)&=&\bar M+a_{1}+2a_{2}-\frac{2}{7}b_{1}+\frac{1}{2}c_{1}.  \label{MM222}
\end{eqnarray}

With the above expressions (\ref{PP121}), (\ref{MM111}), and (\ref{MM222}), we present the estimates of the mass spectra of the ground and excited states in the $\Xi_{QQ^{\prime}}$ and $\Omega_{QQ^{\prime}}$ baryon system in Sec. VI and Sec. VII, respectively.

\section{The scaling relations of doubly heavy baryons}\label{Sec.V}

The study of data set of the spin-coupling parameters $a_{1}$, $a_{2}$, $b_{1}$, and $c_{1}$, as well as the effective mass of the quark, is of special interest to better understand the baryon structure. Under the color configurations, we utilize the scaling relations based on the similarity between a baryon and its partner baryons to study the spin-coupling parameters.

In our previous studies on the mass spectra of the singly heavy baryons $\Sigma_{Q}$, $\Xi^{\prime}_{Q}$, and $\Omega_{Q}$ $(Q=c, b)$ \cite{PP:A11}, we employed the following scaling relations to describe the parameters,
\begin{equation}
\left\{
\begin{array}{rrrr} \vspace{1ex}
a_{1}(B_{a}, (n+1)L)&=&\frac{M^{\prime}_{Q}M^{\prime}_{d}}{M_{Q}M_{d}}\frac{{N^{\prime}_{a_{1}}}}{{N_{a_{1}}}}a_{1}(B_{a}^{\prime}, (n^{\prime}+1)L^{\prime}), \\ \vspace{1ex}
a_{2}(B_{a}, (n+1)L)&=&\frac{M^{\prime}_{Q}M^{\prime}_{d}}{M_{Q}M_{d}}\frac{{N^{\prime}_{a_{2}}}}{{N_{a_{2}}}}a_{2}(B_{a}^{\prime}, (n^{\prime}+1)L^{\prime}), \\  \vspace{1ex}
b_{1}(B_{a}, (n+1)L)&=&\frac{M^{\prime}_{Q}M^{\prime}_{d}}{M_{Q}M_{d}}\frac{{N^{\prime}_{b_{1}}}}{{N_{b_{1}}}}b_{1}(B_{a}^{\prime}, (n^{\prime}+1)L^{\prime}), \\
c_{1}(B_{a}, (n+1)L)&=&\frac{M^{\prime}_{Q}M^{\prime}_{d}}{M_{Q}M_{d}}\frac{{N^{\prime}_{c_{1}}}}{{N_{c_{1}}}}c_{1}(B_{a}^{\prime}, (n^{\prime}+1)L^{\prime}),\\
\end{array}%
\right.   \label{scr:PP888}
\end{equation}%
where $M^{\prime}_{Q}$, $M_{Q}$ are the effective masses of heavy quarks, and $M^{\prime}_{d}$, $M_{d}$ are the effective masses of light diquark in the singly heavy baryon system. $n, n^{\prime}=0, 1 , 2, \cdots$, $L, L^{\prime}= S, P, D, F, \cdots$. $B_{a}, B_{a}^{\prime}$ are baryons with
\begin{eqnarray}
N_{a_{1}} &=& (n+L+1)^{2} = N_{a_{2}}, \notag \\
N_{b_{1}} &=& L(L+1/2)(L+1)(n+L+1)^{3}, \notag \\
N_{c_{1}} &=& (L+\lambda)(n+L+1)^{3}, \label{mm11}
\end{eqnarray}
corresponding to the similar form of $N^{\prime}_{a_{1}}$, $N^{\prime}_{a_{2}}$, $N^{\prime}_{b_{1}}$, $N^{\prime}_{c_{1}}$ with $L^{\prime}$ and $n^{\prime}$, respectively, where the prime denotes the quantities of the baryon $B_{a}^{\prime}$ obtained from experiments, distinguishing them from that of an unobserved baryon $B_{a}$. In addition, the parameters $a_{1}$, $a_{2}$, $b_{1}$, and $c_{1}$ of the $1P$-wave states for the singly heavy $\Sigma_{c}$, $\Xi^{\prime}_{c}$, and $\Omega_{c}$ baryons can be obtained from Ref. \cite{PP:A11}, see Table \ref{Table:PP861}. In particular, the uncertainties in these parameters from the analysis of the experimental values of masses for the $\Omega_{c}$ baryon \cite{Aaij:App11}. In Ref. \cite{Ali:PP888}, the fitting results of the parameters for the $\Omega_{c}$ baryon are performed based on the LHCb data. By applying Eq. (\ref{scr:PP888}), we can obtain the errors of the parameters for the $\Sigma_{c}$ and $\Xi^{\prime}_{c}$ baryons, respectively.
\renewcommand{\tabcolsep}{0.6cm}
\renewcommand{\arraystretch}{1.0}
\begin{table}[tbh]
\caption{ The spin-coupling parameters (in MeV) of singly heavy baryons. \label{Table:PP861}}%
\label{tab:Eff-mass}
\begin{tabular}
[c]{ccccc}\hline\hline
\text{baryon} & $a_{1}$ & $a_{2}$ & $b_{1}$ & $c_{1}$ \\\hline
$\Omega_{c}$ & 26.96 $\pm$ 0.28 & 25.76 $\pm$ 0.76 & 13.51 $\pm$ 0.54 & 4.04 $\pm$ 0.44   \\\hline
$\Sigma_{c}$ & 35.86 $\pm$ 0.37 & 34.27 $\pm$ 1.01 & 17.97 $\pm$ 0.72 & 5.37 $\pm$ 0.59    \\\hline
$\Xi^{\prime}_{c}$  & 30.64 $\pm$ 0.32 & 29.28 $\pm$ 0.86 & 15.35 $\pm$ 0.62 & 4.59 $\pm$ 0.50     \\\hline\hline
\end{tabular}
\end{table}

Even though the dynamics of doubly heavy baryons may differ significantly from those of singly heavy baryons, we can assume that their interiors are very similar in the three body system. Therefore, the scale relationship is utilized in the study of the doubly heavy baryon states. In order to further investigate the spin-coupling parameters $a_{1}$, $a_{2}$, $b_{1}$, and $c_{1}$ of the doubly heavy $\Xi_{QQ^{\prime}}$ and $\Omega_{QQ^{\prime}}$ baryons, we need to generalize Eq. (\ref{scr:PP888}). As these parameters in Refs. \cite{Ebert:A11, EFG:C10} should be related to the running coupling constant $\alpha_{s}$ from the Coulomb potential $V$ = $-4\alpha_{s}/3r$. Thus, the scaling relation (\ref{scr:PP888}) become
\begin{equation}
\left\{
\begin{array}{rrrr} \vspace{1ex}
a_{1}(B_{a}, (n+1)L)&=&\frac{M^{\prime}_{1}M^{\prime}_{2}}{M_{1}M_{2}}\frac{{N^{\prime}_{a_{1}}}}{{N_{a_{1}}}}\frac{\alpha_{s}(B_{a})}{\alpha_{s}(B_{a}^{\prime})}a_{1}(B_{a}^{\prime}, (n^{\prime}+1)L^{\prime}), \\ \vspace{1ex}
a_{2}(B_{a}, (n+1)L)&=&\frac{M^{\prime}_{1}M^{\prime}_{2}}{M_{1}M_{2}}\frac{{N^{\prime}_{a_{2}}}}{{N_{a_{2}}}}\frac{\alpha_{s}(B_{a})}{\alpha_{s}(B_{a}^{\prime})}a_{2}(B_{a}^{\prime}, (n^{\prime}+1)L^{\prime}), \\  \vspace{1ex}
b_{1}(B_{a}, (n+1)L)&=&\frac{M^{\prime}_{1}M^{\prime}_{2}}{M_{1}M_{2}}\frac{{N^{\prime}_{b_{1}}}}{{N_{b_{1}}}}\frac{\alpha_{s}(B_{a})}{\alpha_{s}(B_{a}^{\prime})}b_{1}(B_{a}^{\prime}, (n^{\prime}+1)L^{\prime}), \\
c_{1}(B_{a}, (n+1)L)&=&\frac{M^{\prime}_{1}M^{\prime}_{2}}{M_{1}M_{2}}\frac{{N^{\prime}_{c_{1}}}}{{N_{c_{1}}}}\frac{\alpha_{s}(B_{a})}{\alpha_{s}(B_{a}^{\prime})}c_{1}(B_{a}^{\prime}, (n^{\prime}+1)L^{\prime}).
\end{array}%
\right.   \label{scr:pp888666}
\end{equation}%

Next, it is necessary to estimate the four spin-coupling parameters $a_{1}$, $a_{2}$, $b_{1}$, and $c_{1}$ for the $\Xi_{QQ^{\prime}}$ and $\Omega_{QQ^{\prime}}$ baryons in Eq. (\ref{scr:pp888666}). With the help of the relations (\ref{OPP7}) and (\ref{OPP8}), we begin by calculating the effective masses of the quarks. In general, it is extracted from the experimental values of mass spectra for the baryons that have been discovered. Different from the case of the singly heavy baryons and singly heavy mesons, there are no more experimental values for the $\Xi_{QQ^{\prime}}$ and $\Omega_{QQ^{\prime}}$ baryons apart from the two $\Xi^{++}_{cc}(3621)$ and $\Xi^{+}_{cc}(3519)$ states. Hence, in estimating the quark masses within the $\Xi_{QQ^{\prime}}$ and $\Omega_{QQ^{\prime}}$ baryons, we adopt the current masses of the quarks from the Particle Data Group \cite{Navas:A11}, as follows
\begin{eqnarray}
&m_{c} = 1.273\ \text{GeV}, \ m_{b} = 4.183\ \text{GeV},    \notag \\
&m_{u} = 0.00216\ \text{GeV}, \  m_{d} = 0.0047\ \text{GeV}, \  m_{s} = 0.0935\ \text{GeV}.  \label{VVV11}
\end{eqnarray}
With this information (\ref{VVV11}), we first consider the masses of the heavy quarks $c$ and $b$ for the radial and orbital excitations
in the diquark ($\rho$-mode). The results are
\begin{eqnarray}
cc&:&\ M_{c} = 1381.03 \ \text{MeV},\;  m_{cc} = M_{c}+M_{c}= 2762.06\ \text{MeV}, \notag \\
bc&:&\ M_{c} = 1451.20 \ \text{MeV},\;  M_{b} = 4228.38 \ \text{MeV},\;  m_{bc} = M_{c}+M_{b}= 5679.58\ \text{MeV}, \notag \\
bb&:&\ M_{b} = 4537.99 \ \text{MeV},\;  m_{bb} = M_{b}+M_{b}= 9075.98\ \text{MeV}.
\end{eqnarray}
And then the effective masses of the radial or orbital excitations between the quark and diquark ($\lambda$-mode) for the heavy diquark and light quark are calculated, with the results
\begin{eqnarray}
&M_{cc} = 2763.20\ \text{MeV}, \  M_{bc} = 5680.15\ \text{MeV}, \  M_{bb} = 9076.33\ \text{MeV}, \notag \\
&M_{u} = 4.25\ \text{MeV}, \  M_{d} = 9.24\ \text{MeV}, \  M_{s} = 176.18\ \text{MeV}.  \label{VVV12}
\end{eqnarray}
Here, for simplicity, we present only the result for the  $1S$-wave states for clarity, while other results will be given along the calculations in section VI and VII.

In addition, there are two parameters which should be fixed, $i.e.$, $\lambda$ and $k$ in Eqs. (\ref{mm11}) and (\ref{pptt1}), respectively. In Eq. (\ref{mm11}), the parameter $c_{1}$, which is expected to be negligible in higher excited states of the baryons, because it should be very small. Considering that the parameter $c_1$ becomes dominant in determining the mass splitting in $S$-wave states, we can estimate the parameter $c_1$ based on the hyperfine structure term in the doubly heavy baryon system. Combining the experimental value of the $\Xi^{++}_{cc}(3621)$ state and the following calculations of the mass splitting, we take the parameter $\lambda$ = 0.68. Meanwhile, in order to obtain the spin-average masses in Eq.  (\ref{PP421}) of the the radial and orbital excited states, we need to consider the slope $\alpha$ of Regge trajectory with $k$. Here, we fit the parameter $k$ =1 by calculating the spin-average masses based on the known masses of the doubly heavy baryon states.

By substituting the effective masses (\ref{VVV12}) and the parameters in Table \ref{Table:PP861} into Eq. (\ref{scr:pp888666}), we can calculate the spin-coupling parameters of the $\Xi_{QQ^{\prime}}$ and $\Omega_{QQ^{\prime}}$ baryons. The results are listed in Table \ref{Table:PP86} and \ref{Table:PP87}, respectively. While the uncertainties in the parameters of the $\Xi_{QQ^{\prime}}$ and $\Omega_{QQ^{\prime}}$ baryons are determined from the error results of the singly heavy baryons presented in Table \ref{Table:PP861} by using the scaling relations. In Eq. (\ref{VVV12}), the running coupling constant $\alpha_{s}$ are also taken from the experimental values, $\alpha_{s}(B_{a}^{\prime})$ = 0.593 GeV and $\alpha_{s}(B_{a})$ = 0.557 GeV for the singly heavy baryons and doubly heavy baryons, respectively. According to our model, it can be seen that the value of $\alpha_{s}$ is slightly larger than that of other model \cite{ChenLuo:H11}.

\renewcommand\tabcolsep{0.55cm}
\renewcommand{\arraystretch}{0.7}
\begin{table*}[!htbp]
\caption{Spin coupling parameters (MeV) of $\Xi_{QQ^{\prime}}$ baryons.   \label{Table:PP86}}
\begin{tabular}{ccccccc}
\hline\hline
Baryon & State & $a_{1}$ & $a_{2}$ & $b_{1}$ &$c_{1}$ \\
 \hline
  &1$S$ &      &         &       & 52.08 $\pm$ 5.73             \\
  &2$S$ &      &         &       & 11.57 $\pm$ 1.28           \\
  &3$S$ &      &         &       & 3.64  $\pm$ 0.40               \\
  &4$S$ &      &         &       & 1.56  $\pm$ 0.18             \\
  &5$S$ &      &         &       & 0.81  $\pm$ 0.09               \\

  &1$P$ &31.27 $\pm$ 0.33 &29.88 $\pm$ 0.89   &15.67 $\pm$ 0.63  &4.68 $\pm$ 0.52     \\
  &2$P$ &14.75 $\pm$ 0.16 &14.10 $\pm$ 0.42   &4.93 $\pm$ 0.20   &1.47 $\pm$ 0.17               \\
$\Xi_{cc}$  &3$P$ &8.46 $\pm$ 0.09  &8.08 $\pm$ 0.24     &2.12 $\pm$ 0.09   &0.63 $\pm$ 0.07                \\
  &4$P$ &5.46 $\pm$ 0.06  &5.22 $\pm$ 0.16     &1.10 $\pm$ 0.05   &0.33 $\pm$ 0.04               \\
  &5$P$ &3.81 $\pm$ 0.04  &3.64 $\pm$ 0.11     &0.64 $\pm$ 0.03   &0.19 $\pm$ 0.02                 \\

  &1$D$ &14.75 $\pm$ 0.16 &14.10 $\pm$ 0.42    &0.99 $\pm$ 0.04   &0.92 $\pm$ 0.10              \\
  &2$D$ &8.46 $\pm$ 0.09  &8.08 $\pm$ 0.24     &0.42 $\pm$ 0.02   &0.40  $\pm$ 0.05              \\
  &3$D$ &5.46 $\pm$ 0.06  &5.22 $\pm$ 0.16      &0.22 $\pm$ 0.01   &0.21 $\pm$ 0.03            \\
  &4$D$ &3.81 $\pm$ 0.04  &3.64 $\pm$ 0.11     &0.13 $\pm$ 0.006   &0.12 $\pm$ 0.02              \\
  &5$D$ &2.81 $\pm$ 0.03  &2.68 $\pm$ 0.08     &0.08 $\pm$ 0.004   &0.08 $\pm$ 0.009             \\
\hline
  &1$S$ &       &         &       & 25.30$ \pm$ 2.82       \\
  &2$S$ &       &         &       & 5.62 $\pm$ 0.63       \\
  &3$S$ &       &         &       & 1.77  $\pm$ 0.20      \\
  &4$S$ &       &         &       & 0.76  $\pm$ 0.09      \\
  &5$S$ &       &         &       & 0.39   $\pm$ 0.05      \\

  &1$P$ &15.21 $\pm$ 0.17 &14.53 $\pm$ 0.44   &7.66 $\pm$ 0.31   &2.28 $\pm$ 0.26          \\
  &2$P$ &7.17 $\pm$ 0.08  &6.85 $\pm$ 0.21    &2.41 $\pm$ 0.10   &0.72 $\pm$ 0.08         \\
$\Xi_{bc}$  &3$P$ &4.11 $\pm$ 0.05 &3.93 $\pm$ 0.12    &1.04 $\pm$ 0.05  &0.31 $\pm$ 0.04           \\
  &4$P$ &2.66 $\pm$ 0.03  &2.54 $\pm$ 0.08    &0.54 $\pm$ 0.03  &0.16 $\pm$ 0.02         \\
  &5$P$ &1.85 $\pm$ 0.02 &1.77 $\pm$ 0.06    &0.31 $\pm$ 0.02  &0.09 $\pm$ 0.01         \\

  &1$D$ &7.17 $\pm$ 0.08  &6.85 $\pm$ 0.21    &0.48 $\pm$ 0.02  &0.45 $\pm$ 0.05         \\
  &2$D$ &4.11 $\pm$ 0.05  &3.93 $\pm$ 0.12    &0.21 $\pm$ 0.01  &0.19 $\pm$ 0.02        \\
  &3$D$ &2.66 $\pm$ 0.03 &2.54  $\pm$ 0.08   &0.11  $\pm$ 0.005  &0.10 $\pm$ 0.02        \\
  &4$D$ &1.85 $\pm$ 0.02 &1.77  $\pm$ 0.06   &0.06 $\pm$ 0.003 &0.06  $\pm$ 0.006          \\
  &5$D$ &1.37 $\pm$ 0.02 &1.30  $\pm$ 0.04   &0.04 $\pm$ 0.001 &0.04  $\pm$ 0.004          \\
\hline
  &1$S$ &      &         &       & 15.83 $\pm$ 1.76      \\
  &2$S$ &      &         &       & 3.52 $\pm$ 0.40       \\
  &3$S$ &      &         &       & 1.11 $\pm$ 0.13       \\
  &4$S$ &      &         &       & 0.48 $\pm$ 0.06       \\
  &5$S$ &      &         &       & 0.25 $\pm$ 0.03        \\

  &1$P$ &9.52 $\pm$ 0.11 &9.09 $\pm$ 0.28    &4.80 $\pm$ 0.20  &1.42 $\pm$ 0.16         \\
  &2$P$ &4.49 $\pm$ 0.05 &4.29 $\pm$ 0.13    &1.51 $\pm$ 0.06  &0.45 $\pm$ 0.05         \\
$\Xi_{bb}$  &3$P$ &2.57 $\pm$ 0.03 &2.46 $\pm$ 0.08    &0.65 $\pm$ 0.03  &0.19 $\pm$ 0.03          \\
  &4$P$ &1.66 $\pm$ 0.02 &1.59 $\pm$ 0.05    &0.34 $\pm$ 0.02  &0.10 $\pm$ 0.02          \\
  &5$P$ &1.16 $\pm$ 0.02 &1.11 $\pm$ 0.04    &0.19 $\pm$ 0.008  &0.06 $\pm$ 0.007          \\

  &1$D$ &4.49 $\pm$ 0.05  &4.29 $\pm$ 0.13    &0.30 $\pm$ 0.02  &0.28 $\pm$ 0.04         \\
  &2$D$ &2.57 $\pm$ 0.03  &2.46 $\pm$ 0.08    &0.13 $\pm$ 0.005  &0.12 $\pm$ 0.02         \\
  &3$D$ &1.66 $\pm$ 0.02  &1.59 $\pm$ 0.05    &0.07 $\pm$ 0.003 &0.06 $\pm$ 0.007         \\
  &4$D$ &1.16 $\pm$ 0.02  &1.11 $\pm$ 0.04    &0.04 $\pm$ 0.002  &0.04 $\pm$ 0.005         \\
  &5$D$ &0.85 $\pm$ 0.01  &0.82 $\pm$ 0.03    &0.03 $\pm$ 0.001 &0.02 $\pm$ 0.003         \\
\hline\hline
\end{tabular}
\end{table*}

\renewcommand\tabcolsep{0.55cm}
\renewcommand{\arraystretch}{0.7}
\begin{table*}[!htbp]
\caption{Spin coupling parameters (MeV) of $\Omega_{QQ^{\prime}}$ baryons.   \label{Table:PP87}}
\begin{tabular}{cccccc}
\hline\hline
Baryon & State & $a_{1}$ & $a_{2}$ & $b_{1}$ &$c_{1}$ \\
\hline
 &1$S$ &      &         &       & 40.39 $\pm$ 4.40    \\
 &2$S$ &      &         &       & 8.62  $\pm$ 0.94      \\
 &3$S$ &      &         &       & 2.70  $\pm$ 0.30     \\
 &4$S$ &      &         &       & 1.16  $\pm$ 0.13       \\
 &5$S$ &      &         &       & 0.60  $\pm$ 0.07       \\

 &1$P$ &23.28 $\pm$ 0.25 &22.24 $\pm$ 0.66   &11.67 $\pm$ 0.47  &3.49 $\pm$ 0.38          \\
 &2$P$ &10.96 $\pm$ 0.12 &10.47 $\pm$ 0.31   &3.66 $\pm$ 0.15  &1.09 $\pm$ 0.12       \\
$\Omega_{cc}$ &3$P$ &6.28 $\pm$ 0.07 &6.00 $\pm$ 0.18    &1.57 $\pm$ 0.07  &0.47 $\pm$ 0.06          \\
 &4$P$ &4.05 $\pm$ 0.05  &3.87 $\pm$ 0.12    &0.81 $\pm$ 0.04  &0.24 $\pm$ 0.03          \\
 &5$P$ &2.83 $\pm$ 0.03  &2.70 $\pm$ 0.08    &0.74 $\pm$ 0.02  &0.14 $\pm$ 0.02          \\

 &1$D$ &10.96 $\pm$ 0.12 &10.47 $\pm$ 0.31   &0.73 $\pm$ 0.03  &0.69 $\pm$ 0.08          \\
 &2$D$ &6.28 $\pm$ 0.07  &6.00 $\pm$ 0.18    &0.31 $\pm$ 0.02  &0.29 $\pm$ 0.04            \\
 &3$D$ &4.05 $\pm$ 0.05  &3.87 $\pm$ 0.12    &0.16 $\pm$ 0.007  &0.15 $\pm$ 0.02           \\
 &4$D$ &2.83 $\pm$ 0.03  &2.70  $\pm$ 0.08   &0.10  $\pm$ 0.004 &0.09 $\pm$ 0.01         \\
 &5$D$ &2.08 $\pm$ 0.03  &1.99  $\pm$ 0.06   &0.06  $\pm$ 0.003 &0.06 $\pm$ 0.007          \\
\hline
 &1$S$ &      &         &       & 19.22 $\pm$ 2.10       \\
 &2$S$ &      &         &       & 4.19  $\pm$ 0.46       \\
 &3$S$ &      &         &       & 1.31  $\pm$ 0.15       \\
 &4$S$ &      &         &       & 0.57  $\pm$ 0.07       \\
 &5$S$ &      &         &       & 0.29  $\pm$ 0.04        \\

 &1$P$ &11.30 $\pm$ 0.13 &10.80 $\pm$ 0.33   &5.67 $\pm$ 0.23  &1.69 $\pm$ 0.19         \\
 &2$P$ &5.33  $\pm$ 0.06 &5.09 $\pm$ 0.16    &1.78 $\pm$ 0.07  &0.53 $\pm$ 0.06         \\
$\Omega_{bc}$ &3$P$ &3.05 $\pm$ 0.04  &2.92 $\pm$ 0.09    &0.77 $\pm$ 0.04  &0.23 $\pm$ 0.03          \\
 &4$P$ &1.97 $\pm$ 0.03  &1.88 $\pm$ 0.06    &0.40 $\pm$ 0.02  &0.12 $\pm$ 0.02           \\
 &5$P$ &1.37 $\pm$ 0.02 &1.31 $\pm$ 0.04    &0.23 $\pm$ 0.01  &0.07 $\pm$ 0.008         \\

 &1$D$ &5.33 $\pm$ 0.06 &5.09 $\pm$ 0.16    &0.36 $\pm$ 0.02  &0.33 $\pm$ 0.04         \\
 &2$D$ &3.05 $\pm$ 0.04 &2.92 $\pm$ 0.09    &0.15 $\pm$ 0.007  &0.14 $\pm$ 0.02          \\
 &3$D$ &1.97 $\pm$ 0.03 &1.88 $\pm$ 0.06    &0.08 $\pm$ 0.004 &0.07 $\pm$ 0.009           \\
 &4$D$ &1.37 $\pm$ 0.02 &1.31 $\pm$ 0.04    &0.05 $\pm$ 0.002 &0.04  $\pm$ 0.004       \\
 &5$D$ &1.01 $\pm$ 0.02 &0.97 $\pm$ 0.03    &0.03 $\pm$ 0.001 &0.03  $\pm$ 0.003        \\
\hline
 &1$S$ &      &         &       & 11.93 $\pm$ 1.30      \\
 &2$S$ &      &         &       & 2.62  $\pm$ 0.29      \\
 &3$S$ &      &         &       & 0.82  $\pm$ 0.09      \\
 &4$S$ &      &         &       & 0.35  $\pm$ 0.04      \\
 &5$S$ &      &         &       & 0.18  $\pm$ 0.02       \\

 &1$P$ &7.07 $\pm$ 0.08 &6.75 $\pm$ 0.21    &3.54 $\pm$ 0.15  &1.06 $\pm$ 0.12        \\
 &2$P$ &3.33 $\pm$ 0.04 &3.18 $\pm$ 0.10    &1.11 $\pm$ 0.05  &0.33 $\pm$ 0.04         \\
$\Omega_{bb}$ &3$P$ &1.91 $\pm$ 0.03 &1.82 $\pm$ 0.06    &0.48 $\pm$ 0.02  &0.14 $\pm$ 0.02          \\
 &4$P$ &1.23 $\pm$ 0.02 &1.18 $\pm$ 0.04    &0.25 $\pm$ 0.01  &0.07 $\pm$ 0.009          \\
 &5$P$ &0.86 $\pm$ 0.01 &0.82 $\pm$ 0.03    &0.14 $\pm$ 0.006  &0.04 $\pm$ 0.005          \\

 &1$D$ &3.33 $\pm$ 0.04 &3.18 $\pm$ 0.10    &0.22 $\pm$ 0.01  &0.21 $\pm$ 0.03         \\
 &2$D$ &1.91 $\pm$ 0.03 &1.82 $\pm$ 0.06    &0.10 $\pm$ 0.004  &0.09 $\pm$ 0.01         \\
 &3$D$ &1.23 $\pm$ 0.02 &1.18 $\pm$ 0.04    &0.05$\pm$ 0.002  &0.05 $\pm$ 0.005         \\
 &4$D$ &0.86 $\pm$ 0.01 &0.82 $\pm$ 0.03    &0.03 $\pm$ 0.002 &0.03  $\pm$ 0.003        \\
 &5$D$ &0.63 $\pm$ 0.007 &0.61 $\pm$ 0.02    &0.02 $\pm$ 0.001 &0.02  $\pm$ 0.002         \\
\hline\hline
\end{tabular}
\end{table*}

\section{The $\Xi_{QQ^{\prime}}$ baryons }\label{Sec.VI}

Let us now begin with calculating the masses of the ground and excited states for the doubly heavy $\Xi_{cc}$, $\Xi_{bc}$, and $\Xi_{bb}$ baryons. It was a pleasant surprise that the doubly charmed baryon $\Xi^{++}_{cc}$ is observed by the LHCb Collaboration in the $\Lambda^{+}_{c}K^{-}\pi^{+}\pi^{+}$ mass spectrum \cite{Aaije:A11}. The mass reported for this resonance of $\Xi^{++}_{cc}$ is $M(\Xi^{++}_{cc})$ = $3621.40\pm0.72\pm0.27$ MeV, but its $J^{P}$ value is still unknown. In addition, the LHCb Collaboration has also carried out searches for the doubly heavy baryon $\Xi^{+}_{cc}$ decaying to $D^{0}pK^{-}$ \cite{Aaij:PP86}, and for the doubly heavy baryons $\Xi^{0}_{cc}$ and $\Omega^{0}_{cc}$ decaying to $\Lambda^{+}_{c}\pi^{-}$ and $\Xi^{+}_{c}\pi^{-}$ in Refs. \cite{Aaij:PP8866, Aaij:PP888666}. These baryons have not yet been observed. Many workers try to study their properties and internal structure for the doubly heavy $\Xi_{QQ^{\prime}}$ baryons, we recommend interested readers to see Refs. \cite{Albertu:A11, WengDeng:A11, ChenLuo:H11, Oudichhya:H11, BerezhnoyLL:H11, MHarada:A11}.

Based on the close similarity of the strong interaction binds the heavy diquark and light quark in hadron system, we obtain a relation between the mass splitting of the doubly heavy baryons and that of the singly heavy mesons ($Q \bar q$) \cite{Ebert:A11},
\begin{eqnarray}
\Delta M(\Xi_{QQ^{\prime}}) = \frac{3}{2}\frac{M_{Q}M_{\bar q}}{M_{QQ^{\prime}}M_{q}}\Delta M(B,D),  \label{pwe123}
\end{eqnarray}
with the factor 3/2 being just the ratio of the baryon and meson spin matrix elements.

From PDG \cite{Navas:A11}, the $D^{\pm}$ and $D^{\ast}(2010)^{\pm}$ states, which are regarded as $1S$-wave states corresponding to the masses $M(D^{\pm})$ = 1869.66 MeV and $M(D^{\ast}(2010)^{\pm})$ = 2010.26 MeV with $J^{P}$ = $1/2^{+}$ and $3/2^{+}$, respectively. Thus, the hyperfine splitting between $D^{\pm}$ and $D^{\ast}(2010)^{\pm}$ would be $\Delta M(D)$ = $M(D^{\ast}(2010)^{\pm})$ $-$ $M(D^{\pm})$ = 140.60 MeV. Substituting this value into Eq. (\ref{pwe123}), the predicted hyperfine splitting of $1S$-wave for the $\Xi_{cc}$ baryon is given as follows
\begin{eqnarray}
\Delta M(\Xi_{cc}) &=& \frac{3}{2}\frac{M_{c}M_{\bar u}}{M_{cc}M_{u}}\Delta M(D) \notag \\
                   &=& 78.12 \ \text{MeV},   \label{pww123}
\end{eqnarray}
with $M_{c}$ = 1273.00 MeV, $M_{cc}$ = 2762.07 MeV, $M_{\bar u}$ = 3.42 MeV, and $M_{u}$ = 4.25 MeV. Our predicted values is about 14.42 MeV higher than the hyperfine splitting $\Delta M(\Xi_{cc})$ = $M_{3/2}(\Xi_{cc})$ $-$ $M_{1/2}(\Xi_{cc})$ = 63.7 MeV in the constituent quark model \cite{Karlinere:A11}. Note that no other experiments confirmed the existence of the $\Xi^{++}_{cc}$ \cite{BAubert:H11, RChistov:H11} so far, as they find no evidence for the $\Xi^{+}_{cc}$ baryon in $\Lambda^{+}_{c}K^{-}\pi^{+}$ and $\Xi^{0}_{c}\pi^{+}$ decay, and the $\Xi^{++}_{cc}$ baryon in $\Lambda^{+}_{c}K^{-}\pi^{+}\pi^{+}$ and $\Xi^{0}_{c}\pi^{+}\pi^{+}$ decay, respectively. In our model, the $\Xi^{++}_{cc}$(3621) can be grouped into the $1S$ state. We assign $J^{P}=1/2^{+}$ for $\Xi^{++}_{cc}$(3621). By applying Eq. (\ref{PP421}), one estimates the spin-average mass  of $1S$-wave ($n$ = 0, $L$ = 0) for the $\Xi^{++}_{cc}$ baryon,
\begin{eqnarray}
\bar M(\Xi_{cc}) &=& M_{cc}+\left(\frac{1}{2}(M_{cc})^{1/2}+\left(M_{u}+\frac{M_{cc}m_{u}}{m^{2}_{cc}}\frac{4\alpha_{s}}{3}\mu_{ccu} \right)^{2}\right)^{\frac{1}{2}}  \notag \\
       &=& 3673.65 \ \text{MeV},  \label{PP101}
\end{eqnarray}
with $\mu_{ccu}$ = $m_{cc}m_{u}/(m_{cc}+m_{u})$  =  2.158 MeV, as well as the following rough estimate for the parameter $c_{1}$ by Eq. (\ref{scr:pp888666}),
\begin{eqnarray}
c_{1}(\Xi_{cc}, 1S) &=& \frac{M_{c}M_{uu}}{M_{cc}M_{u}}\frac{{N^{\prime}_{c_{1}}}}{{N_{c_{1}}}}\frac{\alpha_{s}(\Xi_{cc})}{\alpha_{s}(\Sigma_{c})}c_{1}(\Sigma_{c}, 1P) = 52.08 \ \text{MeV}.  \label{PP101}
\end{eqnarray}
The mass spectra of the two $1S$-wave states for the $\Xi_{cc}$ baryon are given by
\begin{equation}
\begin{array}{c}
\text{State}|J,j\rangle: \\ [11pt]
M(\Xi^{++}_{cc},1S\text{):}
\end{array}%
\begin{array}{ccccc}
|1/2,1\rangle \qquad & |3/2,1\rangle  \\ [11pt]
\multicolumn{1}{r}{3621.57 \pm 8.85} \text{MeV} \qquad & \multicolumn{1}{r}{3699.69 \pm 4.59} \text{MeV},
\end{array}
\end{equation}
which are compared with the former theoretical results reported in Refs. \cite{Ebert:A11, Karlinere:A11, Giannuzzi:A11}. The prediction errors of the masses in the ground states are then estimated to be 8.85 MeV and 4.59 MeV in our model. For the ground and excited states of the $\Xi_{cc}$ baryon, we calculate the parameters $a_{1}$, $a_{2}$, $b_{1}$, and $c_{1}$ as shown in Table \ref{Table:PP86}, while the mass spectra are compared to results of other models in Table \ref{ppdm5}.

In 2021, the neutral doubly heavy baryon $\Xi^{0}_{bc}$ was searched by the LHCb experiment in $\Xi^{0}_{bc}$ $\rightarrow$ $\Xi^{+}_{c}\pi^{-}$ decay \cite{Aaij:PP8866}, no significant excess is found for invariant masses between 6700 and 7300 MeV. The LHCb experiment had searched the baryon $\Xi^{+}_{bc}$ in the charged decay mode $\Xi^{+}_{bc}$ $\rightarrow$ $J/\psi\Xi^{+}_{c}$ \cite{Aaij:PP888666} in 2023. The most significant peaks in the mass region considered correspond to local (global) significance of 4.3$\sigma$ (2.8$\sigma$) and 4.1$\sigma$ (2.4$\sigma$) at 6571 MeV and 6694 MeV. Thus, there are no evidence for the $\Xi^{0}_{bc}$ and $\Xi^{+}_{bc}$ baryons with the current data sample. According to this information for the $\Xi_{bc}$ and $\Xi_{bb}$ baryons in our model, the spin-average masses of the ground states are given by using the spin-independent mass (\ref{PP421}) with $n = 0$, $L = 0$, $\mu_{bcu}$ = $m_{bc}m_{u}/(m_{bc}+m_{u})$ = 2.159 MeV, and $\mu_{bbu}$ = $m_{bb}m_{u}/(m_{bb}+m_{u})$ = 2.160 MeV,
\begin{eqnarray}
\bar M(\Xi_{bc}) &=& M_{bc}+\left(\frac{1}{2}(M_{bc})^{1/2}+\left(M_{u}+\frac{M_{bc}m_{u}}{m^{2}_{bc}}\frac{4\alpha_{s}}{3}\mu_{bcu} \right)^{2}\right)^{\frac{1}{2}}  \notag \\
       &=& 6771.19 \ \text{MeV},  \label{PP021}
\end{eqnarray}
and
\begin{eqnarray}
\bar M(\Xi_{bb}) &=& M_{bb}+\left(\frac{1}{2}(M_{bb})^{1/2}+\left(M_{u}+\frac{M_{bb}m_{u}}{m^{2}_{bb}}\frac{4\alpha_{s}}{3}\mu_{bbu} \right) ^{2}\right)^{\frac{1}{2}} \notag \\
       &=& 10303.31 \ \text{MeV}.  \label{PP131}
\end{eqnarray}
Utilizing the scaling relation (\ref{scr:pp888666}) to calculate the spin coupling parameter $c_{1}$, the results are obtained by
\begin{eqnarray}
c_{1}(\Xi_{bc}, 1S) &=& \frac{M_{cc}M^{\prime}_{u}}{M_{bc}M_{u}}\frac{{N^{\prime}_{c_{1}}}}{{N_{c_{1}}}}\frac{\alpha_{s}(\Xi_{bc})}{\alpha_{s}(\Xi_{cc})}c_{1}(\Xi_{cc}, 1P) = 25.30 \ \text{MeV},  \label{PP101}
\end{eqnarray}
\begin{eqnarray}
c_{1}(\Xi_{bb}, 1S) &=& \frac{M_{cc}M^{\prime}_{u}}{M_{bb}M_{u}}\frac{{N^{\prime}_{c_{1}}}}{{N_{c_{1}}}}\frac{\alpha_{s}(\Xi_{bb})}{\alpha_{s}(\Xi_{cc})}c_{1}(\Xi_{cc}, 1P) = 15.83 \ \text{MeV}.  \label{PP101}
\end{eqnarray}
Thus, the predicted masses of the $1S$-wave states for the $\Xi_{bc}$ and $\Xi_{bb}$ baryons are
\begin{equation}
\begin{array}{c}
\text{State}|J,j\rangle: \\ [11pt]
M(\Xi_{bc},1S\text{):}
\end{array}%
\begin{array}{ccccc}
|1/2,1\rangle \qquad & |3/2,1\rangle  \\ [11pt]
\multicolumn{1}{r}{6745.89 \pm 4.05} \text{MeV} \qquad & \multicolumn{1}{r}{6783.84 \pm 2.40} \text{MeV},
\end{array}
\end{equation}
and
\begin{equation}
\begin{array}{c}
\text{State}|J,j\rangle: \\ [11pt]
M(\Xi_{bb},1S\text{):}
\end{array}%
\begin{array}{ccccc}
|1/2,1\rangle \qquad & |3/2,1\rangle  \\ [11pt]
\multicolumn{1}{r}{10287.48 \pm 4.39} \text{MeV} \qquad & \multicolumn{1}{r}{10311.22 \pm 3.90} \text{MeV},
\end{array}
\end{equation}
respectively. In our model calculations, the $1S$-wave level-splitting are predicted to be $\Delta M(\Xi_{bc})$ = 37.95 \text{MeV} and $\Delta M(\Xi_{bb})$ = 23.74 \text{MeV} smaller than 47 \text{MeV} and 35 \text{MeV} in Ref. \cite{Ebert:A11} for the partner $\Xi_{bc}$ and $\Xi_{bb}$ baryons, respectively. By exploiting the mass spectra expressions (\ref{PP121}), (\ref{MM111}), and (\ref{MM222}), one can obtain the mass spectra listed in Table \ref{ppddmm35} and \ref{ppddmm45}. In order to compute the uncertainties of the mass spectra, we considered all possible sources of the uncertainties. Based on the error analysis of the experimental value for $\Xi^{++}_{cc}$ baryon state \cite{Aaije:A11,Aaije:A111}, as well as considering the errors of the spin-coupling parameters $a_{1}$, $a_{2}$, $b_{1}$, and $c_{1}$ for the $\Xi_{QQ^{\prime}}$ baryons and errors of other input parameters, the uncertainties in the mass spectra of the $\Xi_{cc}$, $\Xi_{bc}$, and $\Xi_{bb}$ baryons are estimated by adding these error results in quadrature. Calculation of the mass spectra of the douby heavy $\Xi_{cc}$, $\Xi_{bc}$, and $\Xi_{bb}$ baryons together with a more detailed discussion can be found in Refs. \cite{Shah:A11, Azizi:EH11,Yoshida:A11, GershteinK:A11, WangYZ:A11, AlexandrouD:A11, Cheng:H11, ZhangW:EH11, ChengX:EH11, KiselevK:EH11, ChangW:EH11, WengDeng:A11}.

\renewcommand\tabcolsep{0.8cm}
\renewcommand{\arraystretch}{0.6}
\begin{table}[htbp]
\caption{Mass spectra (MeV) of $\Xi_{cc}$ baryons are given and compared with different quark models.}\label{ppdm5}
\resizebox{\textwidth}{12cm}{\begin{tabular}{ccccccc}
\hline\hline
{\small State }\; $J^{P}$  &{Ours}&   \cite{Ebert:A11}   &  \cite{Zhong:A11}   & \cite{BEakins:A11} & \cite{MKarliner:H11}  &        \\
\hline
$%
\begin{array}{rr}
{\small 1}^{1}{\small S}_{1/2} & {\small 1/2}^{+} \\
{\small 1}^{3}{\small S}_{3/2} & {\small 3/2}^{+}%
\end{array}%
$ & $%
\begin{array}{r}
{\small 3621.57 \pm 8.85} \\
{\small 3699.69 \pm 4.59}%
\end{array}%
$ & $%
\begin{array}{r}
{\small 3620} \\
{\small 3727}%
\end{array}%
$ & $%
\begin{array}{r}
{\small 3606} \\
{\small 3675}%
\end{array}%
$ & $%
\begin{array}{r}
{\small 3678} \\
{\small 3752}%
\end{array}%
$ & $%
\begin{array}{r}
{\small 3627} \\
{\small 3690}%
\end{array}%
$ & $%
\begin{array}{r}
{\small } \\
{\small }%
\end{array}%
$ \\ $%

\begin{array}{rr}
{\small 2}^{1}{\small S}_{1/2} & {\small 1/2}^{+} \\
{\small 2}^{3}{\small S}_{3/2} & {\small 3/2}^{+}%
\end{array}%
$ & $%
\begin{array}{r}
{\small 4153.86 \pm 2.52} \\
{\small 4171.21 \pm 1.79}%
\end{array}%
$ & $%
\begin{array}{r}
{\small } \\
{\small }%
\end{array}%
$ & $%
\begin{array}{r}
{\small 4004} \\
{\small 4036}%
\end{array}%
$ & $%
\begin{array}{r}
{\small 4311} \\
{\small 4368}%
\end{array}%
$ & $%
\begin{array}{r}
{\small } \\
{\small }%
\end{array}%
$ & $%
\begin{array}{r}
{\small } \\
{\small }%
\end{array}%
$ \\ $%

\begin{array}{rr}
{\small 3}^{1}{\small S}_{1/2} & {\small 1/2}^{+} \\
{\small 3}^{3}{\small S}_{3/2} & {\small 3/2}^{+}%
\end{array}%
$ & $%
\begin{array}{r}
{\small 4521.34 \pm 1.87} \\
{\small 4526.79 \pm 1.79}%
\end{array}%
$ & $%
\begin{array}{r}
{\small } \\
{\small }%
\end{array}%
$ & $%
\begin{array}{r}
{\small } \\
{\small }%
\end{array}%
$ & $%
\begin{array}{r}
{\small } \\
{\small }%
\end{array}%
$ & $%
\begin{array}{r}
{\small } \\
{\small }%
\end{array}%
$ & $%
\begin{array}{r}
{\small } \\
{\small }%
\end{array}%
$ \\ $%

\begin{array}{rr}
{\small 4}^{1}{\small S}_{1/2} & {\small 1/2}^{+} \\
{\small 4}^{3}{\small S}_{3/2} & {\small 3/2}^{+}%
\end{array}%
$ & $%
\begin{array}{r}
{\small 4821.15 \pm 1.96} \\
{\small 4823.50 \pm 1.94}%
\end{array}%
$ & $%
\begin{array}{r}
{\small } \\
{\small }%
\end{array}%
$ & $%
\begin{array}{r}
{\small } \\
{\small }%
\end{array}%
$ & $%
\begin{array}{r}
{\small } \\
{\small }%
\end{array}%
$ & $%
\begin{array}{r}
{\small } \\
{\small }%
\end{array}%
$ & $%
\begin{array}{r}
{\small } \\
{\small }%
\end{array}%
$ \\ $%

\begin{array}{rr}
{\small 5}^{1}{\small S}_{1/2} & {\small 1/2}^{+} \\
{\small 5}^{3}{\small S}_{3/2} & {\small 3/2}^{+}%
\end{array}%
$ & $%
\begin{array}{r}
{\small 5081.76 \pm 2.10} \\
{\small 5082.97 \pm 2.09}%
\end{array}%
$ & $%
\begin{array}{r}
{\small } \\
{\small }%
\end{array}%
$ & $%
\begin{array}{r}
{\small } \\
{\small }%
\end{array}%
$ & $%
\begin{array}{r}
{\small } \\
{\small }%
\end{array}%
$ & $%
\begin{array}{r}
{\small } \\
{\small }%
\end{array}%
$ & $%
\begin{array}{r}
{\small } \\
{\small }%
\end{array}%
$ \\
\hline

$
\begin{array}{rr}
{\small 1}^{2}{\small P}_{1/2} & {\small 1/2}^{-} \\
{\small 1}^{4}{\small P}_{1/2} & {\small 1/2}^{-} \\
{\small 1}^{2}{\small P}_{3/2} & {\small 3/2}^{-} \\
{\small 1}^{4}{\small P}_{3/2} & {\small 3/2}^{-} \\
{\small 1}^{4}{\small P}_{5/2} & {\small 5/2}^{-}%
\end{array}%
$ & $%
\begin{array}{r}
{\small 3959.65 \pm 7.78} \\
{\small 4019.53 \pm 3.38}\\
{\small 4035.53 \pm 2.19} \\
{\small 4061.50 \pm 1.94}\\
{\small 4095.96 \pm 4.06}%
\end{array}%
$ & $%
\begin{array}{r}
{\small 4053} \\
{\small 4136} \\
{\small 4101} \\
{\small 4196} \\
{\small 4155}%
\end{array}%
$ & $%
\begin{array}{r}
{\small 3998} \\
{\small 3985} \\
{\small 4014} \\
{\small 4025} \\
{\small 4050}%
\end{array}%
$ & $%
\begin{array}{r}
{\small 4081} \\
{\small 4073} \\
{\small 4077} \\
{\small 4079} \\
{\small 4089}%
\end{array}%
$ & $%
\begin{array}{r}
{\small } \\
{\small } \\
{\small } \\
{\small } \\
{\small }%
\end{array}%
$ & $%
\begin{array}{r}
{\small } \\
{\small }\\
{\small } \\
{\small }\\
{\small }%
\end{array}%
$ \\
\hline

$
\begin{array}{rr}
{\small 2}^{2}{\small P}_{1/2} & {\small 1/2}^{-} \\
{\small 2}^{4}{\small P}_{1/2} & {\small 1/2}^{-} \\
{\small 2}^{2}{\small P}_{3/2} & {\small 3/2}^{-} \\
{\small 2}^{4}{\small P}_{3/2} & {\small 3/2}^{-} \\
{\small 2}^{4}{\small P}_{5/2} & {\small 5/2}^{-}%
\end{array}%
$ & $%
\begin{array}{r}
{\small 4394.95 \pm 3.81} \\
{\small 4420.72 \pm 2.23}\\
{\small 4425.84 \pm 1.93} \\
{\small 4440.99 \pm 1.81}\\
{\small 4456.72 \pm 2.49}%
\end{array}%
$ & $%
\begin{array}{r}
{\small } \\
{\small } \\
{\small } \\
{\small } \\
{\small }%
\end{array}%
$ & $%
\begin{array}{r}
{\small } \\
{\small } \\
{\small } \\
{\small } \\
{\small }%
\end{array}%
$ & $%
\begin{array}{r}
{\small } \\
{\small } \\
{\small } \\
{\small } \\
{\small }%
\end{array}%
$ & $%
\begin{array}{r}
{\small } \\
{\small } \\
{\small } \\
{\small } \\
{\small }%
\end{array}%
$ & $%
\begin{array}{r}
{\small } \\
{\small }\\
{\small } \\
{\small }\\
{\small }%
\end{array}%
$ \\
\hline

$
\begin{array}{rr}
{\small 3}^{2}{\small P}_{1/2} & {\small 1/2}^{-} \\
{\small 3}^{4}{\small P}_{1/2} & {\small 1/2}^{-} \\
{\small 3}^{2}{\small P}_{3/2} & {\small 3/2}^{-} \\
{\small 3}^{4}{\small P}_{3/2} & {\small 3/2}^{-} \\
{\small 3}^{4}{\small P}_{5/2} & {\small 5/2}^{-}%
\end{array}%
$ & $%
\begin{array}{r}
{\small 4724.19 \pm 2.68} \\
{\small 4738.32 \pm 2.06}\\
{\small 4740.51 \pm 1.97} \\
{\small 4750.06 \pm 1.93}\\
{\small 4758.92 \pm 2.68}%
\end{array}%
$ & $%
\begin{array}{r}
{\small } \\
{\small } \\
{\small } \\
{\small } \\
{\small }%
\end{array}%
$ & $%
\begin{array}{r}
{\small } \\
{\small } \\
{\small } \\
{\small } \\
{\small }%
\end{array}%
$ & $%
\begin{array}{r}
{\small } \\
{\small } \\
{\small } \\
{\small } \\
{\small }%
\end{array}%
$ & $%
\begin{array}{r}
{\small } \\
{\small } \\
{\small } \\
{\small } \\
{\small }%
\end{array}%
$ & $%
\begin{array}{r}
{\small } \\
{\small } \\
{\small } \\
{\small } \\
{\small }%
\end{array}%
$ \\
\hline

$
\begin{array}{rr}
{\small 4}^{2}{\small P}_{1/2} & {\small 1/2}^{-} \\
{\small 4}^{4}{\small P}_{1/2} & {\small 1/2}^{-} \\
{\small 4}^{2}{\small P}_{3/2} & {\small 3/2}^{-} \\
{\small 4}^{4}{\small P}_{3/2} & {\small 3/2}^{-} \\
{\small 4}^{4}{\small P}_{5/2} & {\small 5/2}^{-}%
\end{array}%
$ & $%
\begin{array}{r}
{\small 5001.07 \pm 2.38} \\
{\small 5009.97 \pm 2.12}\\
{\small 5011.08 \pm 2.09} \\
{\small 5017.58 \pm 2.07}\\
{\small 5023.24 \pm 2.16}%
\end{array}%
$ & $%
\begin{array}{r}
{\small } \\
{\small } \\
{\small } \\
{\small } \\
{\small }%
\end{array}%
$ & $%
\begin{array}{r}
{\small } \\
{\small } \\
{\small } \\
{\small } \\
{\small }%
\end{array}%
$ & $%
\begin{array}{r}
{\small } \\
{\small } \\
{\small } \\
{\small } \\
{\small }%
\end{array}%
$ & $%
\begin{array}{r}
{\small } \\
{\small } \\
{\small } \\
{\small } \\
{\small }%
\end{array}%
$ & $%
\begin{array}{r}
{\small } \\
{\small } \\
{\small } \\
{\small } \\
{\small }%
\end{array}%
$ \\
\hline

$
\begin{array}{rr}
{\small 5}^{2}{\small P}_{1/2} & {\small 1/2}^{-} \\
{\small 5}^{4}{\small P}_{1/2} & {\small 1/2}^{-} \\
{\small 5}^{2}{\small P}_{3/2} & {\small 3/2}^{-} \\
{\small 5}^{4}{\small P}_{3/2} & {\small 3/2}^{-} \\
{\small 5}^{4}{\small P}_{5/2} & {\small 5/2}^{-}%
\end{array}%
$ & $%
\begin{array}{r}
{\small 5245.36 \pm 2.35} \\
{\small 5251.47 \pm 2.23}\\
{\small 5252.10 \pm 2.22} \\
{\small 5256.80 \pm 2.20}\\
{\small 5260.71 \pm 2.25}%
\end{array}%
$ & $%
\begin{array}{r}
{\small } \\
{\small } \\
{\small } \\
{\small } \\
{\small }%
\end{array}%
$ & $%
\begin{array}{r}
{\small } \\
{\small } \\
{\small } \\
{\small } \\
{\small }%
\end{array}%
$ & $%
\begin{array}{r}
{\small } \\
{\small } \\
{\small } \\
{\small } \\
{\small }%
\end{array}%
$ & $%
\begin{array}{r}
{\small } \\
{\small } \\
{\small } \\
{\small } \\
{\small }%
\end{array}%
$ & $%
\begin{array}{r}
{\small } \\
{\small } \\
{\small } \\
{\small } \\
{\small }%
\end{array}%
$ \\
\hline

$
\begin{array}{rr}
{\small 1}^{4}{\small D}_{1/2} & {\small 1/2}^{+} \\
{\small 1}^{2}{\small D}_{3/2} & {\small 3/2}^{+} \\
{\small 1}^{4}{\small D}_{3/2} & {\small 3/2}^{+} \\
{\small 1}^{2}{\small D}_{5/2} & {\small 5/2}^{+} \\
{\small 1}^{4}{\small D}_{5/2} & {\small 5/2}^{+} \\
{\small 1}^{4}{\small D}_{7/2} & {\small 7/2}^{+}%
\end{array}%
$ & $%
\begin{array}{r}
{\small 4276.03 \pm 5.71} \\
{\small 4298.42 \pm 3.94}\\
{\small 4319.29 \pm 2.60}\\
{\small 4334.96 \pm 6.49}\\
{\small 4353.94 \pm 6.48}\\
{\small 4384.09 \pm 3.99}%
\end{array}%
$ & $%
\begin{array}{r}
{\small } \\
{\small } \\
{\small } \\
{\small } \\
{\small } \\
{\small }%
\end{array}%
$ & $%
\begin{array}{r}
{\small } \\
{\small } \\
{\small } \\
{\small } \\
{\small } \\
{\small }%
\end{array}%
$ & $%
\begin{array}{r}
{\small } \\
{\small } \\
{\small } \\
{\small } \\
{\small } \\
{\small }%
\end{array}%
$ & $%
\begin{array}{r}
{\small } \\
{\small } \\
{\small } \\
{\small } \\
{\small } \\
{\small }%
\end{array}%
$ & $%
\begin{array}{r}
{\small }\\
{\small }\\
{\small }\\
{\small }\\
{\small }\\
{\small }%
\end{array}%
$ \\
\hline

$
\begin{array}{rr}
{\small 2}^{4}{\small D}_{1/2} & {\small 1/2}^{+} \\
{\small 2}^{2}{\small D}_{3/2} & {\small 3/2}^{+} \\
{\small 2}^{4}{\small D}_{3/2} & {\small 3/2}^{+} \\
{\small 2}^{2}{\small D}_{5/2} & {\small 5/2}^{+} \\
{\small 2}^{4}{\small D}_{5/2} & {\small 5/2}^{+} \\
{\small 2}^{4}{\small D}_{7/2} & {\small 7/2}^{+}%
\end{array}%
$ & $%
\begin{array}{r}
{\small 4630.51 \pm 3.64} \\
{\small 4643.22 \pm 2.77}\\
{\small 4655.37 \pm 2.17}\\
{\small 4664.06 \pm 2.82} \\
{\small 4675.25 \pm 2.82}\\
{\small 4692.38 \pm 2.78}%
\end{array}%
$ & $%
\begin{array}{r}
{\small } \\
{\small } \\
{\small } \\
{\small } \\
{\small } \\
{\small }%
\end{array}%
$ & $%
\begin{array}{r}
{\small } \\
{\small } \\
{\small } \\
{\small } \\
{\small } \\
{\small }%
\end{array}%
$ & $%
\begin{array}{r}
{\small } \\
{\small } \\
{\small } \\
{\small } \\
{\small } \\
{\small }%
\end{array}%
$ & $%
\begin{array}{r}
{\small } \\
{\small } \\
{\small } \\
{\small } \\
{\small } \\
{\small }%
\end{array}%
$ & $%
\begin{array}{r}
{\small }\\
{\small }\\
{\small }\\
{\small }\\
{\small }\\
{\small }%
\end{array}%
$ \\
\hline

$
\begin{array}{rr}
{\small 3}^{4}{\small D}_{1/2} & {\small 1/2}^{+} \\
{\small 3}^{2}{\small D}_{3/2} & {\small 3/2}^{+} \\
{\small 3}^{4}{\small D}_{3/2} & {\small 3/2}^{+} \\
{\small 3}^{2}{\small D}_{5/2} & {\small 5/2}^{+} \\
{\small 3}^{4}{\small D}_{5/2} & {\small 5/2}^{+} \\
{\small 3}^{4}{\small D}_{7/2} & {\small 7/2}^{+}%
\end{array}%
$ & $%
\begin{array}{r}
{\small 4922.09 \pm 2.86} \\
{\small 4930.24 \pm 2.41}\\
{\small 4938.16 \pm 2.15} \\
{\small 4943.65 \pm 2.21}\\
{\small 4951.00 \pm 2.21}\\
{\small 4962.00 \pm 2.43}%
\end{array}%
$ & $%
\begin{array}{r}
{\small } \\
{\small } \\
{\small } \\
{\small } \\
{\small } \\
{\small }%
\end{array}%
$ & $%
\begin{array}{r}
{\small } \\
{\small } \\
{\small } \\
{\small } \\
{\small } \\
{\small }%
\end{array}%
$ & $%
\begin{array}{r}
{\small } \\
{\small } \\
{\small } \\
{\small } \\
{\small } \\
{\small }%
\end{array}%
$ & $%
\begin{array}{r}
{\small } \\
{\small } \\
{\small } \\
{\small } \\
{\small } \\
{\small }%
\end{array}%
$ & $%
\begin{array}{r}
{\small }\\
{\small }\\
{\small }\\
{\small }\\
{\small }\\
{\small }%
\end{array}%
$ \\
\hline

$
\begin{array}{rr}
{\small 4}^{4}{\small D}_{1/2} & {\small 1/2}^{+} \\
{\small 4}^{2}{\small D}_{3/2} & {\small 3/2}^{+} \\
{\small 4}^{4}{\small D}_{3/2} & {\small 3/2}^{+} \\
{\small 4}^{2}{\small D}_{5/2} & {\small 5/2}^{+} \\
{\small 4}^{4}{\small D}_{5/2} & {\small 5/2}^{+} \\
{\small 4}^{4}{\small D}_{7/2} & {\small 7/2}^{+}%
\end{array}%
$ & $%
\begin{array}{r}
{\small 5176.05 \pm 2.59} \\
{\small 5181.71 \pm 2.36}\\
{\small 5187.26 \pm 2.23}\\
{\small 5191.05 \pm 2.21} \\
{\small 5196.22 \pm 2.21}\\
{\small 5203.87 \pm 2.36}%
\end{array}%
$ & $%
\begin{array}{r}
{\small } \\
{\small } \\
{\small } \\
{\small } \\
{\small } \\
{\small }%
\end{array}%
$ & $%
\begin{array}{r}
{\small } \\
{\small } \\
{\small } \\
{\small } \\
{\small } \\
{\small }%
\end{array}%
$ & $%
\begin{array}{r}
{\small } \\
{\small } \\
{\small } \\
{\small } \\
{\small } \\
{\small }%
\end{array}%
$ & $%
\begin{array}{r}
{\small } \\
{\small } \\
{\small } \\
{\small } \\
{\small } \\
{\small }%
\end{array}%
$ & $%
\begin{array}{r}
{\small }\\
{\small }\\
{\small }\\
{\small }\\
{\small }\\
{\small }%
\end{array}%
$ \\
\hline

$
\begin{array}{rr}
{\small 5}^{4}{\small D}_{1/2} & {\small 1/2}^{+} \\
{\small 5}^{2}{\small D}_{3/2} & {\small 3/2}^{+} \\
{\small 5}^{4}{\small D}_{3/2} & {\small 3/2}^{+} \\
{\small 5}^{2}{\small D}_{5/2} & {\small 5/2}^{+} \\
{\small 5}^{4}{\small D}_{5/2} & {\small 5/2}^{+} \\
{\small 5}^{4}{\small D}_{7/2} & {\small 7/2}^{+}%
\end{array}%
$ & $%
\begin{array}{r}
{\small 5404.32 \pm 2.53} \\
{\small 5408.48 \pm 2.40}\\
{\small 5412.59 \pm 2.33}\\
{\small 5415.35 \pm 2.32} \\
{\small 5419.19 \pm 2.31}\\
{\small 5424.81 \pm 2.40}%
\end{array}%
$ & $%
\begin{array}{r}
{\small } \\
{\small } \\
{\small } \\
{\small } \\
{\small } \\
{\small }%
\end{array}%
$ & $%
\begin{array}{r}
{\small } \\
{\small } \\
{\small } \\
{\small } \\
{\small } \\
{\small }%
\end{array}%
$ & $%
\begin{array}{r}
{\small } \\
{\small } \\
{\small } \\
{\small } \\
{\small } \\
{\small }%
\end{array}%
$ & $%
\begin{array}{r}
{\small } \\
{\small } \\
{\small } \\
{\small } \\
{\small } \\
{\small }%
\end{array}%
$ & $%
\begin{array}{r}
{\small }\\
{\small }\\
{\small }\\
{\small }\\
{\small }\\
{\small }%
\end{array}%
$ \\
\hline\hline
\end{tabular}}
\end{table}

\begin{table}[htbp]
\caption{Mass spectra (MeV) of $\Xi_{bc}$ baryons are given and compared with different quark models.}\label{ppddmm35}
\resizebox{\textwidth}{12cm}{\begin{tabular}{ccccccc}
\hline\hline
{\small State }\; $J^{P}$ &{Ours}&  \cite{Ebert:A11}  &  \cite{BEakins:A11} & \cite{Giannuzzi:A11} & \cite{Karlinere:A11}  &          \\
\hline
$%
\begin{array}{rr}
{\small 1}^{1}{\small S}_{1/2} & {\small 1/2}^{+} \\
{\small 1}^{3}{\small S}_{3/2} & {\small 3/2}^{+}%
\end{array}%
$ & $%
\begin{array}{r}
{\small 6745.89 \pm 4.05} \\
{\small 6783.84 \pm 2.40}%
\end{array}%
$ & $%
\begin{array}{r}
{\small 6933} \\
{\small 6980}%
\end{array}%
$ & $%
\begin{array}{r}
{\small 7014} \\
{\small 7064}%
\end{array}%
$ & $%
\begin{array}{r}
{\small 6904} \\
{\small 6936}%
\end{array}%
$ & $%
\begin{array}{r}
{\small 6914} \\
{\small 6969}%
\end{array}%
$ & $%
\begin{array}{r}
{\small } \\
{\small }%
\end{array}%
$ \\ $%

\begin{array}{rr}
{\small 2}^{1}{\small S}_{1/2} & {\small 1/2}^{+} \\
{\small 2}^{3}{\small S}_{3/2} & {\small 3/2}^{+}%
\end{array}%
$ & $%
\begin{array}{r}
{\small 7354.46 \pm 1.93} \\
{\small 7362.90 \pm 1.77}%
\end{array}%
$ & $%
\begin{array}{r}
{\small } \\
{\small }%
\end{array}%
$ & $%
\begin{array}{r}
{\small 7634} \\
{\small 7676}%
\end{array}%
$ & $%
\begin{array}{r}
{\small 7478} \\
{\small 7495}%
\end{array}%
$ & $%
\begin{array}{r}
{\small } \\
{\small }%
\end{array}%
$ & $%
\begin{array}{r}
{\small } \\
{\small }%
\end{array}%
$ \\ $%

\begin{array}{rr}
{\small 3}^{1}{\small S}_{1/2} & {\small 1/2}^{+} \\
{\small 3}^{3}{\small S}_{3/2} & {\small 3/2}^{+}%
\end{array}%
$ & $%
\begin{array}{r}
{\small 7788.87 \pm 1.98} \\
{\small 7791.53 \pm 1.96}%
\end{array}%
$ & $%
\begin{array}{r}
{\small } \\
{\small }%
\end{array}%
$ & $%
\begin{array}{r}
{\small } \\
{\small }%
\end{array}%
$ & $%
\begin{array}{r}
{\small 7904} \\
{\small 7917}%
\end{array}%
$ & $%
\begin{array}{r}
{\small } \\
{\small }%
\end{array}%
$ & $%
\begin{array}{r}
{\small } \\
{\small }%
\end{array}%
$ \\ $%

\begin{array}{rr}
{\small 4}^{1}{\small S}_{1/2} & {\small 1/2}^{+} \\
{\small 4}^{3}{\small S}_{3/2} & {\small 3/2}^{+}%
\end{array}%
$ & $%
\begin{array}{r}
{\small 8146.42 \pm 2.17} \\
{\small 8147.56 \pm 2.16}%
\end{array}%
$ & $%
\begin{array}{r}
{\small } \\
{\small }%
\end{array}%
$ & $%
\begin{array}{r}
{\small } \\
{\small }%
\end{array}%
$ & $%
\begin{array}{r}
{\small } \\
{\small }%
\end{array}%
$ & $%
\begin{array}{r}
{\small } \\
{\small }%
\end{array}%
$ & $%
\begin{array}{r}
{\small } \\
{\small }%
\end{array}%
$ \\ $%

\begin{array}{rr}
{\small 5}^{1}{\small S}_{1/2} & {\small 1/2}^{+} \\
{\small 5}^{3}{\small S}_{3/2} & {\small 3/2}^{+}%
\end{array}%
$ & $%
\begin{array}{r}
{\small 8457.95 \pm 2.35} \\
{\small 8458.54 \pm 2.35}%
\end{array}%
$ & $%
\begin{array}{r}
{\small } \\
{\small }%
\end{array}%
$ & $%
\begin{array}{r}
{\small } \\
{\small }%
\end{array}%
$ & $%
\begin{array}{r}
{\small } \\
{\small }%
\end{array}%
$ & $%
\begin{array}{r}
{\small } \\
{\small }%
\end{array}%
$ & $%
\begin{array}{r}
{\small } \\
{\small }%
\end{array}%
$ \\
\hline

$
\begin{array}{rr}
{\small 1}^{2}{\small P}_{1/2} & {\small 1/2}^{-} \\
{\small 1}^{4}{\small P}_{1/2} & {\small 1/2}^{-} \\
{\small 1}^{2}{\small P}_{3/2} & {\small 3/2}^{-} \\
{\small 1}^{4}{\small P}_{3/2} & {\small 3/2}^{-} \\
{\small 1}^{4}{\small P}_{5/2} & {\small 5/2}^{-}%
\end{array}%
$ & $%
\begin{array}{r}
{\small 7178.75 \pm 1.85} \\
{\small 7207.94 \pm 1.83}\\
{\small 7215.74 \pm 1.72} \\
{\small 7228.34 \pm 1.71}\\
{\small 7245.08 \pm 1.73}%
\end{array}%
$ & $%
\begin{array}{r}
{\small } \\
{\small } \\
{\small } \\
{\small } \\
{\small }%
\end{array}%
$ & $%
\begin{array}{r}
{\small 7397} \\
{\small 7390} \\
{\small 7392} \\
{\small 7394} \\
{\small 7399}%
\end{array}%
$ & $%
\begin{array}{r}
{\small } \\
{\small } \\
{\small } \\
{\small } \\
{\small }%
\end{array}%
$ & $%
\begin{array}{r}
{\small } \\
{\small } \\
{\small } \\
{\small } \\
{\small }%
\end{array}%
$ & $%
\begin{array}{r}
{\small } \\
{\small }\\
{\small } \\
{\small }\\
{\small }%
\end{array}%
$ \\
\hline

$
\begin{array}{rr}
{\small 2}^{2}{\small P}_{1/2} & {\small 1/2}^{-} \\
{\small 2}^{4}{\small P}_{1/2} & {\small 1/2}^{-} \\
{\small 2}^{2}{\small P}_{3/2} & {\small 3/2}^{-} \\
{\small 2}^{4}{\small P}_{3/2} & {\small 3/2}^{-} \\
{\small 2}^{4}{\small P}_{5/2} & {\small 5/2}^{-}%
\end{array}%
$ & $%
\begin{array}{r}
{\small 7663.76 \pm 1.94} \\
{\small 7676.31 \pm 1.93}\\
{\small 7678.81 \pm 1.91} \\
{\small 7686.17 \pm 1.91}\\
{\small 7693.82 \pm 1.92}%
\end{array}%
$ & $%
\begin{array}{r}
{\small } \\
{\small } \\
{\small } \\
{\small } \\
{\small }%
\end{array}%
$ & $%
\begin{array}{r}
{\small } \\
{\small } \\
{\small } \\
{\small } \\
{\small }%
\end{array}%
$ & $%
\begin{array}{r}
{\small } \\
{\small } \\
{\small } \\
{\small } \\
{\small }%
\end{array}%
$ & $%
\begin{array}{r}
{\small } \\
{\small } \\
{\small } \\
{\small } \\
{\small }%
\end{array}%
$ & $%
\begin{array}{r}
{\small } \\
{\small }\\
{\small } \\
{\small }\\
{\small }%
\end{array}%
$ \\
\hline

$
\begin{array}{rr}
{\small 3}^{2}{\small P}_{1/2} & {\small 1/2}^{-} \\
{\small 3}^{4}{\small P}_{1/2} & {\small 1/2}^{-} \\
{\small 3}^{2}{\small P}_{3/2} & {\small 3/2}^{-} \\
{\small 3}^{4}{\small P}_{3/2} & {\small 3/2}^{-} \\
{\small 3}^{4}{\small P}_{5/2} & {\small 5/2}^{-}%
\end{array}%
$ & $%
\begin{array}{r}
{\small 8045.21 \pm 2.13} \\
{\small 8052.09 \pm 2.12}\\
{\small 8053.16 \pm 2.11} \\
{\small 8057.80 \pm 2.11}\\
{\small 8062.10 \pm 2.13}%
\end{array}%
$ & $%
\begin{array}{r}
{\small } \\
{\small } \\
{\small } \\
{\small } \\
{\small }%
\end{array}%
$ & $%
\begin{array}{r}
{\small } \\
{\small } \\
{\small } \\
{\small } \\
{\small }%
\end{array}%
$ & $%
\begin{array}{r}
{\small } \\
{\small } \\
{\small } \\
{\small } \\
{\small }%
\end{array}%
$ & $%
\begin{array}{r}
{\small } \\
{\small } \\
{\small } \\
{\small } \\
{\small }%
\end{array}%
$ & $%
\begin{array}{r}
{\small } \\
{\small }\\
{\small } \\
{\small }\\
{\small }%
\end{array}%
$ \\
\hline

$
\begin{array}{rr}
{\small 4}^{2}{\small P}_{1/2} & {\small 1/2}^{-} \\
{\small 4}^{4}{\small P}_{1/2} & {\small 1/2}^{-} \\
{\small 4}^{2}{\small P}_{3/2} & {\small 3/2}^{-} \\
{\small 4}^{4}{\small P}_{3/2} & {\small 3/2}^{-} \\
{\small 4}^{4}{\small P}_{5/2} & {\small 5/2}^{-}%
\end{array}%
$ & $%
\begin{array}{r}
{\small 8370.91 \pm 2.31} \\
{\small 8375.21 \pm 2.31}\\
{\small 8375.78 \pm 2.30} \\
{\small 8378.94 \pm 2.30}\\
{\small 8381.69 \pm 2.30}%
\end{array}%
$ & $%
\begin{array}{r}
{\small } \\
{\small } \\
{\small } \\
{\small } \\
{\small }%
\end{array}%
$ & $%
\begin{array}{r}
{\small } \\
{\small }\\
{\small } \\
{\small }\\
{\small }%
\end{array}%
$ & $%
\begin{array}{r}
{\small } \\
{\small } \\
{\small } \\
{\small } \\
{\small }%
\end{array}%
$ & $%
\begin{array}{r}
{\small } \\
{\small } \\
{\small } \\
{\small } \\
{\small }%
\end{array}%
$ & $%
\begin{array}{r}
{\small } \\
{\small }\\
{\small } \\
{\small }\\
{\small }%
\end{array}%
$ \\
\hline

$
\begin{array}{rr}
{\small 5}^{2}{\small P}_{1/2} & {\small 1/2}^{-} \\
{\small 5}^{4}{\small P}_{1/2} & {\small 1/2}^{-} \\
{\small 5}^{2}{\small P}_{3/2} & {\small 3/2}^{-} \\
{\small 5}^{4}{\small P}_{3/2} & {\small 3/2}^{-} \\
{\small 5}^{4}{\small P}_{5/2} & {\small 5/2}^{-}%
\end{array}%
$ & $%
\begin{array}{r}
{\small 8660.29 \pm 2.48} \\
{\small 8663.26 \pm 2.48}\\
{\small 8663.57 \pm 2.48} \\
{\small 8665.85 \pm 2.48}\\
{\small 8667.75 \pm 2.48}%
\end{array}%
$ & $%
\begin{array}{r}
{\small } \\
{\small } \\
{\small } \\
{\small } \\
{\small }%
\end{array}%
$ & $%
\begin{array}{r}
{\small } \\
{\small } \\
{\small } \\
{\small } \\
{\small }%
\end{array}%
$ & $%
\begin{array}{r}
{\small } \\
{\small } \\
{\small } \\
{\small } \\
{\small }%
\end{array}%
$ & $%
\begin{array}{r}
{\small } \\
{\small } \\
{\small } \\
{\small } \\
{\small }%
\end{array}%
$ & $%
\begin{array}{r}
{\small } \\
{\small }\\
{\small } \\
{\small }\\
{\small }%
\end{array}%
$ \\
\hline

$
\begin{array}{rr}
{\small 1}^{4}{\small D}_{1/2} & {\small 1/2}^{+} \\
{\small 1}^{2}{\small D}_{3/2} & {\small 3/2}^{+} \\
{\small 1}^{4}{\small D}_{3/2} & {\small 3/2}^{+} \\
{\small 1}^{2}{\small D}_{5/2} & {\small 5/2}^{+} \\
{\small 1}^{4}{\small D}_{5/2} & {\small 5/2}^{+} \\
{\small 1}^{4}{\small D}_{7/2} & {\small 7/2}^{+}%
\end{array}%
$ & $%
\begin{array}{r}
{\small 7538.71 \pm 1.94} \\
{\small 7549.60 \pm 1.89}\\
{\small 7559.75 \pm 1.89}\\
{\small 7567.37 \pm 2.01}\\
{\small 7576.60 \pm 2.01}\\
{\small 7591.26 \pm 1.90}%
\end{array}%
$ & $%
\begin{array}{r}
{\small } \\
{\small } \\
{\small } \\
{\small } \\
{\small } \\
{\small }%
\end{array}%
$ & $%
\begin{array}{r}
{\small } \\
{\small } \\
{\small } \\
{\small } \\
{\small } \\
{\small }%
\end{array}%
$ & $%
\begin{array}{r}
{\small } \\
{\small } \\
{\small } \\
{\small } \\
{\small } \\
{\small }%
\end{array}%
$ & $%
\begin{array}{r}
{\small } \\
{\small } \\
{\small } \\
{\small } \\
{\small } \\
{\small }%
\end{array}%
$ & $%
\begin{array}{r}
{\small }\\
{\small }\\
{\small }\\
{\small }\\
{\small }\\
{\small }%
\end{array}%
$ \\
\hline

$
\begin{array}{rr}
{\small 2}^{4}{\small D}_{1/2} & {\small 1/2}^{+} \\
{\small 2}^{2}{\small D}_{3/2} & {\small 3/2}^{+} \\
{\small 2}^{4}{\small D}_{3/2} & {\small 3/2}^{+} \\
{\small 2}^{2}{\small D}_{5/2} & {\small 5/2}^{+} \\
{\small 2}^{4}{\small D}_{5/2} & {\small 5/2}^{+} \\
{\small 2}^{4}{\small D}_{7/2} & {\small 7/2}^{+}%
\end{array}%
$ & $%
\begin{array}{r}
{\small 7943.45 \pm 2.09} \\
{\small 7949.63 \pm 2.07}\\
{\small 7955.54 \pm 2.07}\\
{\small 7959.77 \pm 2.07} \\
{\small 7965.21 \pm 2.07}\\
{\small 7973.54 \pm 2.08}%
\end{array}%
$ & $%
\begin{array}{r}
{\small } \\
{\small } \\
{\small } \\
{\small } \\
{\small } \\
{\small }%
\end{array}%
$ & $%
\begin{array}{r}
{\small } \\
{\small } \\
{\small } \\
{\small } \\
{\small } \\
{\small }%
\end{array}%
$ & $%
\begin{array}{r}
{\small } \\
{\small } \\
{\small } \\
{\small } \\
{\small } \\
{\small }%
\end{array}%
$ & $%
\begin{array}{r}
{\small } \\
{\small } \\
{\small } \\
{\small } \\
{\small } \\
{\small }%
\end{array}%
$ & $%
\begin{array}{r}
{\small }\\
{\small }\\
{\small }\\
{\small }\\
{\small }\\
{\small }%
\end{array}%
$ \\
\hline

$
\begin{array}{rr}
{\small 3}^{4}{\small D}_{1/2} & {\small 1/2}^{+} \\
{\small 3}^{2}{\small D}_{3/2} & {\small 3/2}^{+} \\
{\small 3}^{4}{\small D}_{3/2} & {\small 3/2}^{+} \\
{\small 3}^{2}{\small D}_{5/2} & {\small 5/2}^{+} \\
{\small 3}^{4}{\small D}_{5/2} & {\small 5/2}^{+} \\
{\small 3}^{4}{\small D}_{7/2} & {\small 7/2}^{+}%
\end{array}%
$ & $%
\begin{array}{r}
{\small 8283.22 \pm 2.27} \\
{\small 8287.19 \pm 2.26}\\
{\small 8291.04 \pm 2.26} \\
{\small 8293.71 \pm 2.26}\\
{\small 8297.28 \pm 2.26}\\
{\small 8302.63 \pm 2.26}%
\end{array}%
$ & $%
\begin{array}{r}
{\small } \\
{\small } \\
{\small } \\
{\small } \\
{\small } \\
{\small }%
\end{array}%
$ & $%
\begin{array}{r}
{\small } \\
{\small } \\
{\small } \\
{\small } \\
{\small } \\
{\small }%
\end{array}%
$ & $%
\begin{array}{r}
{\small } \\
{\small } \\
{\small } \\
{\small } \\
{\small } \\
{\small }%
\end{array}%
$ & $%
\begin{array}{r}
{\small } \\
{\small } \\
{\small } \\
{\small } \\
{\small } \\
{\small }%
\end{array}%
$ & $%
\begin{array}{r}
{\small }\\
{\small }\\
{\small }\\
{\small }\\
{\small }\\
{\small }%
\end{array}%
$ \\
\hline

$
\begin{array}{rr}
{\small 4}^{4}{\small D}_{1/2} & {\small 1/2}^{+} \\
{\small 4}^{2}{\small D}_{3/2} & {\small 3/2}^{+} \\
{\small 4}^{4}{\small D}_{3/2} & {\small 3/2}^{+} \\
{\small 4}^{2}{\small D}_{5/2} & {\small 5/2}^{+} \\
{\small 4}^{4}{\small D}_{5/2} & {\small 5/2}^{+} \\
{\small 4}^{4}{\small D}_{7/2} & {\small 7/2}^{+}%
\end{array}%
$ & $%
\begin{array}{r}
{\small 8582.17 \pm 2.44} \\
{\small 8584.92 \pm 2.44}\\
{\small 8587.63 \pm 2.44}\\
{\small 8589.47 \pm 2.43} \\
{\small 8591.98 \pm 2.43}\\
{\small 8595.70 \pm 2.44}%
\end{array}%
$ & $%
\begin{array}{r}
{\small } \\
{\small } \\
{\small } \\
{\small } \\
{\small } \\
{\small }%
\end{array}%
$ & $%
\begin{array}{r}
{\small } \\
{\small } \\
{\small } \\
{\small } \\
{\small } \\
{\small }%
\end{array}%
$ & $%
\begin{array}{r}
{\small } \\
{\small } \\
{\small } \\
{\small } \\
{\small } \\
{\small }%
\end{array}%
$ & $%
\begin{array}{r}
{\small } \\
{\small } \\
{\small } \\
{\small } \\
{\small } \\
{\small }%
\end{array}%
$ & $%
\begin{array}{r}
{\small }\\
{\small }\\
{\small }\\
{\small }\\
{\small }\\
{\small }%
\end{array}%
$ \\
\hline

$
\begin{array}{rr}
{\small 5}^{4}{\small D}_{1/2} & {\small 1/2}^{+} \\
{\small 5}^{2}{\small D}_{3/2} & {\small 3/2}^{+} \\
{\small 5}^{4}{\small D}_{3/2} & {\small 3/2}^{+} \\
{\small 5}^{2}{\small D}_{5/2} & {\small 5/2}^{+} \\
{\small 5}^{4}{\small D}_{5/2} & {\small 5/2}^{+} \\
{\small 5}^{4}{\small D}_{7/2} & {\small 7/2}^{+}%
\end{array}%
$ & $%
\begin{array}{r}
{\small 8852.39 \pm 2.60} \\
{\small 8854.42 \pm 2.60}\\
{\small 8856.42 \pm 2.60}\\
{\small 8857.76 \pm 2.60} \\
{\small 8859.62 \pm 2.60}\\
{\small 8862.36 \pm 2.60}%
\end{array}%
$ & $%
\begin{array}{r}
{\small } \\
{\small } \\
{\small } \\
{\small } \\
{\small } \\
{\small }%
\end{array}%
$ & $%
\begin{array}{r}
{\small } \\
{\small } \\
{\small } \\
{\small } \\
{\small } \\
{\small }%
\end{array}%
$ & $%
\begin{array}{r}
{\small } \\
{\small } \\
{\small } \\
{\small } \\
{\small } \\
{\small }%
\end{array}%
$ \\
\hline\hline
\end{tabular}}
\end{table}

\begin{table}[htbp]
\caption{Mass spectra (MeV) of $\Xi_{bb}$ baryons are given and compared with different quark models.}\label{ppddmm45}
\resizebox{\textwidth}{12cm}{\begin{tabular}{ccccccc}
\hline\hline
{\small State }\; $J^{P}$  &{Ours}&   \cite{Ebert:A11}   &  \cite{Zhong:A11} &   \cite{ChenLuo:H11}    & \cite{BEakins:A11}   &       \\
\hline
$%
\begin{array}{rr}
{\small 1}^{1}{\small S}_{1/2} & {\small 1/2}^{+} \\
{\small 1}^{3}{\small S}_{3/2} & {\small 3/2}^{+}%
\end{array}%
$ & $%
\begin{array}{r}
{\small 10287.48 \pm 4.39} \\
{\small 10311.22 \pm 3.90}%
\end{array}%
$ & $%
\begin{array}{r}
{\small 10202} \\
{\small 10237}%
\end{array}%
$ & $%
\begin{array}{r}
{\small 10138} \\
{\small 10169}%
\end{array}%
$ & $%
\begin{array}{r}
{\small 10171} \\
{\small 10195}%
\end{array}%
$ & $%
\begin{array}{r}
{\small 10322} \\
{\small 10352}%
\end{array}%
$ & $%
\begin{array}{r}
{\small } \\
{\small }%
\end{array}%
$ \\ $%

\begin{array}{rr}
{\small 2}^{1}{\small S}_{1/2} & {\small 1/2}^{+} \\
{\small 2}^{3}{\small S}_{3/2} & {\small 3/2}^{+}%
\end{array}%
$ & $%
\begin{array}{r}
{\small 10961.90 \pm 4.01} \\
{\small 10967.20 \pm 3.98}%
\end{array}%
$ & $%
\begin{array}{r}
{\small 10832} \\
{\small 10860}%
\end{array}%
$ & $%
\begin{array}{r}
{\small 10662} \\
{\small 10675}%
\end{array}%
$ & $%
\begin{array}{r}
{\small 10738} \\
{\small 10753}%
\end{array}%
$ & $%
\begin{array}{r}
{\small 10940} \\
{\small 10972}%
\end{array}%
$ & $%
\begin{array}{r}
{\small } \\
{\small }%
\end{array}%
$ \\ $%

\begin{array}{rr}
{\small 3}^{1}{\small S}_{1/2} & {\small 1/2}^{+} \\
{\small 3}^{3}{\small S}_{3/2} & {\small 3/2}^{+}%
\end{array}%
$ & $%
\begin{array}{r}
{\small 11448.40 \pm 4.15} \\
{\small 11450.10 \pm 4.15}%
\end{array}%
$ & $%
\begin{array}{r}
{\small } \\
{\small }%
\end{array}%
$ & $%
\begin{array}{r}
{\small } \\
{\small }%
\end{array}%
$ & $%
\begin{array}{r}
{\small } \\
{\small }%
\end{array}%
$ & $%
\begin{array}{r}
{\small } \\
{\small }%
\end{array}%
$ & $%
\begin{array}{r}
{\small } \\
{\small }%
\end{array}%
$ \\ $%

\begin{array}{rr}
{\small 4}^{1}{\small S}_{1/2} & {\small 1/2}^{+} \\
{\small 4}^{3}{\small S}_{3/2} & {\small 3/2}^{+}%
\end{array}%
$ & $%
\begin{array}{r}
{\small 11849.90 \pm 4.30} \\
{\small 11850.60 \pm 4.30}%
\end{array}%
$ & $%
\begin{array}{r}
{\small } \\
{\small }%
\end{array}%
$ & $%
\begin{array}{r}
{\small } \\
{\small }%
\end{array}%
$ & $%
\begin{array}{r}
{\small } \\
{\small }%
\end{array}%
$ & $%
\begin{array}{r}
{\small } \\
{\small }%
\end{array}%
$ & $%
\begin{array}{r}
{\small } \\
{\small }%
\end{array}%
$ \\ $%

\begin{array}{rr}
{\small 5}^{1}{\small S}_{1/2} & {\small 1/2}^{+} \\
{\small 5}^{3}{\small S}_{3/2} & {\small 3/2}^{+}%
\end{array}%
$ & $%
\begin{array}{r}
{\small 12200.00 \pm 4.45} \\
{\small 12200.40 \pm 4.45}%
\end{array}%
$ & $%
\begin{array}{r}
{\small } \\
{\small }%
\end{array}%
$ & $%
\begin{array}{r}
{\small } \\
{\small }%
\end{array}%
$ & $%
\begin{array}{r}
{\small } \\
{\small }%
\end{array}%
$ & $%
\begin{array}{r}
{\small } \\
{\small }%
\end{array}%
$ & $%
\begin{array}{r}
{\small } \\
{\small }%
\end{array}%
$ \\
\hline

$
\begin{array}{rr}
{\small 1}^{2}{\small P}_{1/2} & {\small 1/2}^{-} \\
{\small 1}^{4}{\small P}_{1/2} & {\small 1/2}^{-} \\
{\small 1}^{2}{\small P}_{3/2} & {\small 3/2}^{-} \\
{\small 1}^{4}{\small P}_{3/2} & {\small 3/2}^{-} \\
{\small 1}^{4}{\small P}_{5/2} & {\small 5/2}^{-}%
\end{array}%
$ & $%
\begin{array}{r}
{\small 10787.80 \pm 3.96} \\
{\small 10801.40 \pm 3.96}\\
{\small 10804.30 \pm 3.94} \\
{\small 10814.70 \pm 3.94}\\
{\small 10826.00 \pm 3.95}%
\end{array}%
$ & $%
\begin{array}{r}
{\small 10632} \\
{\small 10675} \\
{\small 10647} \\
{\small 10694} \\
{\small 10661}%
\end{array}%
$ & $%
\begin{array}{r}
{\small 10525} \\
{\small 10504} \\
{\small 10526} \\
{\small 10528} \\
{\small 10547}%
\end{array}%
$ & $%
\begin{array}{r}
{\small 10593} \\
{\small 10547} \\
{\small 10606} \\
{\small 10561} \\
{\small 10560}%
\end{array}%
$ & $%
\begin{array}{r}
{\small 10694} \\
{\small 10694} \\
{\small 10691} \\
{\small 10692} \\
{\small 10695}%
\end{array}%
$ & $%
\begin{array}{r}
{\small } \\
{\small }\\
{\small } \\
{\small }\\
{\small }%
\end{array}%
$ \\
\hline

$
\begin{array}{rr}
{\small 2}^{2}{\small P}_{1/2} & {\small 1/2}^{-} \\
{\small 2}^{4}{\small P}_{1/2} & {\small 1/2}^{-} \\
{\small 2}^{2}{\small P}_{3/2} & {\small 3/2}^{-} \\
{\small 2}^{4}{\small P}_{3/2} & {\small 3/2}^{-} \\
{\small 2}^{4}{\small P}_{5/2} & {\small 5/2}^{-}%
\end{array}%
$ & $%
\begin{array}{r}
{\small 11317.90 \pm 4.11} \\
{\small 11324.30 \pm 4.11}\\
{\small 11325.30 \pm 4.10} \\
{\small 11330.70 \pm 4.10}\\
{\small 11335.70 \pm 4.10}%
\end{array}%
$ & $%
\begin{array}{r}
{\small } \\
{\small } \\
{\small } \\
{\small } \\
{\small }%
\end{array}%
$ & $%
\begin{array}{r}
{\small } \\
{\small } \\
{\small } \\
{\small } \\
{\small }%
\end{array}%
$ & $%
\begin{array}{r}
{\small } \\
{\small } \\
{\small } \\
{\small } \\
{\small }%
\end{array}%
$ & $%
\begin{array}{r}
{\small } \\
{\small } \\
{\small } \\
{\small } \\
{\small }%
\end{array}%
$ & $%
\begin{array}{r}
{\small } \\
{\small }\\
{\small } \\
{\small }\\
{\small }%
\end{array}%
$ \\
\hline

$
\begin{array}{rr}
{\small 3}^{2}{\small P}_{1/2} & {\small 1/2}^{-} \\
{\small 3}^{4}{\small P}_{1/2} & {\small 1/2}^{-} \\
{\small 3}^{2}{\small P}_{3/2} & {\small 3/2}^{-} \\
{\small 3}^{4}{\small P}_{3/2} & {\small 3/2}^{-} \\
{\small 3}^{4}{\small P}_{5/2} & {\small 5/2}^{-}%
\end{array}%
$ & $%
\begin{array}{r}
{\small 11741.70 \pm 4.26} \\
{\small 11745.40 \pm 4.26}\\
{\small 11745.80 \pm 4.26} \\
{\small 11749.10 \pm 4.26}\\
{\small 11751.90 \pm 4.26}%
\end{array}%
$ & $%
\begin{array}{r}
{\small } \\
{\small } \\
{\small } \\
{\small } \\
{\small }%
\end{array}%
$ & $%
\begin{array}{r}
{\small } \\
{\small } \\
{\small } \\
{\small } \\
{\small }%
\end{array}%
$ & $%
\begin{array}{r}
{\small } \\
{\small } \\
{\small } \\
{\small } \\
{\small }%
\end{array}%
$ & $%
\begin{array}{r}
{\small } \\
{\small } \\
{\small } \\
{\small } \\
{\small }%
\end{array}%
$ & $%
\begin{array}{r}
{\small } \\
{\small }\\
{\small } \\
{\small }\\
{\small }%
\end{array}%
$ \\
\hline

$
\begin{array}{rr}
{\small 4}^{2}{\small P}_{1/2} & {\small 1/2}^{-} \\
{\small 4}^{4}{\small P}_{1/2} & {\small 1/2}^{-} \\
{\small 4}^{2}{\small P}_{3/2} & {\small 3/2}^{-} \\
{\small 4}^{4}{\small P}_{3/2} & {\small 3/2}^{-} \\
{\small 4}^{4}{\small P}_{5/2} & {\small 5/2}^{-}%
\end{array}%
$ & $%
\begin{array}{r}
{\small 12105.71 \pm 4.41} \\
{\small 12108.06 \pm 4.41}\\
{\small 12108.19 \pm 4.41} \\
{\small 12110.41 \pm 4.41}\\
{\small 12112.19 \pm 4.41}%
\end{array}%
$ & $%
\begin{array}{r}
{\small } \\
{\small } \\
{\small } \\
{\small } \\
{\small }%
\end{array}%
$ & $%
\begin{array}{r}
{\small } \\
{\small } \\
{\small } \\
{\small } \\
{\small }%
\end{array}%
$ & $%
\begin{array}{r}
{\small } \\
{\small } \\
{\small } \\
{\small } \\
{\small }%
\end{array}%
$ & $%
\begin{array}{r}
{\small } \\
{\small } \\
{\small } \\
{\small } \\
{\small }%
\end{array}%
$ & $%
\begin{array}{r}
{\small } \\
{\small }\\
{\small } \\
{\small }\\
{\small }%
\end{array}%
$ \\
\hline

$
\begin{array}{rr}
{\small 5}^{2}{\small P}_{1/2} & {\small 1/2}^{-} \\
{\small 5}^{4}{\small P}_{1/2} & {\small 1/2}^{-} \\
{\small 5}^{2}{\small P}_{3/2} & {\small 3/2}^{-} \\
{\small 5}^{4}{\small P}_{3/2} & {\small 3/2}^{-} \\
{\small 5}^{4}{\small P}_{5/2} & {\small 5/2}^{-}%
\end{array}%
$ & $%
\begin{array}{r}
{\small 12429.85 \pm 4.55} \\
{\small 12431.51 \pm 4.55}\\
{\small 12431.57 \pm 4.55} \\
{\small 12433.14 \pm 4.55}\\
{\small 12434.37 \pm 4.55}%
\end{array}%
$ & $%
\begin{array}{r}
{\small } \\
{\small } \\
{\small } \\
{\small } \\
{\small }%
\end{array}%
$ & $%
\begin{array}{r}
{\small } \\
{\small } \\
{\small } \\
{\small } \\
{\small }%
\end{array}%
$ & $%
\begin{array}{r}
{\small } \\
{\small } \\
{\small } \\
{\small } \\
{\small }%
\end{array}%
$ & $%
\begin{array}{r}
{\small } \\
{\small } \\
{\small } \\
{\small } \\
{\small }%
\end{array}%
$ & $%
\begin{array}{r}
{\small } \\
{\small }\\
{\small } \\
{\small }\\
{\small }%
\end{array}%
$ \\
\hline

$
\begin{array}{rr}
{\small 1}^{4}{\small D}_{1/2} & {\small 1/2}^{+} \\
{\small 1}^{2}{\small D}_{3/2} & {\small 3/2}^{+} \\
{\small 1}^{4}{\small D}_{3/2} & {\small 3/2}^{+} \\
{\small 1}^{2}{\small D}_{5/2} & {\small 5/2}^{+} \\
{\small 1}^{4}{\small D}_{5/2} & {\small 5/2}^{+} \\
{\small 1}^{4}{\small D}_{7/2} & {\small 7/2}^{+}%
\end{array}%
$ & $%
\begin{array}{r}
{\small 11182.28 \pm 4.08} \\
{\small 11188.83 \pm 4.07}\\
{\small 11195.16 \pm 4.07}\\
{\small 11199.74 \pm 4.07}\\
{\small 11205.72 \pm 4.07}\\
{\small 11214.97 \pm 4.07}%
\end{array}%
$ & $%
\begin{array}{r}
{\small } \\
{\small } \\
{\small } \\
{\small } \\
{\small } \\
{\small }%
\end{array}%
$ & $%
\begin{array}{r}
{\small } \\
{\small } \\
{\small } \\
{\small } \\
{\small } \\
{\small }%
\end{array}%
$ & $%
\begin{array}{r}
{\small 10913} \\
{\small 10918} \\
{\small 10798} \\
{\small 10921} \\
{\small 10803} \\
{\small 10805}%
\end{array}%
$ & $%
\begin{array}{r}
{\small } \\
{\small } \\
{\small } \\
{\small } \\
{\small } \\
{\small }%
\end{array}%
$ & $%
\begin{array}{r}
{\small }\\
{\small }\\
{\small }\\
{\small }\\
{\small }\\
{\small }%
\end{array}%
$ \\
\hline

$
\begin{array}{rr}
{\small 2}^{4}{\small D}_{1/2} & {\small 1/2}^{+} \\
{\small 2}^{2}{\small D}_{3/2} & {\small 3/2}^{+} \\
{\small 2}^{4}{\small D}_{3/2} & {\small 3/2}^{+} \\
{\small 2}^{2}{\small D}_{5/2} & {\small 5/2}^{+} \\
{\small 2}^{4}{\small D}_{5/2} & {\small 5/2}^{+} \\
{\small 2}^{4}{\small D}_{7/2} & {\small 7/2}^{+}%
\end{array}%
$ & $%
\begin{array}{r}
{\small 11630.45 \pm 4.23} \\
{\small 11634.20 \pm 4.22}\\
{\small 11637.89 \pm 4.22}\\
{\small 11640.46 \pm 4.22} \\
{\small 11643.95 \pm 4.22}\\
{\small 11649.19 \pm 4.22}%
\end{array}%
$ & $%
\begin{array}{r}
{\small } \\
{\small } \\
{\small } \\
{\small } \\
{\small } \\
{\small }%
\end{array}%
$ & $%
\begin{array}{r}
{\small } \\
{\small } \\
{\small } \\
{\small } \\
{\small } \\
{\small }%
\end{array}%
$ & $%
\begin{array}{r}
{\small } \\
{\small } \\
{\small } \\
{\small } \\
{\small } \\
{\small }%
\end{array}%
$ & $%
\begin{array}{r}
{\small } \\
{\small } \\
{\small } \\
{\small } \\
{\small } \\
{\small }%
\end{array}%
$ & $%
\begin{array}{r}
{\small }\\
{\small }\\
{\small }\\
{\small }\\
{\small }\\
{\small }%
\end{array}%
$ \\
\hline

$
\begin{array}{rr}
{\small 3}^{4}{\small D}_{1/2} & {\small 1/2}^{+} \\
{\small 3}^{2}{\small D}_{3/2} & {\small 3/2}^{+} \\
{\small 3}^{4}{\small D}_{3/2} & {\small 3/2}^{+} \\
{\small 3}^{2}{\small D}_{5/2} & {\small 5/2}^{+} \\
{\small 3}^{4}{\small D}_{5/2} & {\small 5/2}^{+} \\
{\small 3}^{4}{\small D}_{7/2} & {\small 7/2}^{+}%
\end{array}%
$ & $%
\begin{array}{r}
{\small 12009.20 \pm 4.37} \\
{\small 12011.63 \pm 4.37}\\
{\small 12014.03 \pm 4.37} \\
{\small 12015.66 \pm 4.37}\\
{\small 12017.94 \pm 4.37}\\
{\small 12021.30 \pm 4.37}%
\end{array}%
$ & $%
\begin{array}{r}
{\small } \\
{\small } \\
{\small } \\
{\small } \\
{\small } \\
{\small }%
\end{array}%
$ & $%
\begin{array}{r}
{\small } \\
{\small } \\
{\small } \\
{\small } \\
{\small } \\
{\small }%
\end{array}%
$ & $%
\begin{array}{r}
{\small } \\
{\small } \\
{\small } \\
{\small } \\
{\small } \\
{\small }%
\end{array}%
$ & $%
\begin{array}{r}
{\small } \\
{\small } \\
{\small } \\
{\small } \\
{\small } \\
{\small }%
\end{array}%
$ & $%
\begin{array}{r}
{\small }\\
{\small }\\
{\small }\\
{\small }\\
{\small }\\
{\small }%
\end{array}%
$ \\
\hline

$
\begin{array}{rr}
{\small 4}^{4}{\small D}_{1/2} & {\small 1/2}^{+} \\
{\small 4}^{2}{\small D}_{3/2} & {\small 3/2}^{+} \\
{\small 4}^{4}{\small D}_{3/2} & {\small 3/2}^{+} \\
{\small 4}^{2}{\small D}_{5/2} & {\small 5/2}^{+} \\
{\small 4}^{4}{\small D}_{5/2} & {\small 5/2}^{+} \\
{\small 4}^{4}{\small D}_{7/2} & {\small 7/2}^{+}%
\end{array}%
$ & $%
\begin{array}{r}
{\small 12343.54 \pm 4.52} \\
{\small 12345.22 \pm 4.51}\\
{\small 12346.91 \pm 4.51}\\
{\small 12348.04 \pm 4.51} \\
{\small 12349.64 \pm 4.51}\\
{\small 12351.98 \pm 4.51}%
\end{array}%
$ & $%
\begin{array}{r}
{\small } \\
{\small } \\
{\small } \\
{\small } \\
{\small } \\
{\small }%
\end{array}%
$ & $%
\begin{array}{r}
{\small } \\
{\small } \\
{\small } \\
{\small } \\
{\small } \\
{\small }%
\end{array}%
$ & $%
\begin{array}{r}
{\small } \\
{\small } \\
{\small } \\
{\small } \\
{\small } \\
{\small }%
\end{array}%
$ & $%
\begin{array}{r}
{\small } \\
{\small } \\
{\small } \\
{\small } \\
{\small } \\
{\small }%
\end{array}%
$ & $%
\begin{array}{r}
{\small }\\
{\small }\\
{\small }\\
{\small }\\
{\small }\\
{\small }%
\end{array}%
$ \\
\hline

$
\begin{array}{rr}
{\small 5}^{4}{\small D}_{1/2} & {\small 1/2}^{+} \\
{\small 5}^{2}{\small D}_{3/2} & {\small 3/2}^{+} \\
{\small 5}^{4}{\small D}_{3/2} & {\small 3/2}^{+} \\
{\small 5}^{2}{\small D}_{5/2} & {\small 5/2}^{+} \\
{\small 5}^{4}{\small D}_{5/2} & {\small 5/2}^{+} \\
{\small 5}^{4}{\small D}_{7/2} & {\small 7/2}^{+}%
\end{array}%
$ & $%
\begin{array}{r}
{\small 12646.28 \pm 4.65} \\
{\small 12647.52 \pm 4.65}\\
{\small 12648.77 \pm 4.65}\\
{\small 12649.59 \pm 4.65} \\
{\small 12650.78 \pm 4.65}\\
{\small 12652.49 \pm 4.65}%
\end{array}%
$ & $%
\begin{array}{r}
{\small } \\
{\small } \\
{\small } \\
{\small } \\
{\small } \\
{\small }%
\end{array}%
$ & $%
\begin{array}{r}
{\small } \\
{\small } \\
{\small } \\
{\small } \\
{\small } \\
{\small }%
\end{array}%
$ & $%
\begin{array}{r}
{\small } \\
{\small } \\
{\small } \\
{\small } \\
{\small } \\
{\small }%
\end{array}%
$ & $%
\begin{array}{r}
{\small } \\
{\small } \\
{\small } \\
{\small } \\
{\small } \\
{\small }%
\end{array}%
$ & $%
\begin{array}{r}
{\small }\\
{\small }\\
{\small }\\
{\small }\\
{\small }\\
{\small }%
\end{array}%
$ \\
\hline\hline
\end{tabular}}
\end{table}

\section{The $\Omega_{QQ^{\prime}}$ baryons }\label{Sec.IV}

The doubly heavy $\Omega_{QQ^{\prime}}$ baryons are regarded as an important and unique part of the baryons in heavy-light quark system, as they are composed of light strange quark $s$. Up to now, the doubly heavy $\Omega_{QQ^{\prime}}$ baryons have not been reported yet by the experimental, and their nature are still unknown. Thus, the doubly heavy $\Omega_{QQ^{\prime}}$ baryons were studied by using various approaches, we recommend interested readers to see Refs. \cite{HYCheng:A11, LiYang:A11, Shahr:A11, Oudichhyahr:A11, KiselevL:A11, WengG:A11}. In this section, we may apply the similar methods to investigate the mass spectra of the $\Omega_{cc}$, $\Omega_{bc}$, and $\Omega_{bb}$ baryons. The results of the masses are listed in Table \ref{ppddmd55}, \ref{ppddmm50}, and \ref{ppddm5} for the $\Omega_{cc}$, $\Omega_{bc}$, and $\Omega_{bb}$ baryons,  respectively. The calculations are the same as for the the $\Xi_{QQ^{\prime}}$ baryons, the uncertainties in the mass results of the $\Omega_{QQ^{\prime}}$ baryon states are due to the errors in the experiment values as well as the uncertainties in determination of the parameters $a_{1}$, $a_{2}$, $b_{1}$, and $c_{1}$ and other input parameters.

For the $\Omega_{cc}$ baryon, the spin-averaged mass of the $1S$-wave state with $\mu_{ccs}$ = $m_{cc}m_{s}/(m_{cc}+m_{s})$ = 90.44 MeV by applying Eq. (\ref{PP421}) is predicted to be
\begin{eqnarray}
\bar M(\Omega_{cc}) &=& M_{cc}+\left(\frac{1}{2}(M_{cc})^{1/2}+\left( M_{s}+\frac{M_{cc}m_{s}}{m^{2}_{cc}}\frac{4\alpha_{s}}{3}\mu_{ccs}\right) ^{2}\right)^{\frac{1}{2}} \notag\\
            &=& 3692.18 \  \text{MeV}. \label{PPP666}
\end{eqnarray}
Combining $c_{1}(\Omega_{cc}, 1S)$ = 40.39 MeV as show in Table \ref{Table:PP87}, one can obtain the masses with $J^{P}$ = $1/2^{+}$ and $3/2^{+}$ for the $\Omega_{cc}$ states,
\begin{equation}
\begin{array}{c}
\text{State}|J,j\rangle: \\ [11pt]
M(\Omega _{cc},1S\text{):}
\end{array}%
\begin{array}{ccccc}
|1/2,1\rangle \qquad & |3/2,1\rangle  \\ [11pt]
\multicolumn{1}{r}{3651.79 \pm 6.92} \text{MeV}  \qquad & \multicolumn{1}{r}{3712.37 \pm 3.65} \text{MeV}.
\end{array}
\end{equation}
It can be seen that the result of $1S$-wave level-splitting mass is about $\Delta M(\Omega_{cc}, 1S)$ = 60 MeV smaller than $\Delta M(\Omega_{cc}, 1S)$ = 94 MeV in Ref. \cite{Ebert:A11}. Over the past two decades, the properties of the $\Omega_{cc}$ baryon with one strange quark can be explored experimentally. For example, the LHCb experiment had searched the doubly charmed baryon $\Omega^{+}_{cc}$ in the charged decay mode $\Omega^{+}_{cc}$ $\rightarrow$ $\Lambda^{+}_{c}K^{-}\pi^{+}$ in Ref. \cite{ Aaije:AA1O1}. No significant signal is observed within the invariant mass range of 3.6 GeV to 4.0 GeV. In our model, the spin-averaged mass of the $2S$-wave $\Omega_{cc}$ states with $L$ = 0, $n$ = 1 is calculated by using Eq. (\ref{PP421}), one has
\begin{equation}
\bar M(\Omega_{cc}, 2S) = 4169.58 \text{MeV}, \label{cg11}
\end{equation}
which is about 477 MeV higher than the spin-averaged mass (\ref{PPP666}) of the $1S$-wave states. Thus, the results of the masses for the $2S$-wave are
\begin{equation}
\begin{array}{c}
\text{State}|J,j\rangle: \\ [11pt]
M(\Omega _{cc},2S\text{):}
\end{array}%
\begin{array}{ccccc}
|1/2,1\rangle \qquad & |3/2,1\rangle  \\ [11pt]
\multicolumn{1}{r}{4160.96 \pm 2.15} \text{MeV}  \qquad & \multicolumn{1}{r}{4173.89 \pm 1.69} \text{MeV}.
\end{array}
\end{equation}
On the other hand, we also calculated the spin-average of the $1P$-wave ($L$ = 1, $n$ = 0) for $\Omega_{cc}$ states,
\begin{eqnarray}
\bar M(\Omega_{cc}, 1P) &=& M_{cc}+\left(\frac{1}{2} (M_{cc})^{\frac{1}{2}}\times 2+\left( M_{s}+\frac{M_{cc}m_{s}}{m^{2}_{cc}}\frac{4\alpha_{s}}{3}\frac{\mu_{ccs}}{4}\right) ^{2}\right)^{\frac{1}{2}}   \notag \\
&=& 4055.73\ \text{MeV}, \label{weee123}
\end{eqnarray}
the parameters are
\begin{eqnarray}
a_{1}(\Omega_{cc},1P)&=&\frac{M_{c}M_{ss}}{M_{cc}M_{s}}\frac{{N^{\prime}_{a_{1}}}}{{N_{a_{1}}}}\frac{\alpha_{s}(\Omega_{cc})}{\alpha_{s}(\Omega_{c})}a_{1}(\Omega_{c}, 1P)=23.28\ \text{MeV}, \notag \\
a_{2}(\Omega_{cc},1P)&=&\frac{M_{c}M_{ss}}{M_{cc}M_{s}}\frac{{N^{\prime}_{a_{2}}}}{{N_{a_{2}}}}\frac{\alpha_{s}(\Omega_{cc})}{\alpha_{s}(\Omega_{c})}a_{2}(\Omega_{c}, 1P)=22.24\ \text{MeV}, \notag \\
b_{1}(\Omega_{cc},1P)&=&\frac{M_{c}M_{ss}}{M_{cc}M_{s}}\frac{{N^{\prime}_{b_{1}}}}{{N_{b_{1}}}}\frac{\alpha_{s}(\Omega_{cc})}{\alpha_{s}(\Omega_{c})}b_{1}(\Omega_{c}, 1P)=11.67\ \text{MeV}, \notag \\
c_{1}(\Omega_{cc},1P)&=&\frac{M_{c}M_{ss}}{M_{cc}M_{s}}\frac{{N^{\prime}_{c_{1}}}}{{N_{c_{1}}}}\frac{\alpha_{s}(\Omega_{cc})}{\alpha_{s}(\Omega_{c})}c_{1}(\Omega_{c}, 1P)=3.49\ \text{MeV}. \label{we1132}
\end{eqnarray}
The obtained masses with the bases $|J,j\rangle$ are
\begin{equation}
\begin{array}{c}
\text{State}|J,j\rangle: \\ [11pt]
M(\Omega _{cc},1P\text{):}
\end{array}%
\begin{array}{ccccc}
|1/2,0\rangle \ & |1/2,1\rangle \ & |3/2,1\rangle \ & |3/2,2\rangle \ & |5/2,2\rangle  \\ [11pt]
\multicolumn{1}{r}{3987.54} \text{MeV} & \multicolumn{1}{r}{4032.12} \text{MeV} & \multicolumn{1}{r}{4044.04} \text{MeV} & \multicolumn{1}{r}{4063.36} \text{MeV} & \multicolumn{1}{r}{4089.02} \text{MeV}.%
\end{array}
\end{equation}
Thus, the level-splitting mass of $1P$-wave $\Omega_{cc}$ states is expected to be about 50 MeV higher than the $\Xi_{cc}$ states in Table \ref{ppdm5}. More valuable information of the $\Omega_{cc}$ baryon can be provided to the further experimental exploration.

In Ref. \cite{Aaij:PP8866}, the LHCb experiment had searched the baryon $\Omega^{0}_{bc}$ ($bcs$) in the $\Omega^{0}_{bc}$ $\rightarrow$ $\Lambda^{+}_{c}\pi^{-}$ decay. The search of the $\Omega^{0}_{bc}$ baryon is performed in the mass range between 6700 MeV and 7300 MeV, no significant excess is found in the LHCb experiment. Similar to the case for the $\Omega_{bc}$ baryon, the spin-average mass of the ground states with $\mu_{bcs}$ = $m_{bc}m_{s}/(m_{bc}+m_{s})$= 91.99 MeV is obtained as follow,
\begin{eqnarray}
\bar M(\Omega_{bc}) &=& M_{bc}+\left(\frac{1}{2}(M_{bc})^{1/2}+\left( M_{s}+\frac{M_{bc}m_{s}}{m^{2}_{bc}}\frac{4\alpha_{s}}{3}\mu_{bcs}\right) ^{2}\right)^{\frac{1}{2}} \notag\\
            &=& 6786.72 \  \text{MeV}. \label{PP666}
\end{eqnarray}
The predicted masses are
\begin{equation}
\begin{array}{c}
\text{State}|J,j\rangle: \\ [11pt]
M(\Omega _{bc},1S\text{):} \qquad
\end{array}%
\begin{array}{ccccc}
|1/2,1\rangle \qquad & |3/2,1\rangle  \\ [11pt]
\multicolumn{1}{r}{6767.50 \pm 3.19} \text{MeV}  \qquad & \multicolumn{1}{r}{6796.33 \pm 2.03} \text{MeV}.
\end{array}
\end{equation}
The mass difference of the $M(\Omega _{bc}, 1/2^{+})$ and $M(\Omega _{bc}, 3/2^{+})$ states is predicted to be 28 MeV, which is comparable with these results from the relativistic quark model \cite{Ebert:A11}. The computed ground and excited states of the $\Omega_{bc}$ baryons are compared with different theoretical approaches as shown in Table \ref{ppddmm50}.

We also calculate the predicted mass of the $1S$-wave state with $\mu_{bbs}$ = $m_{bb}m_{s}/(m_{bb}+m_{s})$ = 92.55 MeV for the $\Omega_{bb}$ baryon,
\begin{eqnarray}
\bar M(\Omega_{bb}) &=& M_{bb}+\left(\frac{1}{2}(M_{bb})^{1/2}+\left( M_{s}+\frac{M_{bb}m_{s}}{m^{2}_{bb}}\frac{4\alpha_{s}}{3}\mu_{bbs} \right)^{2}\right)^{\frac{1}{2}} \notag\\
            &=& 10317.14 \  \text{MeV}. \label{PP666}
\end{eqnarray}
The predicted masses of the $\Omega_{bb}$ baryon are
\begin{equation}
\begin{array}{c}
\text{State}|J,j\rangle: \\ [11pt]
M(\Omega _{bb},1S\text{):} \;
\end{array}%
\begin{array}{ccccc}
|1/2,1\rangle \qquad & |3/2,1\rangle  \\ [11pt]
\multicolumn{1}{r}{10305.21 \pm 4.17} \text{MeV}  \qquad & \multicolumn{1}{r}{10323.11 \pm 3.89} \text{MeV}.
\end{array}
\end{equation}
According to the analysis of our calculated values, we infer that the mass shift of about 18 MeV between the $\Omega _{bb}(1/2^{+})$ and $\Omega _{bb}(3/2^{+})$ states in $1S$-wave, which is relatively small due to the large mass of heavy diquark {$bb$}. As indicated by our results in Table \ref{ppddm5}, the masses of these five $1P$-wave states with the bases $|J,j\rangle$ are
\begin{equation}
\begin{array}{c}
\text{State}|J,j\rangle: \\ [11pt]
M(\Omega _{bb},1P\text{):}
\end{array}%
\begin{array}{ccccc}
|1/2,0\rangle \ & |1/2,1\rangle \ & |3/2,1\rangle \ & |3/2,2\rangle \ & |5/2,2\rangle  \\ [11pt]
\multicolumn{1}{r}{10794.16} \text{MeV} & \multicolumn{1}{r}{10807.70} \text{MeV} & \multicolumn{1}{r}{10811.31} \text{MeV} & \multicolumn{1}{r}{10817.18} \text{MeV} & \multicolumn{1}{r}{10824.97} \text{MeV}.\label{fff111}%
\end{array}
\end{equation}
As can be seen from the above masses (\ref{fff111}), the $\Omega _{bb}(10807.70)$ and $\Omega _{bb}(10811.31)$ states are the mass degenerate mixed states in $1P$-wave with $J^{P}$ = $1/2^{-}$ and $3/2^{-}$, respectively. They are most likely consistent with a single resonance due to its degeneracy. However, it is actually composed of these two states. For the excited states of the $\Omega_{bb}$ baryon, the results of the predicted masses are listed in Table \ref{ppddm5} and compared with other models (see Refs. \cite{WeiChen:A11, Salehi:A11} for
more details).

\begin{table}[htbp]
\caption{Mass spectra (MeV) of $\Omega_{cc}$ baryons are given and compared with different quark models.}\label{ppddmd55}
\resizebox{\textwidth}{12cm}{\begin{tabular}{ccccccc}
\hline\hline
{\small State }\; $J^{P}$  &{Ours}&   \cite{Ebert:A11}   & \cite{Zhong:A11} &  \cite{Shahr:A11}  &  \cite{KakadiyaR:A11}  & \\
\hline
$%
\begin{array}{rr}
{\small 1}^{1}{\small S}_{1/2} & {\small 1/2}^{+} \\
{\small 1}^{3}{\small S}_{3/2} & {\small 3/2}^{+}%
\end{array}%
$ & $%
\begin{array}{r}
{\small 3651.79 \pm 6.92} \\
{\small 3712.37 \pm 3.65}%
\end{array}%
$ & $%
\begin{array}{r}
{\small 3778} \\
{\small 3872}%
\end{array}%
$ & $%
\begin{array}{r}
{\small 3715} \\
{\small 3772}%
\end{array}%
$ & $%
\begin{array}{r}
{\small 3650} \\
{\small 3810}%
\end{array}%
$ & $%
\begin{array}{r}
{\small 3736} \\
{\small 3837}%
\end{array}%
$ & $%
\begin{array}{r}
{\small } \\
{\small }%
\end{array}%
$ \\ $%

\begin{array}{rr}
{\small 2}^{1}{\small S}_{1/2} & {\small 1/2}^{+} \\
{\small 2}^{3}{\small S}_{3/2} & {\small 3/2}^{+}%
\end{array}%
$ & $%
\begin{array}{r}
{\small 4160.96 \pm 2.15} \\
{\small 4173.89 \pm 1.69}%
\end{array}%
$ & $%
\begin{array}{r}
{\small } \\
{\small }%
\end{array}%
$ & $%
\begin{array}{r}
{\small 4118} \\
{\small 4142}%
\end{array}%
$ & $%
\begin{array}{r}
{\small 4028} \\
{\small 4085}%
\end{array}%
$ & $%
\begin{array}{r}
{\small 4078} \\
{\small 4110}%
\end{array}%
$ & $%
\begin{array}{r}
{\small } \\
{\small }%
\end{array}%
$ \\ $%

\begin{array}{rr}
{\small 3}^{1}{\small S}_{1/2} & {\small 1/2}^{+} \\
{\small 3}^{3}{\small S}_{3/2} & {\small 3/2}^{+}%
\end{array}%
$ & $%
\begin{array}{r}
{\small 4525.12 \pm 1.83} \\
{\small 4529.18 \pm 1.78}%
\end{array}%
$ & $%
\begin{array}{r}
{\small } \\
{\small }%
\end{array}%
$ & $%
\begin{array}{r}
{\small } \\
{\small }%
\end{array}%
$ & $%
\begin{array}{r}
{\small 4317} \\
{\small 4345}%
\end{array}%
$ & $%
\begin{array}{r}
{\small 4320} \\
{\small 4334}%
\end{array}%
$ & $%
\begin{array}{r}
{\small } \\
{\small }%
\end{array}%
$ \\ $%

\begin{array}{rr}
{\small 4}^{1}{\small S}_{1/2} & {\small 1/2}^{+} \\
{\small 4}^{3}{\small S}_{3/2} & {\small 3/2}^{+}%
\end{array}%
$ & $%
\begin{array}{r}
{\small 4823.86 \pm 1.95} \\
{\small 4825.61 \pm 1.94}%
\end{array}%
$ & $%
\begin{array}{r}
{\small } \\
{\small }%
\end{array}%
$ & $%
\begin{array}{r}
{\small } \\
{\small }%
\end{array}%
$ & $%
\begin{array}{r}
{\small 4570} \\
{\small 4586}%
\end{array}%
$ & $%
\begin{array}{r}
{\small 4514} \\
{\small 4521}%
\end{array}%
$ & $%
\begin{array}{r}
{\small } \\
{\small }%
\end{array}%
$ \\ $%

\begin{array}{rr}
{\small 5}^{1}{\small S}_{1/2} & {\small 1/2}^{+} \\
{\small 5}^{3}{\small S}_{3/2} & {\small 3/2}^{+}%
\end{array}%
$ & $%
\begin{array}{r}
{\small 5083.96 \pm 2.10} \\
{\small 5084.86 \pm 2.09}%
\end{array}%
$ & $%
\begin{array}{r}
{\small } \\
{\small }%
\end{array}%
$ & $%
\begin{array}{r}
{\small } \\
{\small }%
\end{array}%
$ & $%
\begin{array}{r}
{\small 4801} \\
{\small 4811}%
\end{array}%
$ & $%
\begin{array}{r}
{\small 4676} \\
{\small 4680}%
\end{array}%
$ & $%
\begin{array}{r}
{\small } \\
{\small }%
\end{array}%
$ \\
\hline

$
\begin{array}{rr}
{\small 1}^{2}{\small P}_{1/2} & {\small 1/2}^{-} \\
{\small 1}^{4}{\small P}_{1/2} & {\small 1/2}^{-} \\
{\small 1}^{2}{\small P}_{3/2} & {\small 3/2}^{-} \\
{\small 1}^{4}{\small P}_{3/2} & {\small 3/2}^{-} \\
{\small 1}^{4}{\small P}_{5/2} & {\small 5/2}^{-}%
\end{array}%
$ & $%
\begin{array}{r}
{\small 3987.54 \pm 5.98} \\
{\small 4032.12 \pm 2.74}\\
{\small 4044.04 \pm 1.93} \\
{\small 4063.36 \pm 1.77}\\
{\small 4089.02 \pm 3.23}%
\end{array}%
$ & $%
\begin{array}{r}
{\small 4208} \\
{\small 4271} \\
{\small 4252} \\
{\small 4325} \\
{\small 4303}%
\end{array}%
$ & $%
\begin{array}{r}
{\small 4087} \\
{\small 4081} \\
{\small 4107} \\
{\small 4114} \\
{\small 4134}%
\end{array}%
$ & $%
\begin{array}{r}
{\small 3964} \\
{\small 3972} \\
{\small 3948} \\
{\small 3981} \\
{\small 3935}%
\end{array}%
$ & $%
\begin{array}{r}
{\small 4011} \\
{\small 4014} \\
{\small 4004} \\
{\small 4007} \\
{\small 3998}%
\end{array}%
$ & $%
\begin{array}{r}
{\small } \\
{\small }\\
{\small } \\
{\small }\\
{\small }%
\end{array}%
$ \\
\hline

$
\begin{array}{rr}
{\small 2}^{2}{\small P}_{1/2} & {\small 1/2}^{-} \\
{\small 2}^{4}{\small P}_{1/2} & {\small 1/2}^{-} \\
{\small 2}^{2}{\small P}_{3/2} & {\small 3/2}^{-} \\
{\small 2}^{4}{\small P}_{3/2} & {\small 3/2}^{-} \\
{\small 2}^{4}{\small P}_{5/2} & {\small 5/2}^{-}%
\end{array}%
$ & $%
\begin{array}{r}
{\small 4408.38 \pm 3.10} \\
{\small 4427.52 \pm 2.00}\\
{\small 4431.32 \pm 1.84} \\
{\small 4442.58 \pm 1.77}\\
{\small 4454.26 \pm 2.19}%
\end{array}%
$ & $%
\begin{array}{r}
{\small } \\
{\small } \\
{\small } \\
{\small } \\
{\small }%
\end{array}%
$ & $%
\begin{array}{r}
{\small } \\
{\small } \\
{\small } \\
{\small } \\
{\small }%
\end{array}%
$ & $%
\begin{array}{r}
{\small 4241} \\
{\small 4248} \\
{\small 4228} \\
{\small 4234} \\
{\small 4216}%
\end{array}%
$ & $%
\begin{array}{r}
{\small 4253} \\
{\small 4256} \\
{\small 4248} \\
{\small 4251} \\
{\small 4244}%
\end{array}%
$ & $%
\begin{array}{r}
{\small } \\
{\small }\\
{\small } \\
{\small }\\
{\small }%
\end{array}%
$ \\
\hline

$
\begin{array}{rr}
{\small 3}^{2}{\small P}_{1/2} & {\small 1/2}^{-} \\
{\small 3}^{4}{\small P}_{1/2} & {\small 1/2}^{-} \\
{\small 3}^{2}{\small P}_{3/2} & {\small 3/2}^{-} \\
{\small 3}^{4}{\small P}_{3/2} & {\small 3/2}^{-} \\
{\small 3}^{4}{\small P}_{5/2} & {\small 5/2}^{-}%
\end{array}%
$ & $%
\begin{array}{r}
{\small 4732.38 \pm 2.37} \\
{\small 4742.88 \pm 1.99}\\
{\small 4744.50 \pm 1.94} \\
{\small 4751.59 \pm 1.91}\\
{\small 4758.17 \pm 2.37}%
\end{array}%
$ & $%
\begin{array}{r}
{\small } \\
{\small } \\
{\small } \\
{\small } \\
{\small }%
\end{array}%
$ & $%
\begin{array}{r}
{\small } \\
{\small } \\
{\small } \\
{\small } \\
{\small }%
\end{array}%
$ & $%
\begin{array}{r}
{\small 4492} \\
{\small 4498} \\
{\small 4479} \\
{\small 4486} \\
{\small 4469}%
\end{array}%
$ & $%
\begin{array}{r}
{\small 4453} \\
{\small 4454} \\
{\small 4450} \\
{\small 4451} \\
{\small 4447}%
\end{array}%
$ & $%
\begin{array}{r}
{\small } \\
{\small }\\
{\small } \\
{\small }\\
{\small }%
\end{array}%
$ \\
\hline

$
\begin{array}{rr}
{\small 4}^{2}{\small P}_{1/2} & {\small 1/2}^{-} \\
{\small 4}^{4}{\small P}_{1/2} & {\small 1/2}^{-} \\
{\small 4}^{2}{\small P}_{3/2} & {\small 3/2}^{-} \\
{\small 4}^{4}{\small P}_{3/2} & {\small 3/2}^{-} \\
{\small 4}^{4}{\small P}_{5/2} & {\small 5/2}^{-}%
\end{array}%
$ & $%
\begin{array}{r}
{\small 5006.81 \pm 2.25} \\
{\small 5013.41 \pm 2.09}\\
{\small 5014.23 \pm 2.07} \\
{\small 5019.06 \pm 2.06}\\
{\small 5023.26 \pm 2.11}%
\end{array}%
$ & $%
\begin{array}{r}
{\small } \\
{\small } \\
{\small } \\
{\small } \\
{\small }%
\end{array}%
$ & $%
\begin{array}{r}
{\small } \\
{\small } \\
{\small } \\
{\small } \\
{\small }%
\end{array}%
$ & $%
\begin{array}{r}
{\small 4723} \\
{\small 4728} \\
{\small 4712} \\
{\small 4717} \\
{\small 4703}%
\end{array}%
$ & $%
\begin{array}{r}
{\small } \\
{\small } \\
{\small } \\
{\small } \\
{\small }%
\end{array}%
$ & $%
\begin{array}{r}
{\small } \\
{\small }\\
{\small } \\
{\small }\\
{\small }%
\end{array}%
$ \\
\hline

$
\begin{array}{rr}
{\small 5}^{2}{\small P}_{1/2} & {\small 1/2}^{-} \\
{\small 5}^{4}{\small P}_{1/2} & {\small 1/2}^{-} \\
{\small 5}^{2}{\small P}_{3/2} & {\small 3/2}^{-} \\
{\small 5}^{4}{\small P}_{3/2} & {\small 3/2}^{-} \\
{\small 5}^{4}{\small P}_{5/2} & {\small 5/2}^{-}%
\end{array}%
$ & $%
\begin{array}{r}
{\small 5249.73 \pm 2.29} \\
{\small 5254.26 \pm 2.22}\\
{\small 5254.73 \pm 2.21} \\
{\small 5258.21 \pm 2.20}\\
{\small 5261.11 \pm 2.23}%
\end{array}%
$ & $%
\begin{array}{r}
{\small } \\
{\small } \\
{\small } \\
{\small } \\
{\small }%
\end{array}%
$ & $%
\begin{array}{r}
{\small } \\
{\small } \\
{\small } \\
{\small } \\
{\small }%
\end{array}%
$ & $%
\begin{array}{r}
{\small 4939} \\
{\small 4944} \\
{\small 4929} \\
{\small 4934} \\
{\small 4921}%
\end{array}%
$ & $%
\begin{array}{r}
{\small } \\
{\small } \\
{\small } \\
{\small } \\
{\small }%
\end{array}%
$ & $%
\begin{array}{r}
{\small } \\
{\small }\\
{\small } \\
{\small }\\
{\small }%
\end{array}%
$ \\
\hline

$
\begin{array}{rr}
{\small 1}^{4}{\small D}_{1/2} & {\small 1/2}^{+} \\
{\small 1}^{2}{\small D}_{3/2} & {\small 3/2}^{+} \\
{\small 1}^{4}{\small D}_{3/2} & {\small 3/2}^{+} \\
{\small 1}^{2}{\small D}_{5/2} & {\small 5/2}^{+} \\
{\small 1}^{4}{\small D}_{5/2} & {\small 5/2}^{+} \\
{\small 1}^{4}{\small D}_{7/2} & {\small 7/2}^{+}%
\end{array}%
$ & $%
\begin{array}{r}
{\small 4295.89 \pm 4.46} \\
{\small 4312.53 \pm 3.18}\\
{\small 4328.03 \pm 2.25}\\
{\small 4339.67 \pm 3.89}\\
{\small 4353.77 \pm 3.88}\\
{\small 4376.17 \pm 3.21}%
\end{array}%
$ & $%
\begin{array}{r}
{\small } \\
{\small } \\
{\small } \\
{\small } \\
{\small } \\
{\small }%
\end{array}%
$ & $%
\begin{array}{r}
{\small } \\
{\small } \\
{\small } \\
{\small } \\
{\small } \\
{\small }%
\end{array}%
$ & $%
\begin{array}{r}
{\small 4156} \\
{\small 4133} \\
{\small 4141} \\
{\small 4113} \\
{\small 4121} \\
{\small 4075}%
\end{array}%
$ & $%
\begin{array}{r}
{\small 4101} \\
{\small 4091} \\
{\small 4094} \\
{\small 4082} \\
{\small 4086} \\
{\small 4075}%
\end{array}%
$ & $%
\begin{array}{r}
{\small }\\
{\small }\\
{\small }\\
{\small }\\
{\small }\\
{\small }%
\end{array}%
$ \\
\hline

$
\begin{array}{rr}
{\small 2}^{4}{\small D}_{1/2} & {\small 1/2}^{+} \\
{\small 2}^{2}{\small D}_{3/2} & {\small 3/2}^{+} \\
{\small 2}^{4}{\small D}_{3/2} & {\small 3/2}^{+} \\
{\small 2}^{2}{\small D}_{5/2} & {\small 5/2}^{+} \\
{\small 2}^{4}{\small D}_{5/2} & {\small 5/2}^{+} \\
{\small 2}^{4}{\small D}_{7/2} & {\small 7/2}^{+}%
\end{array}%
$ & $%
\begin{array}{r}
{\small 4642.58 \pm 3.00} \\
{\small 4652.01 \pm 2.42}\\
{\small 4661.03 \pm 2.04}\\
{\small 4667.49 \pm 2.20} \\
{\small 4675.80 \pm 2.19}\\
{\small 4688.51 \pm 2.42}%
\end{array}%
$ & $%
\begin{array}{r}
{\small } \\
{\small } \\
{\small } \\
{\small } \\
{\small } \\
{\small }%
\end{array}%
$ & $%
\begin{array}{r}
{\small } \\
{\small } \\
{\small } \\
{\small } \\
{\small } \\
{\small }%
\end{array}%
$ & $%
\begin{array}{r}
{\small 4407} \\
{\small 4389} \\
{\small 4395} \\
{\small 4372} \\
{\small 4378} \\
{\small 4358}%
\end{array}%
$ & $%
\begin{array}{r}
{\small 4324} \\
{\small 4318} \\
{\small 4320} \\
{\small 4312} \\
{\small 4314} \\
{\small 4307}%
\end{array}%
$ & $%
\begin{array}{r}
{\small }\\
{\small }\\
{\small }\\
{\small }\\
{\small }\\
{\small }%
\end{array}%
$ \\
\hline

$
\begin{array}{rr}
{\small 3}^{4}{\small D}_{1/2} & {\small 1/2}^{+} \\
{\small 3}^{2}{\small D}_{3/2} & {\small 3/2}^{+} \\
{\small 3}^{4}{\small D}_{3/2} & {\small 3/2}^{+} \\
{\small 3}^{2}{\small D}_{5/2} & {\small 5/2}^{+} \\
{\small 3}^{4}{\small D}_{5/2} & {\small 5/2}^{+} \\
{\small 3}^{4}{\small D}_{7/2} & {\small 7/2}^{+}%
\end{array}%
$ & $%
\begin{array}{r}
{\small 4930.38 \pm 2.53} \\
{\small 4936.43 \pm 2.25}\\
{\small 4942.31 \pm 2.09} \\
{\small 4946.39 \pm 2.08}\\
{\small 4951.84 \pm 2.08}\\
{\small 4960.00 \pm 2.26}%
\end{array}%
$ & $%
\begin{array}{r}
{\small } \\
{\small } \\
{\small } \\
{\small } \\
{\small } \\
{\small }%
\end{array}%
$ & $%
\begin{array}{r}
{\small } \\
{\small } \\
{\small } \\
{\small } \\
{\small } \\
{\small }%
\end{array}%
$ & $%
\begin{array}{r}
{\small 4446} \\
{\small 4425} \\
{\small 4432} \\
{\small 4407} \\
{\small 4414} \\
{\small 4391}%
\end{array}%
$ & $%
\begin{array}{r}
{\small } \\
{\small } \\
{\small } \\
{\small } \\
{\small } \\
{\small }%
\end{array}%
$ & $%
\begin{array}{r}
{\small }\\
{\small }\\
{\small }\\
{\small }\\
{\small }\\
{\small }%
\end{array}%
$ \\
\hline

$
\begin{array}{rr}
{\small 4}^{4}{\small D}_{1/2} & {\small 1/2}^{+} \\
{\small 4}^{2}{\small D}_{3/2} & {\small 3/2}^{+} \\
{\small 4}^{4}{\small D}_{3/2} & {\small 3/2}^{+} \\
{\small 4}^{2}{\small D}_{5/2} & {\small 5/2}^{+} \\
{\small 4}^{4}{\small D}_{5/2} & {\small 5/2}^{+} \\
{\small 4}^{4}{\small D}_{7/2} & {\small 7/2}^{+}%
\end{array}%
$ & $%
\begin{array}{r}
{\small 5182.23 \pm 2.41} \\
{\small 5186.43 \pm 2.28}\\
{\small 5190.55 \pm 2.20}\\
{\small 5193.36 \pm 2.17} \\
{\small 5197.20 \pm 2.17}\\
{\small 5202.87 \pm 2.28}%
\end{array}%
$ & $%
\begin{array}{r}
{\small } \\
{\small } \\
{\small } \\
{\small } \\
{\small } \\
{\small }%
\end{array}%
$ & $%
\begin{array}{r}
{\small } \\
{\small } \\
{\small } \\
{\small } \\
{\small } \\
{\small }%
\end{array}%
$ & $%
\begin{array}{r}
{\small 4863} \\
{\small 4847} \\
{\small 4853} \\
{\small 4833} \\
{\small 4838} \\
{\small 4821}%
\end{array}%
$ & $%
\begin{array}{r}
{\small } \\
{\small } \\
{\small } \\
{\small } \\
{\small } \\
{\small }%
\end{array}%
$ & $%
\begin{array}{r}
{\small }\\
{\small }\\
{\small }\\
{\small }\\
{\small }\\
{\small }%
\end{array}%
$ \\
\hline

$
\begin{array}{rr}
{\small 5}^{4}{\small D}_{1/2} & {\small 1/2}^{+} \\
{\small 5}^{2}{\small D}_{3/2} & {\small 3/2}^{+} \\
{\small 5}^{4}{\small D}_{3/2} & {\small 3/2}^{+} \\
{\small 5}^{2}{\small D}_{5/2} & {\small 5/2}^{+} \\
{\small 5}^{4}{\small D}_{5/2} & {\small 5/2}^{+} \\
{\small 5}^{4}{\small D}_{7/2} & {\small 7/2}^{+}%
\end{array}%
$ & $%
\begin{array}{r}
{\small 5409.19 \pm 2.43} \\
{\small 5412.28 \pm 2.36}\\
{\small 5415.33 \pm 2.32}\\
{\small 5417.38 \pm 2.30} \\
{\small 5420.23 \pm 2.30}\\
{\small 5424.40 \pm 2.36}%
\end{array}%
$ & $%
\begin{array}{r}
{\small } \\
{\small } \\
{\small } \\
{\small } \\
{\small } \\
{\small }%
\end{array}%
$ & $%
\begin{array}{r}
{\small } \\
{\small } \\
{\small } \\
{\small } \\
{\small } \\
{\small }%
\end{array}%
$ & $%
\begin{array}{r}
{\small } \\
{\small } \\
{\small } \\
{\small } \\
{\small } \\
{\small }%
\end{array}%
$ & $%
\begin{array}{r}
{\small } \\
{\small } \\
{\small } \\
{\small } \\
{\small } \\
{\small }%
\end{array}%
$ & $%
\begin{array}{r}
{\small }\\
{\small }\\
{\small }\\
{\small }\\
{\small }\\
{\small }%
\end{array}%
$ \\
\hline\hline
\end{tabular}}
\end{table}

\begin{table}[htbp]
\caption{Mass spectra (MeV) of $\Omega_{bc}$ baryons are given and compared with different quark models.}\label{ppddmm50}
\resizebox{\textwidth}{12cm}{\begin{tabular}{ccccccc}
\hline\hline
{\small State }\; $J^{P}$  &{Ours}&  \cite{Ebert:A11}    &   \cite{Shahr:A11}  & \cite{KakadiyaR:A11}   &  \cite{Giannuzzi:A11} &           \\
\hline
$%
\begin{array}{rr}
{\small 1}^{1}{\small S}_{1/2} & {\small 1/2}^{+} \\
{\small 1}^{3}{\small S}_{3/2} & {\small 3/2}^{+}%
\end{array}%
$ & $%
\begin{array}{r}
{\small 6767.50 \pm 3.19} \\
{\small 6796.33 \pm 2.03}%
\end{array}%
$ & $%
\begin{array}{r}
{\small 7088} \\
{\small 7130}%
\end{array}%
$ & $%
\begin{array}{r}
{\small 7136} \\
{\small 7187}%
\end{array}%
$ & $%
\begin{array}{r}
{\small 7079} \\
{\small 7182}%
\end{array}%
$ & $%
\begin{array}{r}
{\small 6994} \\
{\small 7017}%
\end{array}%
$ & $%
\begin{array}{r}
{\small } \\
{\small }%
\end{array}%
$ \\ $%

\begin{array}{rr}
{\small 2}^{1}{\small S}_{1/2} & {\small 1/2}^{+} \\
{\small 2}^{3}{\small S}_{3/2} & {\small 3/2}^{+}%
\end{array}%
$ & $%
\begin{array}{r}
{\small 7359.24 \pm 1.83} \\
{\small 7365.52 \pm 1.74}%
\end{array}%
$ & $%
\begin{array}{r}
{\small } \\
{\small }%
\end{array}%
$ & $%
\begin{array}{r}
{\small 7473} \\
{\small 7490}%
\end{array}%
$ & $%
\begin{array}{r}
{\small 7435} \\
{\small 7469}%
\end{array}%
$ & $%
\begin{array}{r}
{\small 7559} \\
{\small 7571}%
\end{array}%
$ & $%
\begin{array}{r}
{\small } \\
{\small }%
\end{array}%
$ \\ $%

\begin{array}{rr}
{\small 3}^{1}{\small S}_{1/2} & {\small 1/2}^{+} \\
{\small 3}^{3}{\small S}_{3/2} & {\small 3/2}^{+}%
\end{array}%
$ & $%
\begin{array}{r}
{\small 7791.65 \pm 1.97} \\
{\small 7793.62 \pm 1.96}%
\end{array}%
$ & $%
\begin{array}{r}
{\small } \\
{\small }%
\end{array}%
$ & $%
\begin{array}{r}
{\small 7753} \\
{\small 7761}%
\end{array}%
$ & $%
\begin{array}{r}
{\small 7688} \\
{\small 7703}%
\end{array}%
$ & $%
\begin{array}{r}
{\small 7976} \\
{\small 7985}%
\end{array}%
$ & $%
\begin{array}{r}
{\small } \\
{\small }%
\end{array}%
$ \\ $%

\begin{array}{rr}
{\small 4}^{1}{\small S}_{1/2} & {\small 1/2}^{+} \\
{\small 4}^{3}{\small S}_{3/2} & {\small 3/2}^{+}%
\end{array}%
$ & $%
\begin{array}{r}
{\small 8148.51 \pm 2.17} \\
{\small 8149.36 \pm 2.16}%
\end{array}%
$ & $%
\begin{array}{r}
{\small } \\
{\small }%
\end{array}%
$ & $%
\begin{array}{r}
{\small 8004} \\
{\small 8009}%
\end{array}%
$ & $%
\begin{array}{r}
{\small 7895} \\
{\small 7903}%
\end{array}%
$ & $%
\begin{array}{r}
{\small } \\
{\small }%
\end{array}%
$ & $%
\begin{array}{r}
{\small } \\
{\small }%
\end{array}%
$ \\ $%

\begin{array}{rr}
{\small 5}^{1}{\small S}_{1/2} & {\small 1/2}^{+} \\
{\small 5}^{3}{\small S}_{3/2} & {\small 3/2}^{+}%
\end{array}%
$ & $%
\begin{array}{r}
{\small 8459.70 \pm 2.35} \\
{\small 8460.14 \pm 2.35}%
\end{array}%
$ & $%
\begin{array}{r}
{\small } \\
{\small }%
\end{array}%
$ & $%
\begin{array}{r}
{\small 8236} \\
{\small 8239}%
\end{array}%
$ & $%
\begin{array}{r}
{\small 8073} \\
{\small 8077}%
\end{array}%
$ & $%
\begin{array}{r}
{\small } \\
{\small }%
\end{array}%
$ & $%
\begin{array}{r}
{\small } \\
{\small }%
\end{array}%
$ \\
\hline

$
\begin{array}{rr}
{\small 1}^{2}{\small P}_{1/2} & {\small 1/2}^{-} \\
{\small 1}^{4}{\small P}_{1/2} & {\small 1/2}^{-} \\
{\small 1}^{2}{\small P}_{3/2} & {\small 3/2}^{-} \\
{\small 1}^{4}{\small P}_{3/2} & {\small 3/2}^{-} \\
{\small 1}^{4}{\small P}_{5/2} & {\small 5/2}^{-}%
\end{array}%
$ & $%
\begin{array}{r}
{\small 7193.89 \pm 1.76} \\
{\small 7215.51 \pm 1.75}\\
{\small 7221.30 \pm 1.69} \\
{\small 7230.67 \pm 1.69}\\
{\small 7243.13 \pm 1.70}%
\end{array}%
$ & $%
\begin{array}{r}
{\small } \\
{\small } \\
{\small } \\
{\small } \\
{\small }%
\end{array}%
$ & $%
\begin{array}{r}
{\small 7375} \\
{\small 7381} \\
{\small 7363} \\
{\small 7369} \\
{\small 7353}%
\end{array}%
$ & $%
\begin{array}{r}
{\small 7369} \\
{\small 7371} \\
{\small 7364} \\
{\small 7366} \\
{\small 7360}%
\end{array}%
$ & $%
\begin{array}{r}
{\small } \\
{\small } \\
{\small } \\
{\small } \\
{\small }%
\end{array}%
$ & $%
\begin{array}{r}
{\small } \\
{\small }\\
{\small } \\
{\small }\\
{\small }%
\end{array}%
$ \\
\hline

$
\begin{array}{rr}
{\small 2}^{2}{\small P}_{1/2} & {\small 1/2}^{-} \\
{\small 2}^{4}{\small P}_{1/2} & {\small 1/2}^{-} \\
{\small 2}^{2}{\small P}_{3/2} & {\small 3/2}^{-} \\
{\small 2}^{4}{\small P}_{3/2} & {\small 3/2}^{-} \\
{\small 2}^{4}{\small P}_{5/2} & {\small 5/2}^{-}%
\end{array}%
$ & $%
\begin{array}{r}
{\small 7671.31 \pm 1.92} \\
{\small 7680.61 \pm 1.92}\\
{\small 7682.46 \pm 1.90} \\
{\small 7687.92 \pm 1.90}\\
{\small 7693.60 \pm 1.91}%
\end{array}%
$ & $%
\begin{array}{r}
{\small } \\
{\small } \\
{\small } \\
{\small } \\
{\small }%
\end{array}%
$ & $%
\begin{array}{r}
{\small 7657} \\
{\small 7662} \\
{\small 7647} \\
{\small 7652} \\
{\small 7639}%
\end{array}%
$ & $%
\begin{array}{r}
{\small 7620} \\
{\small 7621} \\
{\small 7616} \\
{\small 7618} \\
{\small 7614}%
\end{array}%
$ & $%
\begin{array}{r}
{\small } \\
{\small } \\
{\small } \\
{\small } \\
{\small }%
\end{array}%
$ & $%
\begin{array}{r}
{\small } \\
{\small }\\
{\small } \\
{\small }\\
{\small }%
\end{array}%
$ \\
\hline

$
\begin{array}{rr}
{\small 3}^{2}{\small P}_{1/2} & {\small 1/2}^{-} \\
{\small 3}^{4}{\small P}_{1/2} & {\small 1/2}^{-} \\
{\small 3}^{2}{\small P}_{3/2} & {\small 3/2}^{-} \\
{\small 3}^{4}{\small P}_{3/2} & {\small 3/2}^{-} \\
{\small 3}^{4}{\small P}_{5/2} & {\small 5/2}^{-}%
\end{array}%
$ & $%
\begin{array}{r}
{\small 8050.01 \pm 2.12} \\
{\small 8055.11 \pm 2.11}\\
{\small 8055.90 \pm 2.11} \\
{\small 8059.34 \pm 2.11}\\
{\small 8062.54 \pm 2.12}%
\end{array}%
$ & $%
\begin{array}{r}
{\small } \\
{\small } \\
{\small } \\
{\small } \\
{\small }%
\end{array}%
$ & $%
\begin{array}{r}
{\small 7912} \\
{\small 7916} \\
{\small 7903} \\
{\small 7908} \\
{\small 7896}%
\end{array}%
$ & $%
\begin{array}{r}
{\small 7832} \\
{\small 7833} \\
{\small 7830} \\
{\small 7831} \\
{\small 7828}%
\end{array}%
$ & $%
\begin{array}{r}
{\small } \\
{\small } \\
{\small } \\
{\small } \\
{\small }%
\end{array}%
$ & $%
\begin{array}{r}
{\small } \\
{\small }\\
{\small } \\
{\small }\\
{\small }%
\end{array}%
$ \\
\hline

$
\begin{array}{rr}
{\small 4}^{2}{\small P}_{1/2} & {\small 1/2}^{-} \\
{\small 4}^{4}{\small P}_{1/2} & {\small 1/2}^{-} \\
{\small 4}^{2}{\small P}_{3/2} & {\small 3/2}^{-} \\
{\small 4}^{4}{\small P}_{3/2} & {\small 3/2}^{-} \\
{\small 4}^{4}{\small P}_{5/2} & {\small 5/2}^{-}%
\end{array}%
$ & $%
\begin{array}{r}
{\small 8374.40 \pm 2.30} \\
{\small 8377.60 \pm 2.30}\\
{\small 8378.01 \pm 2.30} \\
{\small 8380.35 \pm 2.30}\\
{\small 8382.39 \pm 2.30}%
\end{array}%
$ & $%
\begin{array}{r}
{\small } \\
{\small } \\
{\small } \\
{\small } \\
{\small }%
\end{array}%
$ & $%
\begin{array}{r}
{\small 8147} \\
{\small 8151} \\
{\small 8140} \\
{\small 8143} \\
{\small 8133}%
\end{array}%
$ & $%
\begin{array}{r}
{\small } \\
{\small } \\
{\small } \\
{\small } \\
{\small }%
\end{array}%
$ & $%
\begin{array}{r}
{\small } \\
{\small } \\
{\small } \\
{\small } \\
{\small }%
\end{array}%
$ & $%
\begin{array}{r}
{\small } \\
{\small }\\
{\small } \\
{\small }\\
{\small }%
\end{array}%
$ \\
\hline

$
\begin{array}{rr}
{\small 5}^{2}{\small P}_{1/2} & {\small 1/2}^{-} \\
{\small 5}^{4}{\small P}_{1/2} & {\small 1/2}^{-} \\
{\small 5}^{2}{\small P}_{3/2} & {\small 3/2}^{-} \\
{\small 5}^{4}{\small P}_{3/2} & {\small 3/2}^{-} \\
{\small 5}^{4}{\small P}_{5/2} & {\small 5/2}^{-}%
\end{array}%
$ & $%
\begin{array}{r}
{\small 8663.03 \pm 2.48} \\
{\small 8665.24 \pm 2.48}\\
{\small 8665.47 \pm 2.48} \\
{\small 8667.16 \pm 2.48}\\
{\small 8668.57 \pm 2.48}%
\end{array}%
$ & $%
\begin{array}{r}
{\small } \\
{\small } \\
{\small } \\
{\small } \\
{\small }%
\end{array}%
$ & $%
\begin{array}{r}
{\small 8368} \\
{\small 8372} \\
{\small 8361} \\
{\small 8365} \\
{\small 8355}%
\end{array}%
$ & $%
\begin{array}{r}
{\small } \\
{\small } \\
{\small } \\
{\small } \\
{\small }%
\end{array}%
$ & $%
\begin{array}{r}
{\small } \\
{\small } \\
{\small } \\
{\small } \\
{\small }%
\end{array}%
$ & $%
\begin{array}{r}
{\small } \\
{\small }\\
{\small } \\
{\small }\\
{\small }%
\end{array}%
$ \\
\hline

$
\begin{array}{rr}
{\small 1}^{4}{\small D}_{1/2} & {\small 1/2}^{+} \\
{\small 1}^{2}{\small D}_{3/2} & {\small 3/2}^{+} \\
{\small 1}^{4}{\small D}_{3/2} & {\small 3/2}^{+} \\
{\small 1}^{2}{\small D}_{5/2} & {\small 5/2}^{+} \\
{\small 1}^{4}{\small D}_{5/2} & {\small 5/2}^{+} \\
{\small 1}^{4}{\small D}_{7/2} & {\small 7/2}^{+}%
\end{array}%
$ & $%
\begin{array}{r}
{\small 7549.44 \pm 1.89} \\
{\small 7557.52 \pm 1.86}\\
{\small 7565.04 \pm 1.86}\\
{\small 7570.71 \pm 1.89}\\
{\small 7577.56 \pm 1.90}\\
{\small 7588.45 \pm 1.87}%
\end{array}%
$ & $%
\begin{array}{r}
{\small } \\
{\small } \\
{\small } \\
{\small } \\
{\small } \\
{\small }%
\end{array}%
$ & $%
\begin{array}{r}
{\small 7562} \\
{\small 7545} \\
{\small 7551} \\
{\small 7531} \\
{\small 7536} \\
{\small 7518}%
\end{array}%
$ & $%
\begin{array}{r}
{\small 7464} \\
{\small 7458} \\
{\small 7460} \\
{\small 7452} \\
{\small 7454} \\
{\small 7447}%
\end{array}%
$ & $%
\begin{array}{r}
{\small } \\
{\small } \\
{\small } \\
{\small } \\
{\small } \\
{\small }%
\end{array}%
$ & $%
\begin{array}{r}
{\small }\\
{\small }\\
{\small }\\
{\small }\\
{\small }\\
{\small }%
\end{array}%
$ \\
\hline

$
\begin{array}{rr}
{\small 2}^{4}{\small D}_{1/2} & {\small 1/2}^{+} \\
{\small 2}^{2}{\small D}_{3/2} & {\small 3/2}^{+} \\
{\small 2}^{4}{\small D}_{3/2} & {\small 3/2}^{+} \\
{\small 2}^{2}{\small D}_{5/2} & {\small 5/2}^{+} \\
{\small 2}^{4}{\small D}_{5/2} & {\small 5/2}^{+} \\
{\small 2}^{4}{\small D}_{7/2} & {\small 7/2}^{+}%
\end{array}%
$ & $%
\begin{array}{r}
{\small 7950.16 \pm 2.06} \\
{\small 7954.75 \pm 2.05}\\
{\small 7959.13 \pm 2.05}\\
{\small 7962.27 \pm 2.05} \\
{\small 7966.31 \pm 2.05}\\
{\small 7972.49 \pm 2.06}%
\end{array}%
$ & $%
\begin{array}{r}
{\small } \\
{\small } \\
{\small } \\
{\small } \\
{\small } \\
{\small }%
\end{array}%
$ & $%
\begin{array}{r}
{\small 7821} \\
{\small 7807} \\
{\small 7812} \\
{\small 7795} \\
{\small 7799} \\
{\small 7784}%
\end{array}%
$ & $%
\begin{array}{r}
{\small 7700} \\
{\small 7695} \\
{\small 7697} \\
{\small 7691} \\
{\small 7692} \\
{\small 7687}%
\end{array}%
$ & $%
\begin{array}{r}
{\small } \\
{\small } \\
{\small } \\
{\small } \\
{\small } \\
{\small }%
\end{array}%
$ & $%
\begin{array}{r}
{\small }\\
{\small }\\
{\small }\\
{\small }\\
{\small }\\
{\small }%
\end{array}%
$ \\
\hline

$
\begin{array}{rr}
{\small 3}^{4}{\small D}_{1/2} & {\small 1/2}^{+} \\
{\small 3}^{2}{\small D}_{3/2} & {\small 3/2}^{+} \\
{\small 3}^{4}{\small D}_{3/2} & {\small 3/2}^{+} \\
{\small 3}^{2}{\small D}_{5/2} & {\small 5/2}^{+} \\
{\small 3}^{4}{\small D}_{5/2} & {\small 5/2}^{+} \\
{\small 3}^{4}{\small D}_{7/2} & {\small 7/2}^{+}%
\end{array}%
$ & $%
\begin{array}{r}
{\small 8287.98 \pm 2.26} \\
{\small 8290.92 \pm 2.25}\\
{\small 8293.77 \pm 2.25} \\
{\small 8295.76 \pm 2.25}\\
{\small 8298.40 \pm 2.25}\\
{\small 8302.37 \pm 2.26}%
\end{array}%
$ & $%
\begin{array}{r}
{\small } \\
{\small } \\
{\small } \\
{\small } \\
{\small } \\
{\small }%
\end{array}%
$ & $%
\begin{array}{r}
{\small 8060} \\
{\small 8048} \\
{\small 8052} \\
{\small 8037} \\
{\small 8041} \\
{\small 8028}%
\end{array}%
$ & $%
\begin{array}{r}
{\small } \\
{\small } \\
{\small } \\
{\small } \\
{\small } \\
{\small }%
\end{array}%
$ & $%
\begin{array}{r}
{\small } \\
{\small } \\
{\small } \\
{\small } \\
{\small } \\
{\small }%
\end{array}%
$ & $%
\begin{array}{r}
{\small }\\
{\small }\\
{\small }\\
{\small }\\
{\small }\\
{\small }%
\end{array}%
$ \\
\hline

$
\begin{array}{rr}
{\small 4}^{4}{\small D}_{1/2} & {\small 1/2}^{+} \\
{\small 4}^{2}{\small D}_{3/2} & {\small 3/2}^{+} \\
{\small 4}^{4}{\small D}_{3/2} & {\small 3/2}^{+} \\
{\small 4}^{2}{\small D}_{5/2} & {\small 5/2}^{+} \\
{\small 4}^{4}{\small D}_{5/2} & {\small 5/2}^{+} \\
{\small 4}^{4}{\small D}_{7/2} & {\small 7/2}^{+}%
\end{array}%
$ & $%
\begin{array}{r}
{\small 8585.82 \pm 2.43} \\
{\small 8587.86 \pm 2.43}\\
{\small 8589.86 \pm 2.43}\\
{\small 8591.23 \pm 2.43} \\
{\small 8593.10 \pm 2.43}\\
{\small 8595.85 \pm 2.43}%
\end{array}%
$ & $%
\begin{array}{r}
{\small } \\
{\small } \\
{\small } \\
{\small } \\
{\small } \\
{\small }%
\end{array}%
$ & $%
\begin{array}{r}
{\small 8085} \\
{\small 8274} \\
{\small 8277} \\
{\small 8263} \\
{\small 8267} \\
{\small 8254}%
\end{array}%
$ & $%
\begin{array}{r}
{\small } \\
{\small } \\
{\small } \\
{\small } \\
{\small } \\
{\small }%
\end{array}%
$ & $%
\begin{array}{r}
{\small } \\
{\small } \\
{\small } \\
{\small } \\
{\small } \\
{\small }%
\end{array}%
$ & $%
\begin{array}{r}
{\small }\\
{\small }\\
{\small }\\
{\small }\\
{\small }\\
{\small }%
\end{array}%
$ \\
\hline

$
\begin{array}{rr}
{\small 5}^{4}{\small D}_{1/2} & {\small 1/2}^{+} \\
{\small 5}^{2}{\small D}_{3/2} & {\small 3/2}^{+} \\
{\small 5}^{4}{\small D}_{3/2} & {\small 3/2}^{+} \\
{\small 5}^{2}{\small D}_{5/2} & {\small 5/2}^{+} \\
{\small 5}^{4}{\small D}_{5/2} & {\small 5/2}^{+} \\
{\small 5}^{4}{\small D}_{7/2} & {\small 7/2}^{+}%
\end{array}%
$ & $%
\begin{array}{r}
{\small 8855.35 \pm 2.60} \\
{\small 8856.85 \pm 2.60}\\
{\small 8858.33 \pm 2.60}\\
{\small 8859.33 \pm 2.60} \\
{\small 8860.71 \pm 2.60}\\
{\small 8862.74 \pm 2.60}%
\end{array}%
$ & $%
\begin{array}{r}
{\small } \\
{\small } \\
{\small } \\
{\small } \\
{\small } \\
{\small }%
\end{array}%
$ & $%
\begin{array}{r}
{\small } \\
{\small } \\
{\small } \\
{\small } \\
{\small } \\
{\small }%
\end{array}%
$ & $%
\begin{array}{r}
{\small } \\
{\small } \\
{\small } \\
{\small } \\
{\small } \\
{\small }%
\end{array}%
$ & $%
\begin{array}{r}
{\small } \\
{\small } \\
{\small } \\
{\small } \\
{\small } \\
{\small }%
\end{array}%
$ & $%
\begin{array}{r}
{\small }\\
{\small }\\
{\small }\\
{\small }\\
{\small }\\
{\small }%
\end{array}%
$ \\
\hline\hline
\end{tabular}}
\end{table}

\begin{table}[htbp]
\caption{Mass spectra (MeV) of $\Omega_{bb}$ baryons are given and compared with different quark models.}\label{ppddm5}
\resizebox{\textwidth}{12cm}{\begin{tabular}{ccccccc}
\hline\hline
{\small State }\; $J^{P}$  &{Ours}&   \cite{Ebert:A11}     &\cite{Zhong:A11} &   \cite{ChenLuo:H11} & \cite{Shahr:A11} &        \\
\hline
$%
\begin{array}{rr}
{\small 1}^{1}{\small S}_{1/2} & {\small 1/2}^{+} \\
{\small 1}^{3}{\small S}_{3/2} & {\small 3/2}^{+}%
\end{array}%
$ & $%
\begin{array}{r}
{\small 10305.21 \pm 4.17} \\
{\small 10323.11 \pm 3.89}%
\end{array}%
$ & $%
\begin{array}{r}
{\small 10359} \\
{\small 10389}%
\end{array}%
$ & $%
\begin{array}{r}
{\small 10230} \\
{\small 10258}%
\end{array}%
$ & $%
\begin{array}{r}
{\small 10266} \\
{\small 10291}%
\end{array}%
$ & $%
\begin{array}{r}
{\small 10446} \\
{\small 10467}%
\end{array}%
$ & $%
\begin{array}{r}
{\small } \\
{\small }%
\end{array}%
$ \\ $%

\begin{array}{rr}
{\small 2}^{1}{\small S}_{1/2} & {\small 1/2}^{+} \\
{\small 2}^{3}{\small S}_{3/2} & {\small 3/2}^{+}%
\end{array}%
$ & $%
\begin{array}{r}
{\small 10967.74 \pm 4.00} \\
{\small 10969.66 \pm 3.98}%
\end{array}%
$ & $%
\begin{array}{r}
{\small 10970} \\
{\small 10992}%
\end{array}%
$ & $%
\begin{array}{r}
{\small 10751} \\
{\small 10763}%
\end{array}%
$ & $%
\begin{array}{r}
{\small 10816} \\
{\small 10830}%
\end{array}%
$ & $%
\begin{array}{r}
{\small 10730} \\
{\small 10737}%
\end{array}%
$ & $%
\begin{array}{r}
{\small } \\
{\small }%
\end{array}%
$ \\ $%

\begin{array}{rr}
{\small 3}^{1}{\small S}_{1/2} & {\small 1/2}^{+} \\
{\small 3}^{3}{\small S}_{3/2} & {\small 3/2}^{+}%
\end{array}%
$ & $%
\begin{array}{r}
{\small 11450.73 \pm 4.15} \\
{\small 11452.00 \pm 4.15}%
\end{array}%
$ & $%
\begin{array}{r}
{\small } \\
{\small }%
\end{array}%
$ & $%
\begin{array}{r}
{\small } \\
{\small }%
\end{array}%
$ & $%
\begin{array}{r}
{\small } \\
{\small }%
\end{array}%
$ & $%
\begin{array}{r}
{\small 10973} \\
{\small 10976}%
\end{array}%
$ & $%
\begin{array}{r}
{\small } \\
{\small }%
\end{array}%
$ \\ $%

\begin{array}{rr}
{\small 4}^{1}{\small S}_{1/2} & {\small 1/2}^{+} \\
{\small 4}^{3}{\small S}_{3/2} & {\small 3/2}^{+}%
\end{array}%
$ & $%
\begin{array}{r}
{\small 11851.70 \pm 4.30} \\
{\small 11852.23 \pm 4.30}%
\end{array}%
$ & $%
\begin{array}{r}
{\small } \\
{\small }%
\end{array}%
$ & $%
\begin{array}{r}
{\small } \\
{\small }%
\end{array}%
$ & $%
\begin{array}{r}
{\small } \\
{\small }%
\end{array}%
$ & $%
\begin{array}{r}
{\small 11191} \\
{\small 11193}%
\end{array}%
$ & $%
\begin{array}{r}
{\small } \\
{\small }%
\end{array}%
$ \\ $%

\begin{array}{rr}
{\small 5}^{1}{\small S}_{1/2} & {\small 1/2}^{+} \\
{\small 5}^{3}{\small S}_{3/2} & {\small 3/2}^{+}%
\end{array}%
$ & $%
\begin{array}{r}
{\small 12201.51 \pm 4.45} \\
{\small 12201.78 \pm 4.45}%
\end{array}%
$ & $%
\begin{array}{r}
{\small } \\
{\small }%
\end{array}%
$ & $%
\begin{array}{r}
{\small } \\
{\small }%
\end{array}%
$ & $%
\begin{array}{r}
{\small } \\
{\small }%
\end{array}%
$ & $%
\begin{array}{r}
{\small 11393} \\
{\small 11394}%
\end{array}%
$ & $%
\begin{array}{r}
{\small } \\
{\small }%
\end{array}%
$ \\
\hline

$
\begin{array}{rr}
{\small 1}^{2}{\small P}_{1/2} & {\small 1/2}^{-} \\
{\small 1}^{4}{\small P}_{1/2} & {\small 1/2}^{-} \\
{\small 1}^{2}{\small P}_{3/2} & {\small 3/2}^{-} \\
{\small 1}^{4}{\small P}_{3/2} & {\small 3/2}^{-} \\
{\small 1}^{4}{\small P}_{5/2} & {\small 5/2}^{-}%
\end{array}%
$ & $%
\begin{array}{r}
{\small 10794.16 \pm 3.95} \\
{\small 10807.70 \pm 3.94}\\
{\small 10811.31 \pm 3.94} \\
{\small 10817.18 \pm 3.93}\\
{\small 10824.97 \pm 3.94}%
\end{array}%
$ & $%
\begin{array}{r}
{\small 10771} \\
{\small 10804} \\
{\small 10785} \\
{\small 10821} \\
{\small 10798}%
\end{array}%
$ & $%
\begin{array}{r}
{\small 10605} \\
{\small 10591} \\
{\small 10610} \\
{\small 10611} \\
{\small 10625}%
\end{array}%
$ & $%
\begin{array}{r}
{\small 10669} \\
{\small 10641} \\
{\small 10681} \\
{\small 10656} \\
{\small 10655}%
\end{array}%
$ & $%
\begin{array}{r}
{\small 10580} \\
{\small 10581} \\
{\small 10578} \\
{\small 10579} \\
{\small 10576}%
\end{array}%
$ & $%
\begin{array}{r}
{\small } \\
{\small }\\
{\small } \\
{\small }\\
{\small }%
\end{array}%
$ \\
\hline

$
\begin{array}{rr}
{\small 2}^{2}{\small P}_{1/2} & {\small 1/2}^{-} \\
{\small 2}^{4}{\small P}_{1/2} & {\small 1/2}^{-} \\
{\small 2}^{2}{\small P}_{3/2} & {\small 3/2}^{-} \\
{\small 2}^{4}{\small P}_{3/2} & {\small 3/2}^{-} \\
{\small 2}^{4}{\small P}_{5/2} & {\small 5/2}^{-}%
\end{array}%
$ & $%
\begin{array}{r}
{\small 11322.04 \pm 4.11} \\
{\small 11327.86 \pm 4.10}\\
{\small 11329.01 \pm 4.10} \\
{\small 11332.44 \pm 4.10}\\
{\small 11336.00 \pm 4.10}%
\end{array}%
$ & $%
\begin{array}{r}
{\small } \\
{\small } \\
{\small } \\
{\small } \\
{\small }%
\end{array}%
$ & $%
\begin{array}{r}
{\small } \\
{\small } \\
{\small } \\
{\small } \\
{\small }%
\end{array}%
$ & $%
\begin{array}{r}
{\small } \\
{\small } \\
{\small } \\
{\small } \\
{\small }%
\end{array}%
$ & $%
\begin{array}{r}
{\small 10796} \\
{\small 10797} \\
{\small 10795} \\
{\small 10796} \\
{\small 10794}%
\end{array}%
$ & $%
\begin{array}{r}
{\small } \\
{\small }\\
{\small } \\
{\small }\\
{\small }%
\end{array}%
$ \\
\hline

$
\begin{array}{rr}
{\small 3}^{2}{\small P}_{1/2} & {\small 1/2}^{-} \\
{\small 3}^{4}{\small P}_{1/2} & {\small 1/2}^{-} \\
{\small 3}^{2}{\small P}_{3/2} & {\small 3/2}^{-} \\
{\small 3}^{4}{\small P}_{3/2} & {\small 3/2}^{-} \\
{\small 3}^{4}{\small P}_{5/2} & {\small 5/2}^{-}%
\end{array}%
$ & $%
\begin{array}{r}
{\small 11744.70 \pm 4.27} \\
{\small 11747.89 \pm 4.26}\\
{\small 11748.38 \pm 4.26} \\
{\small 11750.54 \pm 4.27}\\
{\small 11752.54 \pm 4.27}%
\end{array}%
$ & $%
\begin{array}{r}
{\small } \\
{\small } \\
{\small } \\
{\small } \\
{\small }%
\end{array}%
$ & $%
\begin{array}{r}
{\small } \\
{\small } \\
{\small } \\
{\small } \\
{\small }%
\end{array}%
$ & $%
\begin{array}{r}
{\small } \\
{\small } \\
{\small } \\
{\small } \\
{\small }%
\end{array}%
$ & $%
\begin{array}{r}
{\small 10982} \\
{\small 10982} \\
{\small 10981} \\
{\small 10981} \\
{\small 10980}%
\end{array}%
$ & $%
\begin{array}{r}
{\small } \\
{\small }\\
{\small } \\
{\small }\\
{\small }%
\end{array}%
$ \\
\hline

$
\begin{array}{rr}
{\small 4}^{2}{\small P}_{1/2} & {\small 1/2}^{-} \\
{\small 4}^{4}{\small P}_{1/2} & {\small 1/2}^{-} \\
{\small 4}^{2}{\small P}_{3/2} & {\small 3/2}^{-} \\
{\small 4}^{4}{\small P}_{3/2} & {\small 3/2}^{-} \\
{\small 4}^{4}{\small P}_{5/2} & {\small 5/2}^{-}%
\end{array}%
$ & $%
\begin{array}{r}
{\small 12108.00 \pm 4.41} \\
{\small 12110.01 \pm 4.41}\\
{\small 12110.26 \pm 4.41} \\
{\small 12111.73 \pm 4.41}\\
{\small 12113.01 \pm 4.41}%
\end{array}%
$ & $%
\begin{array}{r}
{\small } \\
{\small } \\
{\small } \\
{\small } \\
{\small }%
\end{array}%
$ & $%
\begin{array}{r}
{\small } \\
{\small } \\
{\small } \\
{\small } \\
{\small }%
\end{array}%
$ & $%
\begin{array}{r}
{\small } \\
{\small } \\
{\small } \\
{\small } \\
{\small }%
\end{array}%
$ & $%
\begin{array}{r}
{\small } \\
{\small } \\
{\small } \\
{\small } \\
{\small }%
\end{array}%
$ & $%
\begin{array}{r}
{\small } \\
{\small }\\
{\small } \\
{\small }\\
{\small }%
\end{array}%
$ \\
\hline

$
\begin{array}{rr}
{\small 5}^{2}{\small P}_{1/2} & {\small 1/2}^{-} \\
{\small 5}^{4}{\small P}_{1/2} & {\small 1/2}^{-} \\
{\small 5}^{2}{\small P}_{3/2} & {\small 3/2}^{-} \\
{\small 5}^{4}{\small P}_{3/2} & {\small 3/2}^{-} \\
{\small 5}^{4}{\small P}_{5/2} & {\small 5/2}^{-}%
\end{array}%
$ & $%
\begin{array}{r}
{\small 12431.77 \pm 4.55} \\
{\small 12433.15 \pm 4.55}\\
{\small 12433.29 \pm 4.55} \\
{\small 12434.36 \pm 4.72}\\
{\small 12435.24 \pm 4.55}%
\end{array}%
$ & $%
\begin{array}{r}
{\small } \\
{\small } \\
{\small } \\
{\small } \\
{\small }%
\end{array}%
$ & $%
\begin{array}{r}
{\small } \\
{\small } \\
{\small } \\
{\small } \\
{\small }%
\end{array}%
$ & $%
\begin{array}{r}
{\small } \\
{\small } \\
{\small } \\
{\small } \\
{\small }%
\end{array}%
$ & $%
\begin{array}{r}
{\small } \\
{\small } \\
{\small } \\
{\small } \\
{\small }%
\end{array}%
$ & $%
\begin{array}{r}
{\small } \\
{\small }\\
{\small } \\
{\small }\\
{\small }%
\end{array}%
$ \\
\hline

$
\begin{array}{rr}
{\small 1}^{4}{\small D}_{1/2} & {\small 1/2}^{+} \\
{\small 1}^{2}{\small D}_{3/2} & {\small 3/2}^{+} \\
{\small 1}^{4}{\small D}_{3/2} & {\small 3/2}^{+} \\
{\small 1}^{2}{\small D}_{5/2} & {\small 5/2}^{+} \\
{\small 1}^{4}{\small D}_{5/2} & {\small 5/2}^{+} \\
{\small 1}^{4}{\small D}_{7/2} & {\small 7/2}^{+}%
\end{array}%
$ & $%
\begin{array}{r}
{\small 11189.38 \pm 4.07} \\
{\small 11194.45 \pm 4.06}\\
{\small 11199.15 \pm 4.07}\\
{\small 11202.69 \pm 4.06}\\
{\small 11206.98 \pm 4.06}\\
{\small 11213.78 \pm 4.06}%
\end{array}%
$ & $%
\begin{array}{r}
{\small } \\
{\small } \\
{\small } \\
{\small } \\
{\small } \\
{\small }%
\end{array}%
$ & $%
\begin{array}{r}
{\small } \\
{\small } \\
{\small } \\
{\small } \\
{\small } \\
{\small }%
\end{array}%
$ & $%
\begin{array}{r}
{\small 10971} \\
{\small 10975} \\
{\small 10891} \\
{\small 10979} \\
{\small 10896} \\
{\small 10898}%
\end{array}%
$ & $%
\begin{array}{r}
{\small 10662} \\
{\small 10660} \\
{\small 10661} \\
{\small 10657} \\
{\small 10658} \\
{\small 10655}%
\end{array}%
$ & $%
\begin{array}{r}
{\small }\\
{\small }\\
{\small }\\
{\small }\\
{\small }\\
{\small }%
\end{array}%
$ \\
\hline

$
\begin{array}{rr}
{\small 2}^{4}{\small D}_{1/2} & {\small 1/2}^{+} \\
{\small 2}^{2}{\small D}_{3/2} & {\small 3/2}^{+} \\
{\small 2}^{4}{\small D}_{3/2} & {\small 3/2}^{+} \\
{\small 2}^{2}{\small D}_{5/2} & {\small 5/2}^{+} \\
{\small 2}^{4}{\small D}_{5/2} & {\small 5/2}^{+} \\
{\small 2}^{4}{\small D}_{7/2} & {\small 7/2}^{+}%
\end{array}%
$ & $%
\begin{array}{r}
{\small 11635.06 \pm 4.22} \\
{\small 11637.93 \pm 4.22}\\
{\small 11640.67 \pm 4.22}\\
{\small 11642.64 \pm 4.22} \\
{\small 11645.16 \pm 4.22}\\
{\small 11649.03 \pm 4.22}%
\end{array}%
$ & $%
\begin{array}{r}
{\small } \\
{\small } \\
{\small } \\
{\small } \\
{\small } \\
{\small }%
\end{array}%
$ & $%
\begin{array}{r}
{\small } \\
{\small } \\
{\small } \\
{\small } \\
{\small } \\
{\small }%
\end{array}%
$ & $%
\begin{array}{r}
{\small } \\
{\small } \\
{\small } \\
{\small } \\
{\small } \\
{\small }%
\end{array}%
$ & $%
\begin{array}{r}
{\small 10866} \\
{\small 10864} \\
{\small 10865} \\
{\small 10863} \\
{\small 10863} \\
{\small 10861}%
\end{array}%
$ & $%
\begin{array}{r}
{\small }\\
{\small }\\
{\small }\\
{\small }\\
{\small }\\
{\small }%
\end{array}%
$ \\
\hline

$
\begin{array}{rr}
{\small 3}^{4}{\small D}_{1/2} & {\small 1/2}^{+} \\
{\small 3}^{2}{\small D}_{3/2} & {\small 3/2}^{+} \\
{\small 3}^{4}{\small D}_{3/2} & {\small 3/2}^{+} \\
{\small 3}^{2}{\small D}_{5/2} & {\small 5/2}^{+} \\
{\small 3}^{4}{\small D}_{5/2} & {\small 5/2}^{+} \\
{\small 3}^{4}{\small D}_{7/2} & {\small 7/2}^{+}%
\end{array}%
$ & $%
\begin{array}{r}
{\small 12012.57 \pm 4.37} \\
{\small 12014.41 \pm 4.37}\\
{\small 12016.20 \pm 4.37} \\
{\small 12017.44 \pm 4.37}\\
{\small 12019.10 \pm 4.37}\\
{\small 12021.58 \pm 4.37}%
\end{array}%
$ & $%
\begin{array}{r}
{\small } \\
{\small } \\
{\small } \\
{\small } \\
{\small } \\
{\small }%
\end{array}%
$ & $%
\begin{array}{r}
{\small } \\
{\small } \\
{\small } \\
{\small } \\
{\small } \\
{\small }%
\end{array}%
$ & $%
\begin{array}{r}
{\small } \\
{\small } \\
{\small } \\
{\small } \\
{\small } \\
{\small }%
\end{array}%
$ & $%
\begin{array}{r}
{\small } \\
{\small } \\
{\small } \\
{\small } \\
{\small } \\
{\small }%
\end{array}%
$ & $%
\begin{array}{r}
{\small }\\
{\small }\\
{\small }\\
{\small }\\
{\small }\\
{\small }%
\end{array}%
$ \\
\hline

$
\begin{array}{rr}
{\small 4}^{4}{\small D}_{1/2} & {\small 1/2}^{+} \\
{\small 4}^{2}{\small D}_{3/2} & {\small 3/2}^{+} \\
{\small 4}^{4}{\small D}_{3/2} & {\small 3/2}^{+} \\
{\small 4}^{2}{\small D}_{5/2} & {\small 5/2}^{+} \\
{\small 4}^{4}{\small D}_{5/2} & {\small 5/2}^{+} \\
{\small 4}^{4}{\small D}_{7/2} & {\small 7/2}^{+}%
\end{array}%
$ & $%
\begin{array}{r}
{\small 12346.19 \pm 4.51} \\
{\small 12347.46 \pm 4.51}\\
{\small 12348.92 \pm 4.51}\\
{\small 12349.57 \pm 4.51} \\
{\small 12350.74 \pm 4.51}\\
{\small 12352.47 \pm 4.51}%
\end{array}%
$ & $%
\begin{array}{r}
{\small } \\
{\small } \\
{\small } \\
{\small } \\
{\small } \\
{\small }%
\end{array}%
$ & $%
\begin{array}{r}
{\small } \\
{\small } \\
{\small } \\
{\small } \\
{\small } \\
{\small }%
\end{array}%
$ & $%
\begin{array}{r}
{\small } \\
{\small } \\
{\small } \\
{\small } \\
{\small } \\
{\small }%
\end{array}%
$ & $%
\begin{array}{r}
{\small } \\
{\small } \\
{\small } \\
{\small } \\
{\small } \\
{\small }%
\end{array}%
$ & $%
\begin{array}{r}
{\small }\\
{\small }\\
{\small }\\
{\small }\\
{\small }\\
{\small }%
\end{array}%
$ \\
\hline

$
\begin{array}{rr}
{\small 5}^{4}{\small D}_{1/2} & {\small 1/2}^{+} \\
{\small 5}^{2}{\small D}_{3/2} & {\small 3/2}^{+} \\
{\small 5}^{4}{\small D}_{3/2} & {\small 3/2}^{+} \\
{\small 5}^{2}{\small D}_{5/2} & {\small 5/2}^{+} \\
{\small 5}^{4}{\small D}_{5/2} & {\small 5/2}^{+} \\
{\small 5}^{4}{\small D}_{7/2} & {\small 7/2}^{+}%
\end{array}%
$ & $%
\begin{array}{r}
{\small 12648.47 \pm 4.65} \\
{\small 12649.41 \pm 4.65}\\
{\small 12650.34 \pm 4.65}\\
{\small 12650.96 \pm 4.65} \\
{\small 12651.83 \pm 4.65}\\
{\small 12653.10 \pm 4.65}%
\end{array}%
$ & $%
\begin{array}{r}
{\small } \\
{\small } \\
{\small } \\
{\small } \\
{\small } \\
{\small }%
\end{array}%
$ & $%
\begin{array}{r}
{\small } \\
{\small } \\
{\small } \\
{\small } \\
{\small } \\
{\small }%
\end{array}%
$ & $%
\begin{array}{r}
{\small } \\
{\small } \\
{\small } \\
{\small } \\
{\small } \\
{\small }%
\end{array}%
$ & $%
\begin{array}{r}
{\small } \\
{\small } \\
{\small } \\
{\small } \\
{\small } \\
{\small }%
\end{array}%
$ & $%
\begin{array}{r}
{\small }\\
{\small }\\
{\small }\\
{\small }\\
{\small }\\
{\small }%
\end{array}%
$ \\
\hline\hline
\end{tabular}}
\end{table}

\section{conclusion}

The similarity of dynamics between singly heavy baryons and doubly heavy baryons provides us with the possibility of a comprehensive study of their properties in heavy-light hadron framework. In this paper, we use the relativistic effective mass formula under the Coulomb potential to study the effective masses of the heavy diquark and light quark. To estimate the mass splitting $\Delta M(J, j)$, we have proposed a new scheme of states classification, the all-$JLS$ coupling, to uniformly analyze the mass spectra and fundamental structure of the excited $\Xi_{QQ^{\prime}}$ and $\Omega_{QQ^{\prime}}$ baryons.

According to our results in Table \ref{ppdm5}, \ref{ppddmm35}, and \ref{ppddmm45} for the $\Xi_{QQ^{\prime}}$ baryons, the $\Xi^{++}_{cc}$ state can be grouped into the $1S$-wave state with $J^{P} = 1/2^{+}$, the mass is $M(\Xi^{++}_{cc})$ = 3621.57 $\pm$ 8.65 MeV in our model. The hyperfine partner of this $\Xi_{cc}$ state with $J^{P} = 3/2^{+}$, the predicted mass is $3699.69 \pm 4.52$ MeV, which is closer to the constituent quark model \cite{KakadiyaR:A11}. In the relativistic quark model \cite{Ebert:A11}, its mass is much larger than the result 30 MeV of our model calculation. Similarly, the masse spectra of the ground states for $\Xi_{bc}$ and $\Xi_{bb}$ baryons are also predicted and listed in Table \ref{ppddmm35} and \ref{ppddmm45}. These predictions may provide important references for the experiment (such as LHCb, Belle, $BABAR$, and CLEO) to study for the $\Xi_{QQ^{\prime}}$ baryon states in future.

Our predictions for $\Omega_{QQ^{\prime}}$ baryon states in Table \ref{ppddmd55}, \ref{ppddmm50}, and \ref{ppddm5} are given and compared with different quark models. Since the $\Omega_{QQ^{\prime}}$ baryons have a light strange quark $s$, the finite mass effect of the doubly heavy quark may become significant. Therefore, the mass splitting of these baryon states are also relatively small compared to $\Xi_{QQ^{\prime}}$ baryons. We obtain $M(\Omega_{cc})$ = 3651.79 $\pm$ 6.92 MeV, $M(\Omega^{\ast}_{cc})$ = 3712.37 $\pm$ 3.65 MeV, $M(\Omega_{bc})$ = 6767.50 $\pm$ 3.19 MeV, $M(\Omega^{\ast}_{bc})$ = 6796.33 $\pm$ 2.03 MeV, $M(\Omega_{bb})$ = 10305.21 $\pm$ 4.17 MeV, $M(\Omega^{\ast}_{bb})$ = 10323.11 $\pm$ 3.89 MeV for $1S$-wave states. In addition, for the several unobserved $nS$-, $nP$-, and $nD$-wave states of $\Xi_{QQ^{\prime}}$ and $\Omega_{QQ^{\prime}}$ baryons, we present a complete prediction for their masses in Table \ref{ppdm5}, \ref{ppddmm35}, \ref{ppddmm45},  \ref{ppddmd55}, \ref{ppddmm50}, and \ref{ppddm5}, which are useful for the further exploration in the experiment.

Obviously, the $\Xi_{QQ^{\prime}}$ and $\Omega_{QQ^{\prime}}$ baryons are special in three-body system, due to the large mass values of two heavy quarks. In this work, based on chromodynamics similarity between the singly heavy baryons and doubly heavy baryons, we adopt and extend the scaling relationship to calculate the spin-coupling parameters $a_{1}$, $a_{2}$, $b_{1}$, and $c_{1}$. Although the predicted masses of the excited $\Xi_{QQ^{\prime}}$ and $\Omega_{QQ^{\prime}}$ baryon states are by no means rigorous, our estimations for the baryon spin-multiplet splitting have universality and of great significance for research.

\appendix

\section{$S$-wave}

For the $S$-wave system, we consider the spin-dependent Hamiltonian $H^{SD}(L = 0)$ in Eq. (\ref{PP001}). The matrix elements of $\mathbf{S}_{QQ^{\prime}}\cdot \mathbf{S}_{q}$ may be evaluated by the square of the total spin $\mathbf{S}$ = $\mathbf{S}_{QQ^{\prime}}$ + $\mathbf{S}_{q}$,
\begin{equation}
\mathbf{S}_{QQ^{\prime}}\cdot \mathbf{S}_{q}=\left( \mathbf{S}^{2}-\mathbf{S}_{QQ^{\prime}}^{2}- \mathbf{S}_{q}^{2}\right)/2.  \label{PP7}
\end{equation}
Then, the basis states of $L$-$S$ coupling with the third component $S_{3}$ can be constructed as a linear combination of the $|S_{QQ^{\prime}3}, S_{q3}\rangle$ states. The two basis states are
\begin{eqnarray}
|^{2} S_{1/2},S_{3}&=&1/2\rangle =\sqrt{\frac{2}{3}}|1,-\frac{1}{2}\rangle-\sqrt{\frac{1}{3}}|0,\frac{1}{2}\rangle
\notag, \\
|^{4} S_{3/2},S_{3}&=&3/2\rangle =|1,\frac{1}{2}\rangle.  \label{VPP1V}
\end{eqnarray}%
The eigenvalues (two diagonal elements) of $\langle\mathbf{S}_{QQ^{\prime}}\cdot \mathbf{S}_{q}\rangle$ in the basis $[^{2}S_{1/2}, ^{4}S_{3/2}]$ can be given by
\begin{small}
\begin{eqnarray}
\langle\mathbf{S}_{QQ^{\prime}}\cdot \mathbf{S}_{q}\rangle &=& \left[
\begin{array}{cc}
\langle^{2} S_{1/2},S_{3}=1/2|\mathbf{S}_{QQ^{\prime}}\cdot \mathbf{S}_{q}|^{2} S_{1/2},S_{3}=1/2\rangle & \langle^{2} S_{1/2},S_{3}=1/2|\mathbf{S}_{QQ^{\prime}}\cdot \mathbf{S}_{q}|^{4} S_{3/2},S_{3}=3/2\rangle \\
\langle^{4} S_{3/2},S_{3}=3/2|\mathbf{S}_{QQ^{\prime}}\cdot \mathbf{S}_{q}|^{2} S_{1/2},S_{3}=1/2\rangle & \langle^{4} S_{3/2},S_{3}=3/2|\mathbf{S}_{QQ^{\prime}}\cdot \mathbf{S}_{q}|^{4} S_{3/2},S_{3}=3/2\rangle
\end{array} \notag
\right]\\
&=& \left[
\begin{array}{cc}
-1 & 0 \\
0 & \frac{1}{2} \label{pp8}
\end{array}
\right].
\end{eqnarray}
\end{small}

\section{$P$-wave}

For the $P$-wave ststes of the doubly heavy baryons, the expectation value of $\mathbf{L}\cdot \mathbf{S}$ in any coupling scheme is
\begin{eqnarray}
\langle\mathbf{L}\cdot \mathbf{S}\rangle = [J(J+1)-L(L+1)-S(S+1)]/2.
\end{eqnarray}
The calculation of the operator $\mathbf{L}\cdot\mathbf{S}_{i}$ $(i=QQ^{\prime}, q)$ with raising and lowering operator $L_{\pm}$, $S_{i\pm}$ results in
\begin{eqnarray}
\mathbf{L}\cdot\mathbf{S}_{i}=L_{3}S_{i3}+\left(L_{+}S_{i-}+L_{-}S_{i+}\right)/2.
\end{eqnarray}
In the $L$-$S$ basis can be constructed as linear combinations of the states $|S_{QQ^{\prime}3}, S_{q3}, L_{3}\rangle$ of the third components of the respective angular momenta. Thus, these five $P$-wave states of the doubly heavy baryons may be classified as $^{2S+1}P_{J}$ = $^{2}P_{1/2}$, $^{4}P_{1/2}$, $^{2}P_{3/2}$, $^{4}P_{3/2}$, $^{4}P_{5/2}$ with the third
components of the total angular momenta $J_{3}$,
\begin{eqnarray}
|^{2} P_{1/2},J_{3}&=&1/2\rangle =\frac{\sqrt{2}}{3}|1,-\frac{1}{2},0\rangle-\frac{1}{3}|0,\frac{1}{2},0\rangle-\frac{\sqrt{2}}{3}|0,-\frac{1}{2},1\rangle+\frac{2}{3}|-1,\frac{1}{2},1\rangle
\notag, \\
|^{4} P_{1/2},J_{3}&=&1/2\rangle =\frac{1}{\sqrt{2}}|1,\frac{1}{2},-1\rangle-\frac{1}{3}|1,-\frac{1}{2},0\rangle-\frac{\sqrt{2}}{3}|0,\frac{1}{2},0\rangle+\frac{1}{3}|0,-\frac{1}{2},1\rangle+\frac{1}{3\sqrt{{2}}}|-1,\frac{1}{2},1\rangle
\notag, \\
|^{2} P_{3/2},J_{3}&=&3/2\rangle =\sqrt{\frac{2}{3}}|1,-\frac{1}{2},1\rangle-\sqrt{\frac{1}{3}}|0,\frac{1}{2},1\rangle
\notag, \\
|^{4} P_{3/2},J_{3}&=&3/2\rangle =\sqrt{\frac{3}{5}}|1,\frac{1}{2},0\rangle-\sqrt{\frac{2}{15}}|1,-\frac{1}{2},1\rangle-\frac{2}{\sqrt{15}}|0,\frac{1}{2},1\rangle
\notag, \\
|^{4} P_{5/2},J_{3}&=&5/2\rangle =|1,\frac{1}{2},1\rangle.   \label{VV}
\end{eqnarray}%
The expectation values of $\langle\mathbf{L}\cdot \mathbf{S}_{i}\rangle$, $\langle S_{12}\rangle$ and $\langle\mathbf{S}_{QQ^{\prime}}\cdot \mathbf{S}_{q}\rangle$ are given by
\begin{eqnarray}
\langle\mathbf{L\cdot S}_{q}\rangle_{J=\frac{1}{2}} &=&\left[
\begin{array}{cc}
\frac{1}{3} & \frac{\sqrt{2}}{3} \\ \frac{\sqrt{2}}{3} & -\frac{5}{6}
\end{array}%
\right],
\langle\mathbf{L\cdot S}_{QQ^{\prime}}\rangle_{J=\frac{1}{2}} =\left[
\begin{array}{cc}
-\frac{4}{3}& -\frac{\sqrt{2}}{3} \\-\frac{\sqrt{2}}{3} & -\frac{5}{3}
\end{array}%
\right],
\langle S_{12}\rangle_{J=\frac{1}{2}} =\left[
\begin{array}{cc}
0 &\frac{1}{\sqrt{2}} \\ \frac{1}{\sqrt{2}} & -1%
\end{array}%
\right],\notag \\
&&\langle\mathbf{S}_{QQ^{\prime}}\cdot \mathbf{S}_{q}\rangle_{J=\frac{1}{2}} =\left[
\begin{array}{cc}
-1 &0 \\ 0 & \frac{1}{2}
\end{array}%
\right],\notag \\
\langle\mathbf{L\cdot S}_{q}\rangle_{J=\frac{3}{2}} &=& \left[
\begin{array}{cc}
-\frac{1}{6} &\frac{\sqrt{5}}{3} \\ \frac{\sqrt{5}}{3} & -\frac{1}{3}%
\end{array}%
\right],
\langle\mathbf{L\cdot S}_{QQ^{\prime}}\rangle_{J=\frac{3}{2}} = \left[
\begin{array}{cc}
\frac{2}{3} &-\frac{\sqrt{5}}{3} \\ -\frac{\sqrt{5}}{3} & -\frac{2}{3}%
\end{array}%
\right],
\langle S_{12}\rangle_{J=\frac{3}{2}} =\left[
\begin{array}{cc}
0 &-\frac{\sqrt{5}}{10} \\ -\frac{\sqrt{5}}{10} & \frac{4}{5}%
\end{array}%
\right],\notag\\
&&\langle\mathbf{S}_{QQ^{\prime}}\cdot \mathbf{S}_{q}\rangle_{J=\frac{3}{2}} =\left[
\begin{array}{cc}
-1 & 0 \\ 0 & \frac{1}{2}
\end{array}%
\right],\notag \\
\langle\mathbf{L\cdot S}_{q}\rangle_{J=\frac{5}{2}} &=& \frac{1}{2},\quad
\langle\mathbf{L\cdot S}_{QQ^{\prime}}\rangle_{J=\frac{5}{2}} = 1,\quad
\langle S_{12}\rangle_{J=\frac{5}{2}} =-\frac{1}{5},\quad
\langle\mathbf{S}_{QQ^{\prime}}\cdot \mathbf{S}_{q}\rangle_{J=\frac{5}{2}} =\frac{1}{2}.\quad
\label{In0}
\end{eqnarray}
The matrix forms of these mass shifts are
\begin{eqnarray}
\Delta \mathcal{M}_{J=1/2}&=&\left[
\begin{array}{cc}
\frac{1}{3}a_{1}-\frac{4}{3}a_{2}-c_{1} & \frac{\sqrt{2}}{3}a_{1}-\frac{\sqrt{2}}{3}a_{2}+\frac{1}{\sqrt{2}}b_{1} \\
\frac{\sqrt{2}}{3}a_{1}-\frac{\sqrt{2}}{3}a_{2}+\frac{1}{\sqrt{2}}b_{1} & -\frac{5}{6}a_{1}-\frac{5}{3}a_{2}-b_{1}+\frac{1}{2}c_{1}%
\end{array}%
\right], \notag \\
\Delta \mathcal{M}_{J=3/2}&=&\left[
\begin{array}{cc}
-\frac{1}{6}a_{1}+\frac{2}{3}a_{2}-c_{1} & \frac{\sqrt{5}}{3}a_{1}-\frac{\sqrt{5}}{3}a_{2}-\frac{\sqrt{5}}{10}b_{1} \\
\frac{\sqrt{5}}{3}a_{1}-\frac{\sqrt{5}}{3}a_{2}-\frac{\sqrt{5}}{10}b_{1} & -\frac{1}{3}a_{1}-\frac{2}{3}a_{2}+\frac{4}{5}b_{1}+\frac{1}{2}c_{1}%
\end{array}%
\right], \notag \\
\Delta \mathcal{M}_{J=5/2}&=&\frac{1}{2}a_{1}+a_{2}-\frac{1}{5}b_{1}+\frac{1}{2}c_{1}. \label{M5}
\end{eqnarray}
Diagonalizing the above matrices Eq. (\ref{M5}), we can obtain the mass shifts $\Delta M(J,j)$ with the $j=0, 1, 2$,
\begin{eqnarray}
\Delta M(1/2,0)&=&\frac{1}{4}\left(-a_{1}-6a_{2}-2b_{1}-c_{1}\right) \notag \\
               &-&\frac{1}{12}\sqrt{(9a_{1}-2a_{2}+10b_{1}-7c_{1})^{2}+8(2a_{2}-b_{1}-2c_{1})^{2}},  \notag \\
\Delta M(1/2,1)&=&\frac{1}{4}\left(-a_{1}-6a_{2}-2b_{1}-c_{1}\right) \notag \\
               &+&\frac{1}{12}\sqrt{(9a_{1}-2a_{2}+10b_{1}-7c_{1})^{2}+8(2a_{2}-b_{1}-2c_{1})^{2}},  \notag \\
\Delta M(3/2,1)&=&\frac{1}{20}\left(-5a_{1}+8b_{1}-5c_{1}\right) \notag \\
               &-&\frac{1}{60}\sqrt{(45a_{1}-40a_{2}-16b_{1}-5c_{1})^{2}+5(20a_{2}-10b_{1}-20c_{1})^{2}},  \notag \\
\Delta M(3/2,2)&=&\frac{1}{20}\left(-5a_{1}+8b_{1}-5c_{1}\right) \notag \\
               &+&\frac{1}{60}\sqrt{(45a_{1}-40a_{2}-16b_{1}-5c_{1})^{2}+5(20a_{2}-10b_{1}-20c_{1})^{2}},  \notag \\
\Delta M(5/2,2)&=&\frac{1}{2}a_{1}+a_{2}-\frac{1}{5}b_{1}+\frac{1}{2}c_{1}.  \label{MM121}
\end{eqnarray}%

\section{$D$-wave}

For analyzing the $D$-wave system, the relevant linear combinations of six basis states $^{2S+1}D_{J}$ = $^{4}D_{1/2}$, $^{2}D_{3/2}$, $^{4}D_{3/2}$, $^{2}D_{5/2}$, $^{4}D_{5/2}$, $^{4}D_{7/2}$ are given by
\begin{eqnarray}
|^{4} D_{1/2},J_{3}&=&1/2\rangle =\frac{1}{\sqrt{10}}|1,\frac{1}{2},-1\rangle-\frac{1}{\sqrt{15}}|1,-\frac{1}{2},0\rangle-\sqrt{\frac{2}{15}}|0,\frac{1}{2},0\rangle+\frac{1}{\sqrt{5}}|0,-\frac{1}{2},1\rangle+\frac{1}{\sqrt{10}}|-1,\frac{1}{2},1\rangle \notag \\
&-&\sqrt{\frac{2}{5}}|-1,-\frac{1}{2},2\rangle,
\notag \\
|^{2} D_{3/2},J_{3}&=&3/2\rangle =\sqrt{\frac{2}{15}}|1,-\frac{1}{2},1\rangle-\frac{1}{\sqrt{15}}|0,\frac{1}{2},1\rangle-\frac{2}{\sqrt{15}}|0,-\frac{1}{2},2\rangle+\sqrt{\frac{8}{15}}|-1,\frac{1}{2},2\rangle,
\notag \\
|^{4} D_{3/2},J_{3}&=&3/2\rangle =\frac{1}{\sqrt{5}}|1,\frac{1}{2},0\rangle-\sqrt{\frac{2}{15}}|1,\frac{1}{2},1\rangle-\frac{2}{\sqrt{15}}|0,\frac{1}{2},1\rangle+\frac{2}{\sqrt{15}}|0,-\frac{1}{2},2\rangle+\sqrt{\frac{2}{15}}|-1,\frac{1}{2},2\rangle,
\notag \\
|^{2} D_{5/2},J_{3}&=&5/2\rangle =\sqrt{\frac{2}{3}}|1,-\frac{1}{2},2\rangle-\sqrt{\frac{1}{3}}|0,\frac{1}{2},2\rangle,
\notag \\
|^{4} D_{5/2},J_{3}&=&5/2\rangle =\frac{3}{\sqrt{21}}|1,\frac{1}{2},1\rangle-\frac{2}{\sqrt{21}}|1,-\frac{1}{2},2\rangle-\frac{2\sqrt{2}}{\sqrt{21}}|0,\frac{1}{2},2\rangle,
\notag \\
|^{4} D_{7/2},J_{3}&=&7/2\rangle =|1,\frac{1}{2},2\rangle. \label{VV}
\end{eqnarray}%
The expectation values of $\langle\mathbf{L}\cdot \mathbf{S}_{i}\rangle$ $(i=QQ^{\prime}, q)$, $\langle S_{12}\rangle$ and $\langle\mathbf{S}_{QQ^{\prime}}\cdot \mathbf{S}_{q}\rangle$ are
\begin{eqnarray}
\langle\mathbf{L\cdot S}_{q}\rangle_{J=\frac{1}{2}} &=& -\frac{3}{2},\quad
\langle\mathbf{L\cdot S}_{QQ^{\prime}}\rangle_{J=\frac{1}{2}} = -3,\quad
\langle S_{12}\rangle_{J=\frac{1}{2}} =-1,\quad
\langle\mathbf{S}_{QQ^{\prime}}\cdot \mathbf{S}_{q}\rangle_{J=\frac{1}{2}} =\frac{1}{2},\quad \notag \\
\langle\mathbf{L\cdot S}_{q}\rangle_{J=\frac{3}{2}} &=& \left[
\begin{array}{cc}
\frac{1}{2} & 1 \\ 1 &  -1
\end{array}%
\right],
\langle\mathbf{L\cdot S}_{QQ^{\prime}}\rangle_{J=\frac{3}{2}} = \left[
\begin{array}{cc}
-2 & -1 \\-1 & -2
\end{array}%
\right],
\langle S_{12}\rangle_{J=\frac{3}{2}} =\left[
\begin{array}{cc}
0 &\frac{1}{2} \\ \frac{1}{2} &0%
\end{array}%
\right],\notag \\
&&\langle\mathbf{S}_{QQ^{\prime}}\cdot \mathbf{S}_{q}\rangle_{J=\frac{3}{2}} =\left[
\begin{array}{cc}
-1 &0 \\ 0 & \frac{1}{2}
\end{array}%
\right],\notag \\
\langle\mathbf{L\cdot S}_{q}\rangle_{J=\frac{5}{2}} &=& \left[
\begin{array}{cc}
-\frac{1}{3} &\frac{\sqrt{14}}{3} \\ \frac{\sqrt{14}}{3} & -\frac{1}{6}%
\end{array}%
\right],
\langle\mathbf{L\cdot S}_{QQ^{\prime}}\rangle_{J=\frac{5}{2}} = \left[
\begin{array}{cc}
\frac{4}{3} &-\frac{\sqrt{14}}{3} \\ -\frac{\sqrt{14}}{3} & -\frac{1}{3}%
\end{array}%
\right],
\langle S_{12}\rangle_{J=\frac{5}{2}} =\left[
\begin{array}{cc}
0 &-\frac{\sqrt{14}}{14} \\ -\frac{\sqrt{14}}{14} & \frac{5}{7}%
\end{array}%
\right],\notag\\
&&\langle\mathbf{S}_{QQ^{\prime}}\cdot \mathbf{S}_{q}\rangle_{J=\frac{5}{2}} =\left[
\begin{array}{cc}
-1 & 0 \\ 0 & \frac{1}{2}
\end{array}%
\right],\notag \\
\langle\mathbf{L\cdot S}_{q}\rangle_{J=\frac{7}{2}} &=& 1,\quad
\langle\mathbf{L\cdot S}_{QQ^{\prime}}\rangle_{J=\frac{7}{2}} = 2,\quad
\langle S_{12}\rangle_{J=\frac{7}{2}} =-\frac{2}{7},\quad
\langle\mathbf{S}_{QQ^{\prime}}\cdot \mathbf{S}_{q}\rangle_{J=\frac{7}{2}} =\frac{1}{2}.\quad \label{In0}
\end{eqnarray}%
The matrix forms of these mass shifts are
\begin{eqnarray}
\Delta \mathcal{M}_{J=1/2}&=&-\frac{3}{2}a_{1}-3a_{2} -
b_{1}+\frac{1}{2}c_{1},    \notag \\
\Delta \mathcal{M}_{J=3/2} &=&\left[
\begin{array}{cc}
\frac{1}{2}a_{1}-2a_{2}-c_{1} & a_{1}-a_{2}+\frac{1}{2}b_{1} \\
a_{1}-a_{2}+\frac{1}{2}b_{1} & -a_{1}-2a_{2}+\frac{1}{2}c_{1}%
\end{array}%
\right],\notag \\
\Delta \mathcal{M}_{J=5/2} &=&\left[
\begin{array}{cc}
-\frac{1}{3}a_{1}+\frac{4}{3}a_{2}-c_{1}  & \frac{\sqrt{14}}{3}a_{1}-\frac{\sqrt{14}}{3}a_{2}-\frac{\sqrt{14}}{14}b_{1} \\
\frac{\sqrt{14}}{3}a_{1}-\frac{\sqrt{14}}{3}a_{2}-\frac{\sqrt{14}}{14}b_{1} & -\frac{1}{6}a_{1}-\frac{1}{3}a_{2}+\frac{5}{7}b_{1}+\frac{1}{2}c_{1}
\end{array}%
\right], \notag \\ \label{M3}
\Delta \mathcal{M}_{J=7/2}&=&a_{1}+2a_{2}-\frac{2}{7}b_{1}+\frac{1}{2}c_{1}. \label{Mm5}
\end{eqnarray}%
Diagonalizing the above matrices Eq. (\ref{Mm5}), we can obtain the six mass shifts $\Delta M(J, j)$ with $j=1, 2, 3$,
\begin{eqnarray}
\Delta M(1/2,1)&=&-\frac{3}{2}a_{1}-3a_{2}-b_{1}+\frac{1}{2}c_{1},    \notag \\
\Delta M(3/2,1)&=&\frac{1}{4}\left(-a_{1}-8a_{2}-c_{1}\right) \notag \\
               &-&\frac{1}{20}\sqrt{(25a_{1}-16a_{2}+8b_{1}-9c_{1})^{2}+36(2a_{2}-b_{1}-2c_{1})^{2}}, \notag \\
\Delta M(3/2,2)&=&\frac{1}{4}\left(-a_{1}-8a_{2}-c_{1}\right) \notag \\
               &+&\frac{1}{20}\sqrt{(25a_{1}-16a_{2}+8b_{1}-9c_{1})^{2}+36(2a_{2}-b_{1}-2c_{1})^{2}}, \notag \\
\Delta M(5/2,2)&=&\frac{1}{28}\left(-7a_{1}+14a_{2}+10b_{1}-7c_{1}\right) \notag \\
               &-&\frac{1}{140}\sqrt{(175a_{1}-182a_{2}-34b_{1}+7c_{1})^{2}+2744(2a_{2}-b_{1}-2c_{1})^{2}},  \notag \\
\Delta M(5/2,3)&=&\frac{1}{28}\left(-7a_{1}+14a_{2}+10b_{1}-7c_{1}\right) \notag \\
               &+&\frac{1}{140}\sqrt{(175a_{1}-182a_{2}-34b_{1}+7c_{1})^{2}+2744(2a_{2}-b_{1}-2c_{1})^{2}},  \notag \\
\Delta M(7/2,3)&=&a_{1}+2a_{2}-\frac{2}{7}b_{1}+\frac{c_{1}}{2}.  \label{MM212}
\end{eqnarray}%

\end{document}